\documentclass[12pt]{article}

\usepackage[T1]{fontenc}
\usepackage[utf8]{inputenc}
\usepackage{amsmath,amssymb,mathtools}
\usepackage{amsthm}
\usepackage{graphicx}
\usepackage{xcolor}
\usepackage{cite}
\usepackage{enumitem}
\usepackage{booktabs}
\usepackage{tikz}
\usepackage{pgfplots}
\pgfplotsset{compat=1.18}
\usepackage[colorlinks=true,allcolors=black]{hyperref}

\theoremstyle{plain}
\newtheorem{theorem}{Theorem}
\newtheorem{proposition}{Proposition}
\newtheorem{lemma}{Lemma}
\newtheorem{corollary}{Corollary}
\theoremstyle{definition}
\newtheorem{definition}{Definition}
\newtheorem{remark}{Remark}

\newcommand{\tr}{\operatorname{Tr}}
\newcommand{\spn}{\operatorname{span}}
\newcommand{\ran}{\operatorname{ran}}
\newcommand{\diag}{\operatorname{diag}}
\newcommand{\cl}{\operatorname{cl}}
\newcommand{\bbC}{\mathbb{C}}
\newcommand{\bbR}{\mathbb{R}}
\newcommand{\one}{\mathbf{1}}
\newcommand{\rhon}{\rho^{\otimes n}}
\newcommand{\sigman}{\sigma^{\otimes n}}
\newcommand{\cA}{\mathcal{A}}
\newcommand{\cZ}{\mathcal{Z}}
\newcommand{\cR}{\mathcal{R}}
\newcommand{\cH}{\mathcal{H}}
\newcommand{\cD}{\mathcal{D}}
\newcommand{\cS}{\mathcal{S}}
\newcommand{\cC}{\mathcal{C}}
\newcommand{\cX}{\mathcal{X}}
\newcommand{\cV}{\mathcal{V}}
\newcommand{\cY}{\mathcal{Y}}
\newcommand{\fA}{\mathfrak{A}}
\newcommand{\bv}{\mathbf{v}}
\newcommand{\bn}{\mathbf{n}}
\newcommand{\bs}{\mathbf{s}}
\newcommand{\drho}{d_{\rho}}
\newcommand{\dsig}{d_{\sigma}}
\newcommand{\pos}[1]{\left(#1\right)_{+}}
\newcommand{\abs}[1]{\left|#1\right|}
\newcommand{\norm}[1]{\left\|#1\right\|}
\newcommand{\ip}[2]{\left\langle #1,#2\right\rangle}
\newcommand{\ket}[1]{|#1\rangle}
\newcommand{\bra}[1]{\langle #1|}
\newcommand{\ketbra}[1]{|#1\rangle\langle #1|}
\newcommand{\lmax}{\ell_{\max}}
\newcommand{\Dsand}{\widetilde{D}}
\newcommand{\eqdef}{\coloneqq}
\newcommand{\Pet}{\mathrm{P}}
\newcommand{\Fm}{\mathsf{F}}
\newcommand{\Gm}{\mathsf{G}}

\allowdisplaybreaks[2]
\begin{document}

\title{Binary Quantum Hypothesis Testing with a Rejection Option: The Exact Error-Exponent Trade-off}

\author{Yonglong~Li\\[0.4em]\normalsize School of Mathematics and Statistics, Xi'an Jiaotong University\\\normalsize\href{mailto:liyonglong@xjtu.edu.cn}{liyonglong@xjtu.edu.cn}}

\date{}

\maketitle

\begin{abstract}
We characterize the exact error-exponent trade-off in binary quantum hypothesis testing with a rejection option for faithful states on a finite-dimensional Hilbert space. The probabilities of rejection under the two hypotheses are required to vanish exponentially, with exponents at least $K$ and $L$. We show that a quadruple $(a,b,K,L)$ of positive error and rejection exponents is achievable if and only if it satisfies three single-letter conditions: two conditions involve Petz--R\'enyi quantities and coincide with the converse bound conjectured to be tight by Ji and Regula; the third involves the $\alpha$--$z$ R\'enyi quantities on the triangle $\max\{\alpha,1-\alpha\}\le z\le1$. The proof combines a sector decomposition inside Schur--Weyl blocks, an explicit one-shot measurement construction, and an exact minimax argument.

\end{abstract}

\noindent\textbf{Keywords.}
Quantum hypothesis testing, rejection option, inconclusive outcome, error exponents, Hoeffding bound, $\alpha$--$z$ R\'enyi divergence, Schur--Weyl duality.
\par

\section{Introduction}\label{sec:intro}

\subsection{Background}

In binary quantum hypothesis testing one is given $n$ copies of a quantum system prepared either in the state $\rho$ (null hypothesis) or in the state $\sigma$ (alternative hypothesis), and has to decide between $\rhon$ and $\sigman$ by means of a two-outcome measurement. For two simple i.i.d.\ hypotheses, the optimal performance is known in several distinct asymptotic regimes. The quantum Stein lemma identifies the optimal type-II error exponent at a fixed type-I error tolerance in $(0,1)$ \cite{HiaiPetz91,OgawaNagaoka00}; its strong converse shows that exceeding this exponent forces the type-I error to one \cite{OgawaNagaoka00}. The quantum Chernoff bound characterizes the optimal exponent of the average error for fixed positive priors \cite{Audenaert07,NussbaumSzkola09}, while the quantum Hoeffding bound gives the asymmetric trade-off when both errors vanish exponentially \cite{Hayashi07,Nagaoka06,ANSV08}, extending the classical results of \cite{Hoeffding65,CsiszarLongo71,Blahut74}. The strong converse exponent quantifies the decay of the probability of correctly accepting the null above the Stein threshold \cite{MosonyiOgawa15}. At fixed type-I error tolerance, the optimal negative logarithm of the type-II error has leading term $nD(\rho\|\sigma)$ and a Gaussian correction of order $\sqrt n$, governed by the relative-entropy variance \cite{Li14,TomamichelHayashi13}.

The exponentially vanishing-error regime is the one extended in this paper. For binary tests, the Hoeffding bound states that a positive exponent pair $(a,b)$ is achievable if and only if
\begin{equation}\label{eq:intro-hoeffding}
 s\,a+(1-s)\,b\;\le\;-\log\tr\rho^{s}\sigma^{1-s}\qquad\text{for some } s\in[0,1].
\end{equation}
This trade-off is governed by the Petz--R\'enyi divergences of orders below one \cite{Petz86}, whereas the sandwiched R\'enyi divergences \cite{MuellerLennert13,WWY14} of orders above one govern the strong converse exponent \cite{MosonyiOgawa15}. Our converse excludes unattainable quadruples of positive exponents; it does not assert a strong converse in which an error probability tends to one, and our polynomial finite-blocklength factor is not a second-order expansion.

In many situations a third decision is natural: the tester may \emph{reject} both hypotheses, i.e., abstain, declare an erasure, or return an inconclusive answer. In classical statistics and information theory, decision rules with such an option go back at least to Chow's error--reject trade-off in pattern recognition \cite{Chow70} and to Forney's analysis of erasure decoding \cite{Forney68}, and hypothesis testing with a rejection option has been studied, e.g., in \cite{Nikulin89,Gutman89,Grigoryan11,Sason12,LalithaJavidi16}. In quantum theory an inconclusive outcome is the defining feature of unambiguous state discrimination \cite{Ivanovic87,Dieks88,Peres88} and of its relaxations, in which a fixed probability of inconclusive outcomes is traded against the error probability \cite{CheflesBarnett98,FiurasekJezek03,Eldar03,Bagan12,Herzog12}; see also \cite{RST03,RaynalLutkenhaus05,Zhuang20} and the survey \cite{Chefles00}. Recent systematic asymptotic analyses include Regula, Lami and Wilde \cite{RLW24}, who study postselected testing without the exponential rejection constraints imposed here, and Ji and Regula \cite{JiRegula25}, who show that an inconclusive outcome allows one to beat the Chernoff and Hoeffding trade-offs even when its probability vanishes exponentially.

The present paper is concerned with the regime that \cite{JiRegula25} calls \emph{exponentially low inconclusiveness}: the probability of rejection is required to vanish exponentially under both hypotheses, with exponents at least $K$ under $\rho$ and at least $L$ under $\sigma$, and one asks for the optimal trade-off between the exponents $a$ and $b$ of the two error probabilities. For classical hypotheses, i.e., probability distributions $P$ and $Q$, the answer is known \cite[Lemma~5]{JiRegula25} (see also \cite{Grigoryan11,LalithaJavidi16} for related rejection formulations): with the Hoeffding function $H_r(P\|Q)\eqdef\sup_{0<s<1}\frac{s-1}{s}\bigl(r-D_s(P\|Q)\bigr)$, where $D_s$ is the R\'enyi divergence of order $s$, the quadruple $(a,b,K,L)$ is achievable if and only if
\begin{equation}\label{eq:intro-classical}
 a\le\max\{H_b(P\|Q),H_L(P\|Q)\},\qquad b\le\max\{H_a(Q\|P),H_K(Q\|P)\}.
\end{equation}
For quantum states, Ji and Regula proved that \eqref{eq:intro-classical}, with the Hoeffding functions defined through the Petz--R\'enyi divergences, is necessary \cite[Proposition~S17]{JiRegula25}; we refer to this bound as the \emph{Petz-type converse}. They also obtained an achievability result by asymptotic pinching, in which the Petz--R\'enyi divergences are replaced by sandwiched and reverse sandwiched ones \cite[Proposition~6]{JiRegula25}. They conjectured that the Petz-type converse is tight \cite[Conjecture~S18]{JiRegula25}, while pointing out that the truth might instead be ``governed by a combination of Petz--R\'enyi and sandwiched R\'enyi divergences'', and that a complete description of this regime ``appears to require different techniques''.

\subsection{Contributions}

We solve this problem for faithful states on a finite-dimensional Hilbert space. Let
\begin{equation}\label{eq:intro-psi}
 \psi(\alpha,z)\eqdef-\log\tr\Bigl(\rho^{\frac{\alpha}{2z}}\sigma^{\frac{1-\alpha}{z}}\rho^{\frac{\alpha}{2z}}\Bigr)^{z}
\end{equation}
be the negative logarithm of the $\alpha$--$z$ R\'enyi quantity of \cite{JOPP12,AudenaertDatta15}. Then $\psi(s,1)$ is the Petz quantity appearing in \eqref{eq:intro-hoeffding}, $\psi(s,s)$ and $\psi(1-s,s)$ are the sandwiched quantities of the pairs $(\rho,\sigma)$ and $(\sigma,\rho)$, and $\psi(\frac12,\frac12)=-\log F(\rho,\sigma)$ with $F$ the fidelity. Our main result (Theorem~\ref{thm:main}) states that a quadruple $(a,b,K,L)$ of positive numbers is achievable if and only if it satisfies the following three conditions:
\begin{align*}
&\text{(P1)}\quad s\min\{a,K\}+(1-s)\,b\le\psi(s,1)\ \text{ for some } s\in[0,1];\\
&\text{(P2)}\quad s\,a+(1-s)\min\{b,L\}\le\psi(s,1)\ \text{ for some } s\in[0,1];\\
&\text{(F)}\quad\ \,(\alpha+z-1)\,a+(z-\alpha)\,b+(1-z)(K+L)\le\psi(\alpha,z)\\
&\phantom{\text{(F)}\quad\ \,}\text{for some } (\alpha,z) \text{ with } \max\{\alpha,1-\alpha\}\le z\le1 .
\end{align*}
Conditions (P1) and (P2) are equivalent to the Petz-type converse \eqref{eq:intro-classical}. Condition (F) is new; the parameter set over which it is formulated is the triangle with vertices $(0,1)$, $(1,1)$ and $(\frac12,\frac12)$ in the $(\alpha,z)$-plane, whose edges are the Petz line $z=1$, the sandwiched line $z=\alpha$ and the reverse sandwiched line $z=1-\alpha$ (Fig.~\ref{fig:triangle}). 

Theorem~\ref{thm:fixed} treats the formulation in which the rejection probabilities must not exceed $e^{-nK}$ and $e^{-nL}$ for every $n$, and shows that the resulting region coincides with the corresponding slice of the set described by these three conditions unless $K+L=-2\log F(\rho,\sigma)$.

 The main consequences are the following.
\begin{enumerate}[leftmargin=*,label=\arabic*)]
\item \emph{Status of the conjecture of \cite{JiRegula25}.} At the vertex $(\frac12,\frac12)$ condition (F) reads $K+L\le-2\log F(\rho,\sigma)=\Dsand_{1/2}(\rho\|\sigma)$. Hence the Petz-type converse is tight whenever the sum $K+L$ of the two rejection exponents does not exceed the sandwiched R\'enyi divergence of order $\frac12$ (Remark~\ref{rem:petz}). Above this threshold it is in general not tight, and (F) is the correction: for a qubit pair we exhibit quadruples that satisfy \eqref{eq:intro-classical} but are not achievable (Section~\ref{sec:example}).
\item \emph{An operational role for $\alpha$--$z$ R\'enyi divergences.} The whole two-parameter family $\{Q_{\alpha,z}:\max\{\alpha,1-\alpha\}\le z\le1\}$ enters the answer, and in the example the optimum is attained at interior points of the triangle, where $Q_{\alpha,z}$ is neither a Petz nor a (reverse) sandwiched quantity. For $\alpha\in(0,1)$ this triangle is precisely the part of the data-processing region of the $\alpha$--$z$ divergences \cite{Zhang20} with $z\le1$. 
\item \emph{A sector interpretation.} In an equivalent form of the theorem (Theorem~\ref{thm:pair}) the three conditions are replaced by one inequality that must hold for every point $(\alpha,z)$ of the triangle and that involves the gradient of $\psi$. This inequality is the condition, on the exponential scale, for the existence of a test with rejection for a single pair of pure states, namely a pair of eigenvectors of $\rhon$ and $\sigman$ inside a Schur--Weyl block.
\end{enumerate}

\subsection{Techniques}
The proof has three main ingredients.

\emph{(i) Sectors.} Schur--Weyl duality reduces $\rhon$ and $\sigman$ simultaneously to polynomially many matrix blocks of polynomial dimension, each accompanied by a multiplicity factor. Inside a block we consider pairs consisting of an eigenspace of $\rhon$ and an eigenspace of $\sigman$, and attach to each pair three exponents: the two probability exponents $\drho,\dsig$ and the exponent $\chi$ of the largest overlap between unit vectors of the two eigenspaces. This refines the classical-like picture of Nussbaum and Szko\l a \cite{NussbaumSzkola09}, in which a pair of eigenvectors is weighted by its squared overlap, by keeping the overlap as a separate coordinate. We show that $\psi(\alpha,z)$ is bounded above by $\alpha\,\drho+(1-\alpha)\,\dsig+2z\,\chi$ for every pair and every $n$, and that it equals the minimum of this expression over all pairs up to $O(n^{-1}\log n)$ (Lemma~\ref{lem:support}); in other words, $-\psi$ is the support function of the set of limiting sector exponents, evaluated in the directions $-(\alpha,1-\alpha,2z)$. An immediate by-product is a proof of the joint concavity of $\psi$ in $(\alpha,z)$; this property also follows from a log-convexity theorem of Bourin and Lee \cite{BourinLee16} (see the discussion following Proposition~\ref{prop:psi}).

\emph{(ii) Convex geometry.} The sector constraints depend on the three coordinates $\drho+2\chi$, $\dsig+2\chi$ and $\drho+\dsig+2\chi$. The set of transformed sectors, enlarged by the nonnegative orthant, is a closed convex upper image (Proposition~\ref{prop:upper-image}). This is proved by an exact minimax identity whose essential input is the continuous family of unique minimizing sectors supplied by the gradient of $\psi$ (Lemmas~\ref{lem:exposed} and~\ref{lem:separation}). Violations of the pair condition become three downward open boxes. Excluding these boxes by supporting half-spaces gives (P1), (P2), and (F); every violation has an exposed-sector witness, which gives Theorem~\ref{thm:pair}. These geometric statements are proved together in Section~\ref{sec:sectors}.

\emph{(iii) One-shot lemmas for two orthonormal bases.} For the converse we only need the monotonicity of the fidelity under measurements \cite{FuchsCaves95}, applied to a pair of eigenvectors: the overlap of the pair is bounded by the sum of three terms, which are controlled by the type-I error probability, by the type-II error probability, and by the product of the two rejection probabilities. For the achievability we prove a new lemma (Lemma~\ref{lem:suff}): given two orthonormal bases of $\bbC^D$ and, for every basis vector, a budget for the wrong decision and a budget for rejection, with the former not exceeding the latter, a three-outcome measurement meeting all budgets exists as soon as, for every pair of basis vectors such that the rejection budget of one of them and the error budget of the other are smaller than one, the squared overlap is at most $1/(400D^4)$ times the largest of the three corresponding quantities (the two error budgets and the product of the two rejection budgets); no condition is needed for the other pairs. The construction is explicit. It is related in spirit to the reductions of multiple to pairwise state discrimination in \cite{Li16,ChengLiu26,LeeLami26}, but has to accommodate rejection budgets, which enter through products of two probabilities.

In the achievability direction, the finite-blocklength support bound already implies the required sector inequalities for every $n$ (Lemma~\ref{lem:sector-ineq}). A shift of order $n^{-1}\log n$ then supplies the polynomial margin needed by the one-shot construction.

Since binary tests are contained in the theorem as the case $K\ge a$ or $L\ge b$, the argument also gives an independent proof of the quantum Hoeffding bound \cite{Hayashi07,Nagaoka06,ANSV08} for faithful states, including a non-asymptotic achievability bound; in this binary case, the known bound obtained from the trace inequality of \cite{Audenaert07} is sharper by a polynomial factor.

\subsection{Further related work}

Quantum hypothesis testing also extends beyond two simple i.i.d.\ states. Composite i.i.d.\ hypotheses can lead to regularized relative-entropy formulas \cite{BertaBrandaoHirche21}; correlated-state extensions require structural assumptions, such as factorization for error exponents \cite{HiaiMosonyiOgawa08} or analytic regularity for second-order asymptotics \cite{DattaPautratRouze16}. A relative-entropy chain rule characterizes the asymptotic asymmetric discrimination rate of quantum channels through a regularized channel relative entropy, and shows that adaptive strategies do not improve that rate over nonadaptive ones \cite{FangFawziRennerSutter20}.

The generalized quantum Stein lemma concerns an i.i.d.\ null state against a family of alternative states and identifies a regularized relative-entropy exponent. Following the identification of a gap in the original proof of Brand\~ao and Plenio \cite{BrandaoPlenio10,BertaEtAl23}, complete proofs were given by Lami \cite{Lami25} and by Hayashi and Yamasaki \cite{HayashiYamasaki25}. The latter work needs fewer closure assumptions on the alternative family; in particular, its state theorem does not require closure under partial trace or permutations. Generalized quantum asymptotic equipartition also gives set-to-set testing results under a different set of structural assumptions \cite{FangFawziFawzi26}. Recent preprints investigate robustness to almost-i.i.d.\ null states in entanglement testing \cite{MazzolaSutterRenner26} and a generalized Stein lemma on von Neumann algebras \cite{Gao26}. These results provide context for the scope of quantum testing theory; they concern different constraints from the four simultaneously prescribed error and rejection exponents studied here.

The $\alpha$--$z$ R\'enyi relative entropies were introduced in \cite{JOPP12,AudenaertDatta15}; their data-processing region was determined in \cite{Zhang20} (see also \cite{CFL18}), and further properties can be found in \cite{LinTomamichel15,MosonyiHiai24}. Operational interpretations in which $z$ is not tied to $\alpha$ by the Petz or sandwiched relations are recent: Verhagen, Tomamichel and Haapasalo \cite{VTH26} give $\alpha$--$z$ criteria for large-sample and catalytic conversions of pairs of flat states, including characterizations in approximated settings; Berta, Cheng, Rubboli and Tomamichel \cite{BCRT26} express the strong converse exponent for the entanglement cost of quantum state merging through optimized $\alpha$--$z$ conditional R\'enyi entropies with $z=\alpha/2$; and Hayashi and Fang \cite{HayashiFang26} show that reverse sandwiched divergences determine the Hoeffding exponent of a composite thermal-state versus phase-corrupted-state problem through a variational formula. Sandwiched divergences of orders $\alpha\in(\frac12,1)$ also govern strong converse exponents for smoothing the max-relative entropy, privacy amplification, and decoupling, when performance is measured by fidelity or purified distance \cite{LiYao24}. Representation-theoretic block decompositions have been used in quantum hypothesis testing before, e.g.\ in \cite{Hayashi02,HMH09}, and one-shot pairwise bounds for binary and multiple hypothesis testing can be found in \cite{Audenaert07,AudenaertMosonyi14,Li16,ChengLiu26,LeeLami26}.

\subsection{Organization and notation}

In Section~\ref{sec:main} we state the problem, the main results, their consequences, and a qubit example. In Section~\ref{sec:sectors} we develop the sector support representation, prove the exact minimax identity and the convex upper-image description, and conclude with the proof of Theorem~\ref{thm:pair}. In Section~\ref{sec:oneshot} we present the two one-shot measurement lemmas. In Section~\ref{sec:proofs} we use the finite-blocklength sector bound to prove achievability, derive the converse, and prove Theorem~\ref{thm:fixed} and the symmetric-pair specialization. In Section~\ref{sec:discussion} we discuss the resulting structure and open questions. The appendix proves Lemma~\ref{lem:Hform}; the commuting reduction and the numerical comparisons appear in Remark~\ref{rem:commuting} and Section~\ref{sec:example}, respectively.

Logarithms are natural. For $y\in\bbR$ we write $y_+\eqdef\max\{y,0\}$ and adopt the convention $-\log0=+\infty$. The inner product $\ip{u}{v}$ of $\bbC^{D}$ is linear in the second argument; $\norm{v}$ is the Euclidean norm of a vector, and $\norm{A}_\infty$, $\norm{A}_2$ and $\norm{A}_p=(\tr|A|^p)^{1/p}$ are the operator norm, the Hilbert--Schmidt norm and the Schatten $p$-(quasi)norm of a matrix, where $|A|=(A^*A)^{1/2}$ and $p>0$. $M_D$ is the algebra of complex $D\times D$ matrices and $\one_m$ the identity of $M_m$. For Hermitian matrices, $A\le B$ means that $B-A$ is positive semidefinite. A POVM is a tuple of positive semidefinite matrices summing to the identity. Closures of subsets of $(0,\infty)^k$ are taken relative to $(0,\infty)^k$.

\section{Problem Formulation and Main Results}\label{sec:main}

\subsection{Tests with a rejection option}

Throughout the whole paper the states $\rho$ and $\sigma$ are density matrices on $\bbC^d$, $d\ge2$, that are \emph{faithful} (positive definite) and distinct. The null hypothesis is that $n$ copies of the system are in the state $\rhon$, the alternative hypothesis is that they are in the state $\sigman$.

\begin{definition}[Tests]\label{def:test}
An \emph{$n$-copy test with a rejection option} is a POVM $T_n=(\Pi_{0},\Pi_{1},\Pi_{2})$ on $(\bbC^d)^{\otimes n}$. Outcome $0$ means that $\rho$ is accepted, outcome $1$ that $\sigma$ is accepted, and outcome $2$ that both hypotheses are rejected. The \emph{type-I} and \emph{type-II error probabilities} and the two \emph{rejection probabilities} of $T_n$ are
\begin{equation}\label{eq:four-prob}
\begin{aligned}
 \alpha_n(T_n)&\eqdef\tr\bigl[\rhon\,\Pi_{1}\bigr], &\qquad \beta_n(T_n)&\eqdef\tr\bigl[\sigman\,\Pi_{0}\bigr],\\
 u_n(T_n)&\eqdef\tr\bigl[\rhon\,\Pi_{2}\bigr], &\qquad v_n(T_n)&\eqdef\tr\bigl[\sigman\,\Pi_{2}\bigr].
\end{aligned}
\end{equation}
A test is \emph{binary} if $\Pi_{2}=0$.
\end{definition}

\begin{definition}[Achievable exponents]\label{def:achievable}
A quadruple $(a,b,K,L)\in(0,\infty)^4$ is \emph{achievable} if there is a sequence of tests $(T_n)_{n\ge1}$ such that
\begin{equation}\label{eq:achievable}
\begin{aligned}
 \liminf_{n\to\infty}-\frac1n\log\alpha_n(T_n)&\ge a, &\qquad \liminf_{n\to\infty}-\frac1n\log\beta_n(T_n)&\ge b,\\
 \liminf_{n\to\infty}-\frac1n\log u_n(T_n)&\ge K, &\qquad \liminf_{n\to\infty}-\frac1n\log v_n(T_n)&\ge L .
\end{aligned}
\end{equation}
The set of achievable quadruples is denoted by $\cA$. For $K,L>0$, the \emph{exponent region with rejection exponents $(K,L)$} is the set $\cR(K,L)$ of all pairs $(a,b)\in(0,\infty)^2$ for which there is a sequence of tests with
\begin{equation}\label{eq:hard}
 u_n(T_n)\le e^{-nK}\quad\text{and}\quad v_n(T_n)\le e^{-nL}\quad\text{for all } n\ge1,
\end{equation}
and with $\liminf_{n}-\frac1n\log\alpha_n(T_n)\ge a$ and $\liminf_{n}-\frac1n\log\beta_n(T_n)\ge b$.
\end{definition}

The region $\cR(K,L)$ describes the trade-off between the two error exponents when the rejection probabilities are constrained as in \eqref{eq:hard} for every $n$. In the set $\cA$ the rejection constraints are imposed on the exponential scale only. Since \eqref{eq:hard} implies the second line of \eqref{eq:achievable},
\begin{equation}\label{eq:R-in-A}
 (a,b)\in\cR(K,L)\quad\Longrightarrow\quad(a,b,K,L)\in\cA ,
\end{equation}
and both $\cA$ and $\cR(K,L)$ are decreasing sets: if a point belongs to the set, so does every point with componentwise smaller positive coordinates.

In \cite{JiRegula25} the performance of a test is measured by the error probabilities conditioned on a conclusive outcome, $\alpha_n/(1-u_n)$ and $\beta_n/(1-v_n)$. Since $u_n,v_n\to0$ for every sequence satisfying \eqref{eq:achievable}, these have the same exponents as $\alpha_n$ and $\beta_n$, so that $\cA$ coincides with the set of quadruples for which there are tests satisfying the asymptotic conditions in the hypothesis of \cite[Proposition~S17]{JiRegula25}.

\subsection{The \texorpdfstring{$\alpha$--$z$}{alpha-z} R\'enyi quantities}

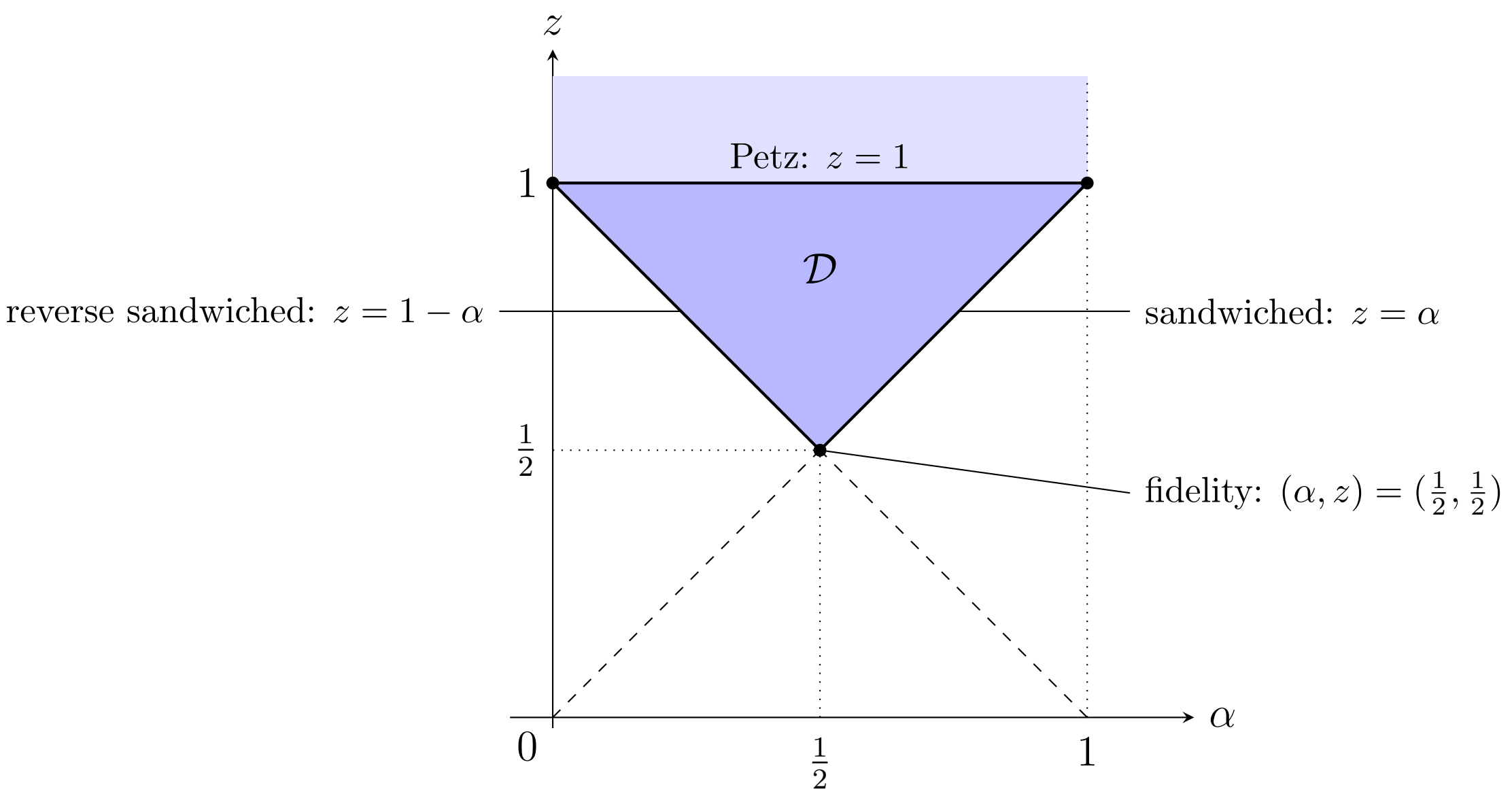
\begin{figure}[t]
\centering
\begin{tikzpicture}[scale=5.2,>=stealth]
 \draw[->] (-0.08,0)--(1.2,0) node[right]{$\alpha$};
 \draw[->] (0,-0.02)--(0,1.25) node[above]{$z$};
 \fill[blue!12] (0,1)--(1,1)--(1,1.2)--(0,1.2)--cycle;
 \fill[blue!28] (0,1)--(1,1)--(0.5,0.5)--cycle;
 \draw[thick] (0,1)--(1,1);
 \draw[thick] (0.5,0.5)--(1,1);
 \draw[thick] (0.5,0.5)--(0,1);
 \draw[dashed] (0,0)--(0.5,0.5);
 \draw[dashed] (1,0)--(0.5,0.5);
 \draw[dotted] (1,0)--(1,1.2);
 \draw[dotted] (0,0.5)--(0.5,0.5);
 \draw[dotted] (0.5,0)--(0.5,0.5);
 \fill (0.5,0.5) circle (0.012);
 \fill (0,1) circle (0.012);
 \fill (1,1) circle (0.012);
 \node[below] at (0.5,-0.01) {$\frac12$};
 \node[below] at (1,-0.01) {$1$};
 \node[left] at (0,1) {$1$};
 \node[left] at (0,0.5) {$\frac12$};
 \node[below left] at (0,0) {$0$};
 \node at (0.5,0.84) {$\cD$};
 \node[above] at (0.5,1.0) {\footnotesize Petz: $z=1$};
 \draw[thin] (0.76,0.76)--(1.08,0.76);
 \node[right] at (1.08,0.76) {\footnotesize sandwiched: $z=\alpha$};
 \draw[thin] (0.24,0.76)--(-0.10,0.76);
 \node[left] at (-0.10,0.76) {\footnotesize reverse sandwiched: $z=1-\alpha$};
 \draw[thin] (0.5,0.5)--(1.08,0.42);
 \node[right] at (1.08,0.42) {\footnotesize fidelity: $(\alpha,z)=(\frac12,\frac12)$};
\end{tikzpicture}
\caption{The triangle $\cD$ of \eqref{eq:triangle} in the $(\alpha,z)$-plane. For $0<\alpha<1$ the $\alpha$--$z$ R\'enyi divergence satisfies the data-processing inequality if and only if $z\ge\max\{\alpha,1-\alpha\}$ \cite{Zhang20} (shaded); $\cD$ (dark) is the part of this region with $z\le1$.}
\label{fig:triangle}
\end{figure}

For $\alpha\in\bbR$ and $z>0$ let
\begin{equation}\label{eq:Qaz}
 Q_{\alpha,z}(\rho\|\sigma)\eqdef\tr\Bigl(\rho^{\frac{\alpha}{2z}}\sigma^{\frac{1-\alpha}{z}}\rho^{\frac{\alpha}{2z}}\Bigr)^{z},
 \qquad
 \psi(\alpha,z)\eqdef-\log Q_{\alpha,z}(\rho\|\sigma).
\end{equation}
For $\alpha\ne1$, $D_{\alpha,z}(\rho\|\sigma)=\frac{1}{\alpha-1}\log Q_{\alpha,z}(\rho\|\sigma)$ is the $\alpha$--$z$ R\'enyi relative entropy \cite{JOPP12,AudenaertDatta15}, so that $\psi(\alpha,z)=(1-\alpha)D_{\alpha,z}(\rho\|\sigma)$. With $R\eqdef\rho^{\frac{\alpha}{2z}}\sigma^{\frac{1-\alpha}{2z}}$ we have $RR^*=\rho^{\frac{\alpha}{2z}}\sigma^{\frac{1-\alpha}{z}}\rho^{\frac{\alpha}{2z}}$, and since $R$ and $R^*$ have the same singular values,
\begin{equation}\label{eq:Qnorm}
\begin{aligned}
 Q_{\alpha,z}(\rho\|\sigma)&=\tr\,(RR^*)^{z}=\norm{R^*}_{2z}^{2z}=\norm{R}_{2z}^{2z}=\Bigl\|\rho^{\frac{\alpha}{2z}}\sigma^{\frac{1-\alpha}{2z}}\Bigr\|_{2z}^{2z},\\
 Q_{\alpha,z}(\rho\|\sigma)&=\norm{R^*}_{2z}^{2z}=\Bigl\|\sigma^{\frac{1-\alpha}{2z}}\rho^{\frac{\alpha}{2z}}\Bigr\|_{2z}^{2z}=Q_{1-\alpha,z}(\sigma\|\rho).
\end{aligned}
\end{equation}
For $s\in(0,1)$ let
\begin{equation*}
 D_s(\rho\|\sigma)\eqdef\frac{\log\tr\rho^s\sigma^{1-s}}{s-1},\qquad
 \Dsand_s(\rho\|\sigma)\eqdef\frac{\log\tr\bigl(\sigma^{\frac{1-s}{2s}}\rho\,\sigma^{\frac{1-s}{2s}}\bigr)^s}{s-1}
\end{equation*}
be the Petz--R\'enyi divergence \cite{Petz86} and the sandwiched R\'enyi divergence \cite{MuellerLennert13,WWY14}, and let $F(\rho,\sigma)\eqdef\norm{\sqrt\rho\sqrt\sigma}_1$ be the fidelity. By \eqref{eq:Qnorm} and these definitions, for $s\in(0,1)$,
\begin{equation}\label{eq:special}
\begin{aligned}
 \psi(s,1)&=(1-s)\,D_s(\rho\|\sigma), &\qquad \psi(s,s)&=(1-s)\,\Dsand_s(\rho\|\sigma),\\
 \psi(1-s,s)&=(1-s)\,\Dsand_s(\sigma\|\rho), &\qquad \psi(\tfrac12,\tfrac12)&=-\log F(\rho,\sigma).
\end{aligned}
\end{equation}
In particular $-2\log F(\rho,\sigma)=\Dsand_{1/2}(\rho\|\sigma)$. The third identity concerns the sandwiched divergence with the two arguments exchanged. In \cite{JiRegula25} the same quantities are expressed through the \emph{reverse sandwiched} R\'enyi divergence $\overleftarrow{D}_s(\rho\|\sigma)\eqdef\frac{s}{1-s}\Dsand_{1-s}(\sigma\|\rho)$, for which \eqref{eq:special} gives
\begin{equation}\label{eq:reverse}
 \psi(s,1-s)=(1-s)\,\overleftarrow{D}_s(\rho\|\sigma),\qquad\psi(1-s,1-s)=(1-s)\,\overleftarrow{D}_s(\sigma\|\rho)\qquad(0<s<1).
\end{equation}
The parameter set that will be relevant is the closed triangle shown in Fig.~\ref{fig:triangle},
\begin{equation}\label{eq:triangle}
\begin{aligned}
 \cD&\eqdef\bigl\{(\alpha,z)\in\bbR^2:\ \tfrac12\le z\le1,\ 1-z\le\alpha\le z\bigr\}\\
 &=\bigl\{(\alpha,z)\in\bbR^2:\ 0\le\alpha\le1,\ \max\{\alpha,1-\alpha\}\le z\le1\bigr\}.
\end{aligned}
\end{equation}

\begin{proposition}[Properties of $\psi$]\label{prop:psi}
For fixed faithful states $\rho,\sigma$:
\begin{enumerate}[label=(\alph*),leftmargin=*]
\item $\psi$ is real-analytic on $\bbR\times(0,\infty)$;
\item $\psi$ is jointly concave in the parameters $(\alpha,z)$ on $\bbR\times(0,\infty)$;
\item $\partial_z\psi(\alpha,z)\ge0$ for all $(\alpha,z)\in\bbR\times(0,\infty)$;
\item $\psi(0,z)=\psi(1,z)=0$ for all $z>0$, and $\psi(\alpha,z)\ge0$ for $\alpha\in[0,1]$ and $z>0$.
\end{enumerate}
\end{proposition}

Item~(a) is a standard consequence of the analytic functional calculus for positive definite matrices. Item~(b) follows from a log-convexity theorem of Bourin and Lee \cite[Corollary~2.1]{BourinLee16} by restricting $\psi$ to straight lines in the $(\alpha,z)$-plane. Item~(c) is \cite[Proposition~6]{LinTomamichel15}: for fixed $\alpha\ne1$, $D_{\alpha,z}(\rho\|\sigma)$ is non-decreasing in $z$ if $\alpha<1$ and non-increasing in $z$ if $\alpha>1$, so that $\psi(\alpha,z)=(1-\alpha)D_{\alpha,z}(\rho\|\sigma)$ is non-decreasing in $z$ in both cases, while $\psi(1,z)=0$ for all $z>0$ by item~(d). Item~(d) follows from the normalization of $\rho$ and $\sigma$ together with item~(b). In Section~\ref{sec:sectors} we give self-contained proofs of all four items; there, items~(b) and~(c) are by-products of the sector representation of $\psi$ (Lemmas~\ref{lem:support}, \ref{lem:limit} and~\ref{lem:exposed}).

By (a), the following three functions are well defined and continuous on $\bbR\times(0,\infty)$:
\begin{equation}\label{eq:gradient}
 \chi^\ast\eqdef\tfrac12\,\partial_z\psi,\qquad
 \dsig^\ast\eqdef\psi-\alpha\,\partial_\alpha\psi-z\,\partial_z\psi,\qquad
 \drho^\ast\eqdef\dsig^\ast+\partial_\alpha\psi .
\end{equation}
Equivalently, $(\drho^\ast,\dsig^\ast,\chi^\ast)$ at $(\alpha,z)$ is the unique solution of the three linear equations
\begin{equation}\label{eq:gradient-eq}
 \alpha\,\drho^\ast+(1-\alpha)\,\dsig^\ast+2z\,\chi^\ast=\psi(\alpha,z),\qquad
 \drho^\ast-\dsig^\ast=\partial_\alpha\psi(\alpha,z),\qquad
 2\chi^\ast=\partial_z\psi(\alpha,z);
\end{equation}
indeed, substituting the second and third equations into the first gives $\dsig^\ast=\psi-\alpha\,\partial_\alpha\psi-z\,\partial_z\psi$.

Finally we recall the Hoeffding functions and the Hoeffding region. For $r>0$ let
\begin{equation}\label{eq:Hoeffding-fn}
 H_r(\rho\|\sigma)\eqdef\sup_{0<s<1}\frac{\psi(s,1)-(1-s)\,r}{s},\qquad
 H_r(\sigma\|\rho)\eqdef\sup_{0<s<1}\frac{\psi(s,1)-s\,r}{1-s}.
\end{equation}
By \eqref{eq:Qnorm} and \eqref{eq:special}, this is the notation of \cite{JiRegula25}, where $H_r(\rho\|\sigma)=\sup_{0<s<1}\frac{s-1}{s}\bigl(r-D_s(\rho\|\sigma)\bigr)$. For fixed $s$ the two quotients in \eqref{eq:Hoeffding-fn} are decreasing in $r$, so both functions are non-increasing in $r$. The \emph{Hoeffding region} is
\begin{equation}\label{eq:H-region}
\begin{aligned}
 \cH&\eqdef\bigl\{(a,b)\in(0,\infty)^2:\ s\,a+(1-s)\,b\le\psi(s,1)\ \text{for some } s\in[0,1]\bigr\}\\
 &=\bigl\{(a,b)\in(0,\infty)^2:\ b\le H_a(\sigma\|\rho)\bigr\},
\end{aligned}
\end{equation}
where the second description follows from Lemma~\ref{lem:Hform} below (applied with $K\ge a$).

\subsection{Main results}

\begin{theorem}[Exact trade-off]\label{thm:main}
Let $\rho,\sigma$ be faithful states on $\bbC^d$, and let $\cZ$ be the set of all quadruples $(a,b,K,L)\in(0,\infty)^4$ that satisfy the following three conditions:
\begin{align}
 &\text{\emph{(P1)}}\qquad s\min\{a,K\}+(1-s)\,b\ \le\ \psi(s,1)\quad\text{for some } s\in[0,1];\label{eq:P1}\\
 &\text{\emph{(P2)}}\qquad s\,a+(1-s)\min\{b,L\}\ \le\ \psi(s,1)\quad\text{for some } s\in[0,1];\label{eq:P2}\\
 &\text{\emph{(F)}}\qquad\ \,(\alpha+z-1)\,a+(z-\alpha)\,b+(1-z)\,(K+L)\ \le\ \psi(\alpha,z)\notag\\
 &\phantom{\text{\emph{(F)}}\qquad\ \,}\text{for some } (\alpha,z)\in\cD.\label{eq:F}
\end{align}
Then a quadruple is achievable if and only if it belongs to $\cZ$, i.e., $\cA=\cZ$. More precisely:
\begin{enumerate}[label=(\roman*),leftmargin=*]
\item \emph{(Converse)} $\cA\subseteq\cZ$.
\item \emph{(Achievability)} If $(a,b,K,L)\in\cZ$, then for every $n\ge1$ there is an $n$-copy test $T_n$ with
\begin{equation}\label{eq:finite-n}
 \alpha_n(T_n)\le c_n\,e^{-na},\quad\beta_n(T_n)\le c_n\,e^{-nb},\quad u_n(T_n)\le c_n\,e^{-nK},\quad v_n(T_n)\le c_n\,e^{-nL},
\end{equation}
where $c_n\eqdef800\,(n+1)^{3(d^2-1)}$. If moreover $K\ge a$ or $L\ge b$, the test $T_n$ can be chosen binary.
\end{enumerate}
\end{theorem}

Since $\frac1n\log c_n\to0$, the tests of part (ii) satisfy \eqref{eq:achievable}; thus (ii) implies $\cZ\subseteq\cA$. The factor $c_n$ depends on $\rho$ and $\sigma$ only through the dimension $d$. Note that neither $\cA=\cZ$ nor \eqref{eq:finite-n} involves a closure or a loss in the exponents: every quadruple of $\cZ$, including those on its boundary, is achievable, with error and rejection probabilities that exceed the target values $e^{-na}$, $e^{-nb}$, $e^{-nK}$, $e^{-nL}$ by at most a polynomial factor, for every blocklength $n$.

For $K,L>0$ let $\cZ(K,L)\eqdef\{(a,b)\in(0,\infty)^2:(a,b,K,L)\in\cZ\}$ be the slice of $\cZ$. The next theorem describes the region $\cR(K,L)$, in which the rejection constraints \eqref{eq:hard} are imposed for every $n$.

\begin{theorem}[Fixed rejection exponents]\label{thm:fixed}
For all $K,L>0$,
\begin{equation}\label{eq:sandwich}
 \bigcup_{\eta>0}\cZ(K+\eta,L+\eta)\ \subseteq\ \cR(K,L)\ \subseteq\ \cZ(K,L),
\end{equation}
and both $\cZ(K,L)$ and $\cR(K,L)$ are closed in $(0,\infty)^2$. If $K+L\ne-2\log F(\rho,\sigma)$, then $\cZ(K,L)$ is the closure (relative to $(0,\infty)^2$) of the left-hand side of \eqref{eq:sandwich}, and consequently
\begin{equation}\label{eq:fixed-equality}
 \cR(K,L)=\cZ(K,L)\qquad\text{whenever } K+L\ne\Dsand_{1/2}(\rho\|\sigma).
\end{equation}
\end{theorem}

Thus, away from the hypersurface $K+L=-2\log F(\rho,\sigma)$, the fixed-budget region $\cR(K,L)$ is determined exactly, including its boundary. On this hypersurface the theorem gives the inclusions \eqref{eq:sandwich}; equality still holds for commuting states (Remark~\ref{rem:commuting}).

The third result gives an equivalent description of $\cZ$ in terms of the functions \eqref{eq:gradient}. For real numbers $a,b,K,L$ and $(\drho,\dsig)\in\bbR^2$ let
\begin{equation}\label{eq:pairfn}
 \Phi_{a,b,K,L}(\drho,\dsig)\eqdef\min\Bigl\{\pos{a-\drho},\ \pos{b-\dsig},\ \pos{K-\drho}+\pos{L-\dsig}\Bigr\}.
\end{equation}
The function $\Phi$ is jointly continuous in all six real arguments, since positive part, addition, and minimum are continuous.

\begin{theorem}[Pair form via exposed sectors]\label{thm:pair}
A quadruple $(a,b,K,L)\in(0,\infty)^4$ belongs to $\cZ$ if and only if
\begin{equation}\label{eq:pair}
 \Phi_{a,b,K,L}\bigl(\drho^\ast(\alpha,z),\dsig^\ast(\alpha,z)\bigr)\ \le\ 2\chi^\ast(\alpha,z)\qquad\text{for all }(\alpha,z)\in\cD .
\end{equation}
\end{theorem}

The triple $(\drho^\ast,\dsig^\ast,\chi^\ast)(\alpha,z)$ is a limiting triple of sector exponents, obtained from a sequence of pairs of eigenspaces in common Schur--Weyl blocks, with limiting probability exponents $\drho^\ast,\dsig^\ast$ and maximal-overlap exponent $\chi^\ast$. It is the unique exponent triple minimizing $\alpha\drho+(1-\alpha)\dsig+2z\chi$ (Lemmas~\ref{lem:limit} and~\ref{lem:exposed}). The pair inequality expresses the condition for testing the corresponding sectors on the exponential scale; the two error terms and the product of the rejection probabilities provide the three terms in \eqref{eq:pairfn}.

The equivalence has a convex-analytic explanation. Under the coordinate map $\mathsf T(\drho,\dsig,\chi)=(\drho+2\chi,\dsig+2\chi,\drho+\dsig+2\chi)$, the upper image $\mathsf T(\cS)+\bbR_+^3$ of the limiting sector set $\cS$ is closed and convex. Its supporting half-spaces are specified by $\psi$, and it is generated by the exposed sectors in \eqref{eq:gradient}. Failure of the pair inequality amounts to entering one of three downward open boxes in these coordinates. Such a failure always has an exposed-sector witness. Thus Theorem~\ref{thm:pair} relates two descriptions of the same geometric obstruction: supporting inequalities in (P1), (P2), and (F), and inequalities evaluated at exposed sectors. Section~\ref{sec:sectors} proves this equivalence before the measurement construction.
\subsection{Remarks on the three conditions}

\begin{lemma}[Hoeffding form of (P1) and (P2)]\label{lem:Hform}
For $a,b,K,L>0$, condition \eqref{eq:P1} is equivalent to $b\le\max\{H_a(\sigma\|\rho),H_K(\sigma\|\rho)\}$, and condition \eqref{eq:P2} is equivalent to $a\le\max\{H_b(\rho\|\sigma),H_L(\rho\|\sigma)\}$.
\end{lemma}

The proof is given in Appendix~\ref{app:Hform}. Thus (P1) and (P2) are exactly the inequalities \eqref{eq:intro-classical} of the converse \cite[Proposition~S17]{JiRegula25}. We call
\begin{equation}\label{eq:Petz-region}
 \cC^{\Pet}(K,L)\eqdef\bigl\{(a,b)\in(0,\infty)^2:\ \text{\eqref{eq:P1} and \eqref{eq:P2} hold}\bigr\}
\end{equation}
the \emph{Petz region}.

\begin{remark}[Condition (F) on the boundary of the triangle]\label{rem:edges}
By \eqref{eq:special}, condition \eqref{eq:F} reads as follows on the edges (with $\frac12\le s<1$ on the slanted edges) and at the lower vertex of $\cD$:
\begin{align*}
 &z=1,\ \alpha=s: && s\,a+(1-s)\,b\le\psi(s,1), \text{ i.e., } (a,b)\in\cH;\\
 &z=\alpha=s: && (2s-1)\,a+(1-s)(K+L)\le(1-s)\Dsand_s(\rho\|\sigma);\\
 &z=1-\alpha=s: && (2s-1)\,b+(1-s)(K+L)\le(1-s)\Dsand_s(\sigma\|\rho);\\
 &(\alpha,z)=(\tfrac12,\tfrac12): && K+L\le-2\log F(\rho,\sigma)=\Dsand_{1/2}(\rho\|\sigma).
\end{align*}
At the lower vertex the coefficients of $a$ and $b$ vanish. Therefore (F) holds for all $(a,b)$ if $K+L\le-2\log F(\rho,\sigma)$. If $K+L>-2\log F(\rho,\sigma)$, then (F) fails whenever $\min\{a,b\}$ is sufficiently large: by continuity $(1-z)(K+L)>\psi(\alpha,z)$ in a neighborhood of the lower vertex, and outside this neighborhood the sum $2z-1$ of the coefficients of $a$ and $b$ is bounded away from zero. Whether (F) is then more restrictive than (P1) and (P2) depends on the states; see Corollary~\ref{cor:symmetric} and Section~\ref{sec:example}.
\end{remark}

\begin{remark}[Inclusions]\label{rem:inclusions}
If $(a,b)\in\cH$ with witness $s$, then
\begin{equation*}
\begin{aligned}
 s\min\{a,K\}+(1-s)\,b&\le s\,a+(1-s)\,b\le\psi(s,1),\\
 s\,a+(1-s)\min\{b,L\}&\le s\,a+(1-s)\,b\le\psi(s,1),
\end{aligned}
\end{equation*}
and \eqref{eq:F} holds at $(\alpha,z)=(s,1)$, where it reads $s\,a+(1-s)\,b\le\psi(s,1)$. Hence, for all $K,L>0$,
\begin{equation}\label{eq:inclusions}
 \cH\ \subseteq\ \cZ(K,L)\ \subseteq\ \cC^{\Pet}(K,L).
\end{equation}
Conversely, if $K\ge a$, then $\min\{a,K\}=a$ and \eqref{eq:P1} says $(a,b)\in\cH$; if $L\ge b$, then \eqref{eq:P2} says $(a,b)\in\cH$. Hence
\begin{equation}\label{eq:no-help}
 (a,b,K,L)\in\cZ\ \text{ and }\ \bigl(K\ge a\ \text{or}\ L\ge b\bigr)\quad\Longrightarrow\quad(a,b)\in\cH:
\end{equation}
a point outside the Hoeffding region can only be reached if the rejection probability under each hypothesis decays strictly more slowly than the corresponding error probability.
\end{remark}

\begin{remark}[The triangle and data processing]\label{rem:dpi}
For $\alpha\in(0,1)$, $D_{\alpha,z}$ is monotone under quantum channels if and only if $z\ge\max\{\alpha,1-\alpha\}$ \cite[Theorem~1.2]{Zhang20}, and $\cD$ is the part of this region with $z\le1$. The general data-processing inequality for $D_{\alpha,z}$ is not used in the proofs of the main theorems; the converse uses the elementary fidelity inequality of Lemma~\ref{lem:nec}. Applying the data-processing inequality for $D_{\alpha,z}$ to the quantum-to-classical channel defined by a test yields, for achievable quadruples, the necessary condition 
\begin{align}\label{eq:DPI}
\min\{\alpha a,(1-\alpha)b,\alpha K+(1-\alpha)L\}\le\psi(\alpha,z)\quad \mbox{for}\quad (\alpha,z)\in\cD.
\end{align}
This condition is implied by \eqref{eq:pair}: if, say, $\pos{a-\drho^\ast}\le2\chi^\ast$, then $\alpha a\le\alpha\drho^\ast+2\alpha\chi^\ast\le\alpha\drho^\ast+(1-\alpha)\dsig^\ast+2z\chi^\ast=\psi(\alpha,z)$ by $\alpha\le z$, $\dsig^\ast\ge0$, $\chi^\ast\ge0$ and \eqref{eq:gradient-eq}; the other two terms are treated in the same way, using $\max\{\alpha,1-\alpha\}\le z$.

This condition is strictly weaker in general. Consider the qubit pair
\eqref{eq:qubit} with $p=19/20$, $q=1/20$, and put $e=2/5$.
Since every entry of $V$ has squared modulus $1/2$,
\begin{equation*}
 \begin{aligned}
 \tr\rho^s\sigma^{1-s}
 &=\tfrac12\bigl(1+p^s q^{1-s}+q^s p^{1-s}\bigr),
 \\
 C\eqdef\max_{0\le s\le1}\psi(s,1)
 &=-\log(\tfrac12+\sqrt{pq})=\log\frac{20}{10+\sqrt{19}},
 \end{aligned}
\end{equation*}
where the maximum is attained at $s=\frac12$ by the
arithmetic--geometric mean inequality. Moreover,
$F(\rho,\sigma)^2=\tr\rho\sigma+2\sqrt{\det\rho\det\sigma}
=\frac12+2pq=\frac{119}{200}$, so
\begin{equation*}
 C=0.3313623881\ldots<e=0.4
 <\Dsand_{1/2}(\rho\|\sigma)
 =\log\tfrac{200}{119}=0.5191938734\ldots.
\end{equation*}

For $0<\alpha<1$ put $s=\max\{\alpha,1-\alpha\}$.
Symmetry, monotonicity of $\psi$ in $z$, and monotonicity of the
sandwiched divergence in its order
\cite[Theorem~7]{MuellerLennert13} give, for every $(\alpha,z)\in\cD$,
\begin{equation*}
 \begin{aligned}
 \psi(\alpha,z)&=\psi(s,z)\ge\psi(s,s)=(1-s)\Dsand_s(\rho\|\sigma)\\
 &\ge(1-s)\Dsand_{1/2}(\rho\|\sigma)>(1-s)e
 =\min\{\alpha e,(1-\alpha)e,e\}.
 \end{aligned}
\end{equation*}
At $\alpha=0,1$ both sides of \eqref{eq:DPI} vanish. Hence
$(e,e,e,e)$ satisfies the entire family \eqref{eq:DPI}.
It is not achievable, however: since $K=a=e$ and $L=b=e$,
either \eqref{eq:P1} or \eqref{eq:P2} would require $e\le C$.
The gap is not specific to $p=19/20$. For every $p\in(\frac12,1)$,
putting $x=\sqrt{p(1-p)}\in(0,\frac12)$ gives
$\frac12+x>\frac12+2x^2$, and therefore
$C<\Dsand_{1/2}(\rho\|\sigma)$. Every
$e\in(C,\Dsand_{1/2}(\rho\|\sigma)]$ yields the same separation.

\end{remark}

\begin{remark}[Commuting states]\label{rem:commuting}
If $\rho$ and $\sigma$ commute, $Q_{\alpha,z}$ does not depend on $z$, and (F) is implied by (P1) and (P2). Thus $\cZ(K,L)=\cC^{\Pet}(K,L)$ and $\cR(K,L)=\cC^{\Pet}(K,L)$ for all $K,L>0$, and Theorem~\ref{thm:main} recovers the classical result \cite[Lemma~5]{JiRegula25} for probability distributions with full support.

\end{remark}

\subsection{Specializations}

\begin{remark}[Quantum Hoeffding bound \cite{Hayashi07,Nagaoka06,ANSV08}]\label{rem:hoeffding}
Theorem~\ref{thm:main} recovers the quantum Hoeffding bound. Let $\cA_{\rm bin}$ be the set of pairs $(a,b)\in(0,\infty)^2$ for which there is a sequence of binary tests with $\liminf_n-\frac1n\log\alpha_n\ge a$ and $\liminf_n-\frac1n\log\beta_n\ge b$. Then $\cA_{\rm bin}=\cH$. Moreover, for every $(a,b)\in\cH$ and every $n\ge1$ there is a binary test with $\alpha_n\le c_n\,e^{-na}$ and $\beta_n\le c_n\,e^{-nb}$.
Indeed, binary tests have zero rejection probabilities, so the converse follows from Theorem~\ref{thm:main}(i) with $K=a$, $L=b$ and \eqref{eq:no-help}. Conversely, \eqref{eq:inclusions} and Theorem~\ref{thm:main}(ii), again with $K=a$, $L=b$, give the asserted binary tests; $\frac1n\log c_n\to0$ gives the exponent statement.

The binary finite-blocklength assertion is a recovery of an established bound, rather than a new refinement: the standard weighted Chernoff trace inequality \cite[Theorem~1]{Audenaert07}, applied to $e^{n(a-b)}\rhon$ and $\sigman$, gives a binary test with $\alpha_n\le e^{-na}$ and $\beta_n\le e^{-nb}$ whenever $(a,b)\in\cH$. The common factor $c_n$ is retained here to reflect the unified construction with rejection.
\end{remark}

\begin{remark}[The Petz region below the fidelity threshold]\label{rem:petz}
If $K+L\le-2\log F(\rho,\sigma)$, then $\cZ(K,L)=\cC^{\Pet}(K,L)$; that is, a quadruple $(a,b,K,L)$ with $K+L\le\Dsand_{1/2}(\rho\|\sigma)$ is achievable if and only if
\begin{equation}\label{eq:petz-threshold}
 a\le\max\{H_b(\rho\|\sigma),H_L(\rho\|\sigma)\}\qquad\text{and}\qquad b\le\max\{H_a(\sigma\|\rho),H_K(\sigma\|\rho)\}.
\end{equation}
If $K+L<-2\log F(\rho,\sigma)$, then moreover $\cR(K,L)=\cC^{\Pet}(K,L)$.
Indeed, under the non-strict inequality, (F) holds at $(\alpha,z)=(\frac12,\frac12)$ for every $(a,b)$, so Theorem~\ref{thm:main} and Lemma~\ref{lem:Hform} give the first assertion. Under the strict inequality, Theorem~\ref{thm:fixed} gives $\cR(K,L)=\cZ(K,L)$.

Hence \cite[Conjecture~S18]{JiRegula25} holds for all quadruples with $K+L\le\Dsand_{1/2}(\rho\|\sigma)$. This condition is sufficient but not necessary for $\cZ(K,L)=\cC^{\Pet}(K,L)$ (for commuting states, for instance, equality holds for all $K,L$). For larger $K+L$ the Petz region is in general not achievable; an explicit example is given in Section~\ref{sec:example}.
\end{remark}

The following specialization concerns equal error exponents and equal rejection exponents for pairs of states that are exchanged by a unitary. Its proof is given in Section~\ref{sec:proof-symmetric}.

\begin{corollary}[Symmetric pairs]\label{cor:symmetric}
Assume that there is a unitary $V$ with $\sigma=V\rho V^*$ and $\rho=V\sigma V^*$. Then $\psi(\alpha,z)=\psi(1-\alpha,z)$ and $H_r(\rho\|\sigma)=H_r(\sigma\|\rho)\eqqcolon H_r$. Put $\psi_{1/2}(z)\eqdef\psi(\frac12,z)$,
\begin{equation}\label{eq:sym-constants}
 C\eqdef\psi_{1/2}(1),\qquad r_0\eqdef\psi_{1/2}(\tfrac12)=-\log F(\rho,\sigma),\qquad r_{\rm c}\eqdef\psi_{1/2}(1)-\tfrac12\psi_{1/2}'(1),
\end{equation}
and for $r>0$ let $e^\ast(r)\eqdef\sup\{e>0:(e,e,r,r)\in\cA\}$ be the optimal equal-error exponent for rejection exponents $K=L=r$. Then $r_0\le r_{\rm c}\le C$, where $C=\max_{s\in[0,1]}\psi(s,1)$ is the Chernoff exponent, and:
\begin{enumerate}[label=(\alph*),leftmargin=*]
\item for $0<r\le r_0$: $e^\ast(r)=\max\{C,H_r\}$;
\item for $r>r_0$: $e^\ast(r)=\min\bigl\{\max\{C,H_r\},\,e_F(r)\bigr\}$, where
\begin{equation}\label{eq:eF}
 e_F(r)\eqdef\sup_{\frac12<z\le1}\frac{\psi_{1/2}(z)-2(1-z)\,r}{2z-1};
\end{equation}
\item for $r>r_0$ with $r\ge r_{\rm c}$: $e_F(r)=C$; for $r_0<r<r_{\rm c}$:
\begin{equation}\label{eq:eF-formula}
 e_F(r)=\psi_{1/2}(z_r)+(1-z_r)\,\psi_{1/2}'(z_r)>C,
\end{equation}
where $z_r\in(\frac12,1)$ is any solution of $\psi_{1/2}(z)-(z-\frac12)\psi_{1/2}'(z)=r$.
\end{enumerate}
In particular $e^\ast(r)=C$ for every $r\ge r_{\rm c}$, and $e^\ast(r)>C$ for every $0<r<r_{\rm c}$.
\end{corollary}

By Theorem~\ref{thm:fixed}, and since $\cR(r,r)$ is a decreasing set, for $r\ne r_0$ the number $e^\ast(r)$ is also the supremum of all $e$ with $(e,e)\in\cR(r,r)$, i.e., the optimal exponent of $\max\{\alpha_n,\beta_n\}$ under the constraints $u_n,v_n\le e^{-nr}$ for all $n$. If $r_0<r_{\rm c}$, Corollary~\ref{cor:symmetric} says that rejection improves the equal-error exponent beyond the Chernoff exponent if and only if $r<r_{\rm c}$; here $r_{\rm c}=C-\chi^\ast(\frac12,1)$ is strictly smaller than $C$ unless $\partial_z\psi(\frac12,1)=0$, and $r_0<r_{\rm c}$ unless $\rho$ and $\sigma$ commute, by \eqref{eq:sym-threshold-equality} in the proof in Section~\ref{sec:proof-symmetric}. Moreover, if $r_0<r_{\rm c}$, then $\lim_{r\downarrow r_0}e_F(r)=r_0+\frac12\psi_{1/2}'(\frac12)$ (see the end of the proof in Section~\ref{sec:proof-symmetric}), so that $e^\ast$ is discontinuous at $r_0$ if and only if
\begin{equation}\label{eq:jump}
 r_0+\tfrac12\psi_{1/2}'(\tfrac12)<\max\{C,H_{r_0}\}.
\end{equation}
If the reverse strict inequality holds, then $e^\ast(r)=\max\{C,H_r\}$ also for $r$ slightly above $r_0$, i.e., the Petz-type converse remains tight for equal exponents slightly above the fidelity threshold.

\subsection{Example}\label{sec:example}
\begin{table}[t]
\centering
\small
\caption{The qubit pair \eqref{eq:qubit} with $p=0.95$, $q=0.05$: optimal equal-error exponent $e^\ast(r)$ for rejection exponents $K=L=r$. Decimal entries are rounded; $z_r$ is shown when (F) is binding. The row $r_0^+$ denotes the right-hand limit, not the value at $r_0$.}
\label{tab:example}

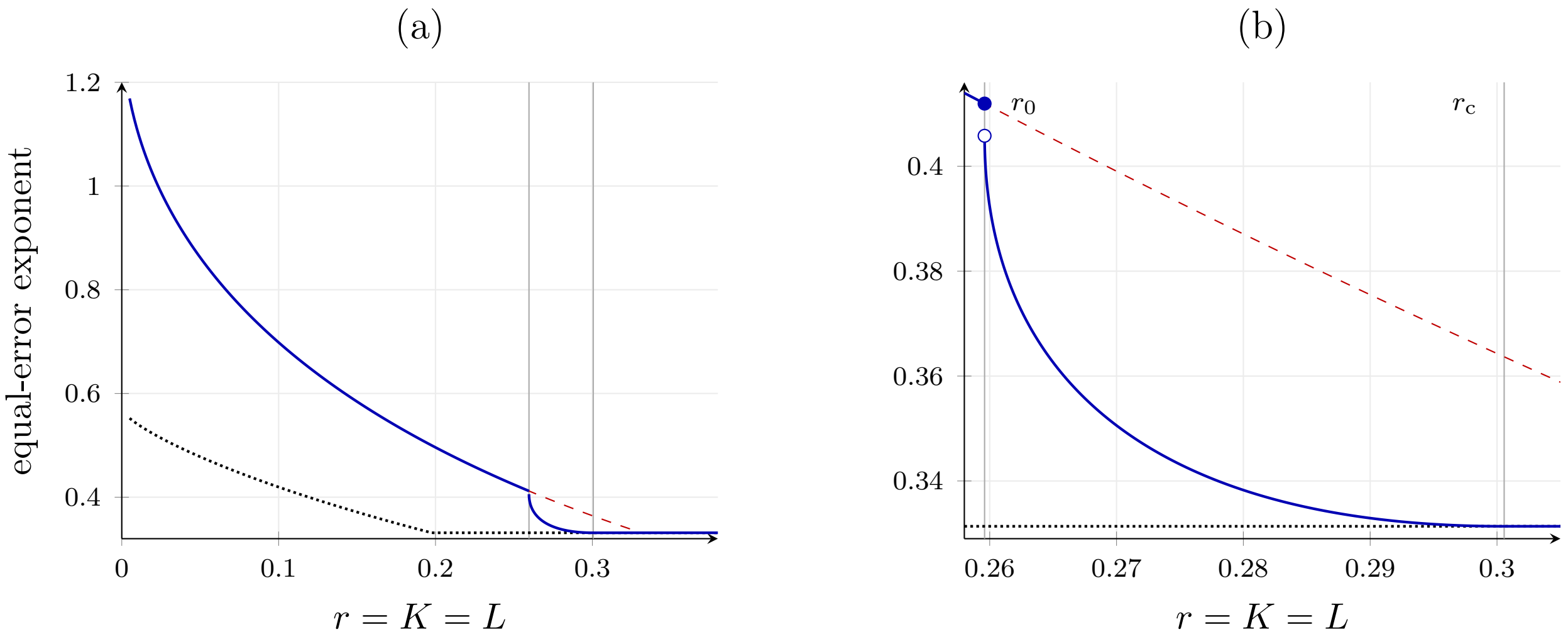
\caption{The qubit pair \eqref{eq:qubit} with $p=0.95$, $q=0.05$. (a) Optimal equal-error exponent $e^\ast(r)$ (solid blue), the Petz-type converse $\max\{C,H_r\}$ (dashed red), and the exponent achievable by \cite[Proposition~6]{JiRegula25} (dotted black). (b) Enlargement near the fidelity threshold. The vertical lines mark $r_0$ and $r_{\rm c}$. In (b), the filled and open circles mark $e^\ast(r_0)$ and its smaller right-hand limit, respectively. The two blue branches are not joined across the jump. Condition (F) is binding for $r_0<r<r_{\rm c}$.}
\label{fig:equal}
\end{figure}

Fig.~\ref{fig:regions} shows the regions $\cZ(K,L)$ for $K=L=0.2$ and $K=L=0.28$. In the first case $K+L<\Dsand_{1/2}(\rho\|\sigma)$ and
$\cZ(K,L)=\cC^{\Pet}(K,L)=\cH\cup(0,H_{0.2}]^2$, with $H_{0.2}\approx0.49598612$. In the second case the fidelity threshold is exceeded and (F) cuts away part of the Petz region; the point $(0.35,0.35)$ excluded above is marked in the figure. The boundaries are computed directly from (P1), (P2), and (F).

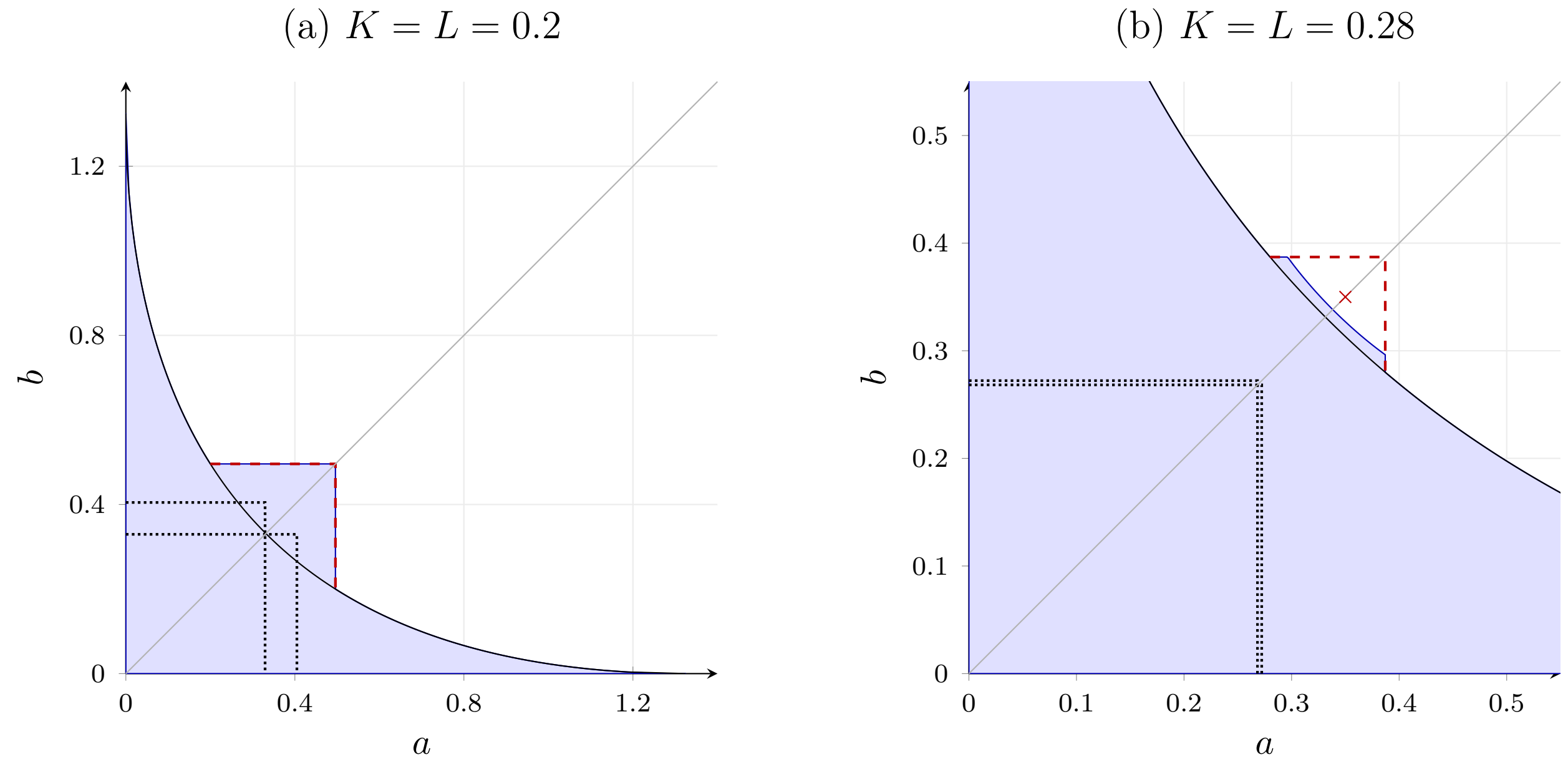
\begin{figure}[t]
\centering
\begin{tikzpicture}
\begin{axis}[width=.38\textwidth,height=.38\textwidth,scale only axis,axis lines=left,grid=major,grid style={gray!15},tick label style={font=\scriptsize},label style={font=\small},scaled ticks=false,ticklabel style={/pgf/number format/fixed},clip=true,xmin=0,xmax=1.4,ymin=0,ymax=1.4,xlabel={$a$},ylabel={$b$},title={(a) $K=L=0.2$},xtick={0,.4,.8,1.2},ytick={0,.4,.8,1.2}]
\addplot[draw=blue!70!black,fill=blue!12,thin] coordinates {(0.000000000,0.000000000) (0.000000000,1.324997541) (0.007361097,1.136584878) (0.014722195,1.062421524) (0.022083292,1.007100100) (0.029444390,0.961547814) (0.036805487,0.922247727) (0.044166585,0.887394817) (0.051527682,0.855916041) (0.058888780,0.827111749) (0.066249877,0.800495608) (0.073610974,0.775713269) (0.080972072,0.752497030) (0.088333169,0.730638762) (0.095694267,0.709972832) (0.103055364,0.690364846) (0.110416462,0.671703962) (0.117777559,0.653897480) (0.125138657,0.636866930) (0.132499754,0.620545184) (0.139860852,0.604874280) (0.147221949,0.589803755) (0.154583046,0.575289356) (0.161944144,0.561292008) (0.169305241,0.547777011) (0.176666339,0.534713375) (0.184027436,0.522073288) (0.191388534,0.509831678) (0.198749631,0.497965847) (0.200000000,0.495986121) (0.206110729,0.495986121) (0.213471826,0.495986121) (0.220832923,0.495986121) (0.228194021,0.495986121) (0.235555118,0.495986121) (0.242916216,0.495986121) (0.250277313,0.495986121) (0.257638411,0.495986121) (0.264999508,0.495986121) (0.272360606,0.495986121) (0.279721703,0.495986121) (0.287082800,0.495986121) (0.294443898,0.495986121) (0.301804995,0.495986121) (0.309166093,0.495986121) (0.316527190,0.495986121) (0.323888288,0.495986121) (0.331249385,0.495986121) (0.331362388,0.495986121) (0.338610483,0.495986121) (0.345971580,0.495986121) (0.353332677,0.495986121) (0.360693775,0.495986121) (0.368054872,0.495986121) (0.375415970,0.495986121) (0.382777067,0.495986121) (0.390138165,0.495986121) (0.397499262,0.495986121) (0.404860360,0.495986121) (0.412221457,0.495986121) (0.419582555,0.495986121) (0.426943652,0.495986121) (0.434304749,0.495986121) (0.441665847,0.495986121) (0.449026944,0.495986121) (0.456388042,0.495986121) (0.463749139,0.495986121) (0.471110237,0.495986121) (0.478471334,0.495986121) (0.485832432,0.495986121) (0.493193529,0.495986121) (0.495986121,0.495986121) (0.495986131,0.199999994) (0.500554626,0.197124271) (0.507915724,0.192561845) (0.515276821,0.188085540) (0.522637919,0.183693487) (0.529999016,0.179383885) (0.537360114,0.175154995) (0.544721211,0.171005136) (0.552082309,0.166932688) (0.559443406,0.162936082) (0.566804503,0.159013802) (0.574165601,0.155164383) (0.581526698,0.151386406) (0.588887796,0.147678498) (0.596248893,0.144039328) (0.603609991,0.140467608) (0.610971088,0.136962091) (0.618332186,0.133521565) (0.625693283,0.130144856) (0.633054381,0.126830827) (0.640415478,0.123578371) (0.647776575,0.120386416) (0.655137673,0.117253920) (0.662498770,0.114179871) (0.669859868,0.111163285) (0.677220965,0.108203208) (0.684582063,0.105298709) (0.691943160,0.102448886) (0.699304258,0.099652858) (0.706665355,0.096909771) (0.714026452,0.094218792) (0.721387550,0.091579111) (0.728748647,0.088989938) (0.736109745,0.086450504) (0.743470842,0.083960061) (0.750831940,0.081517878) (0.758193037,0.079123243) (0.765554135,0.076775464) (0.772915232,0.074473864) (0.780276329,0.072217783) (0.787637427,0.070006576) (0.794998524,0.067839616) (0.802359622,0.065716290) (0.809720719,0.063635999) (0.817081817,0.061598159) (0.824442914,0.059602198) (0.831804012,0.057647560) (0.839165109,0.055733699) (0.846526207,0.053860083) (0.853887304,0.052026192) (0.861248401,0.050231518) (0.868609499,0.048475563) (0.875970596,0.046757841) (0.883331694,0.045077876) (0.890692791,0.043435204) (0.898053889,0.041829369) (0.905414986,0.040259926) (0.912776084,0.038726439) (0.920137181,0.037228481) (0.927498278,0.035765635) (0.934859376,0.034337492) (0.942220473,0.032943652) (0.949581571,0.031583721) (0.956942668,0.030257316) (0.964303766,0.028964061) (0.971664863,0.027703587) (0.979025961,0.026475532) (0.986387058,0.025279541) (0.993748155,0.024115269) (1.001109253,0.022982374) (1.008470350,0.021880522) (1.015831448,0.020809386) (1.023192545,0.019768645) (1.030553643,0.018757984) (1.037914740,0.017777094) (1.045275838,0.016825673) (1.052636935,0.015903422) (1.059998032,0.015010049) (1.067359130,0.014145269) (1.074720227,0.013308801) (1.082081325,0.012500367) (1.089442422,0.011719697) (1.096803520,0.010966526) (1.104164617,0.010240591) (1.111525715,0.009541636) (1.118886812,0.008869409) (1.126247910,0.008223663) (1.133609007,0.007604153) (1.140970104,0.007010642) (1.148331202,0.006442895) (1.155692299,0.005900680) (1.163053397,0.005383770) (1.170414494,0.004891944) (1.177775592,0.004424981) (1.185136689,0.003982667) (1.192497787,0.003564789) (1.199858884,0.003171140) (1.207219981,0.002801514) (1.214581079,0.002455710) (1.221942176,0.002133531) (1.229303274,0.001834781) (1.236664371,0.001559269) (1.244025469,0.001306806) (1.251386566,0.001077207) (1.258747664,0.000870290) (1.266108761,0.000685874) (1.273469858,0.000523784) (1.280830956,0.000383846) (1.288192053,0.000265887) (1.295553151,0.000169741) (1.302914248,0.000095241) (1.310275346,0.000042224) (1.317636443,0.000010530) (1.324997541,0.000000000) (1.324997541,0.000000000) (0.000000000,0.000000000)} --cycle;
\addplot[black,thin] coordinates {(1.324997541,0.000000000) (1.188552913,0.003785707) (1.134006840,0.007571415) (1.092952676,0.011357122) (1.058888838,0.015142829) (1.029296013,0.018928536) (1.002881666,0.022714244) (0.978877686,0.026499951) (0.956783427,0.030285658) (0.936251012,0.034071365) (0.917027059,0.037857073) (0.898920219,0.041642780) (0.881781775,0.045428487) (0.865493420,0.049214194) (0.849959193,0.052999902) (0.835099982,0.056785609) (0.820849644,0.060571316) (0.807152213,0.064357023) (0.793959825,0.068142731) (0.781231170,0.071928438) (0.768930295,0.075714145) (0.757025678,0.079499852) (0.745489502,0.083285560) (0.734297067,0.087071267) (0.723426322,0.090856974) (0.712857486,0.094642681) (0.702572730,0.098428389) (0.692555918,0.102214096) (0.682792390,0.105999803) (0.673268779,0.109785511) (0.663972857,0.113571218) (0.654893403,0.117356925) (0.646020089,0.121142632) (0.637343385,0.124928340) (0.628854475,0.128714047) (0.620545184,0.132499754) (0.612407913,0.136285461) (0.604435585,0.140071169) (0.596621597,0.143856876) (0.588959773,0.147642583) (0.581444330,0.151428290) (0.574069843,0.155213998) (0.566831211,0.158999705) (0.559723638,0.162785412) (0.552742598,0.166571119) (0.545883824,0.170356827) (0.539143280,0.174142534) (0.532517150,0.177928241) (0.526001815,0.181713948) (0.519593846,0.185499656) (0.513289985,0.189285363) (0.507087134,0.193071070) (0.500982348,0.196856777) (0.494972820,0.200642485) (0.489055873,0.204428192) (0.483228956,0.208213899) (0.477489630,0.211999606) (0.471835563,0.215785314) (0.466264526,0.219571021) (0.460774384,0.223356728) (0.455363093,0.227142436) (0.450028691,0.230928143) (0.444769296,0.234713850) (0.439583102,0.238499557) (0.434468374,0.242285265) (0.429423443,0.246070972) (0.424446703,0.249856679) (0.419536610,0.253642386) (0.414691676,0.257428094) (0.409910468,0.261213801) (0.405191602,0.264999508) (0.400533745,0.268785215) (0.395935610,0.272570923) (0.391395953,0.276356630) (0.386913576,0.280142337) (0.382487315,0.283928044) (0.378116050,0.287713752) (0.373798695,0.291499459) (0.369534198,0.295285166) (0.365321542,0.299070873) (0.361159741,0.302856581) (0.357047840,0.306642288) (0.352984914,0.310427995) (0.348970063,0.314213702) (0.345002418,0.317999410) (0.341081132,0.321785117) (0.337205385,0.325570824) (0.333374378,0.329356532) (0.329587339,0.333142239) (0.325843514,0.336927946) (0.322142171,0.340713653) (0.318482599,0.344499361) (0.314864106,0.348285068) (0.311286017,0.352070775) (0.307747679,0.355856482) (0.304248451,0.359642190) (0.300787714,0.363427897) (0.297364860,0.367213604) (0.293979301,0.370999311) (0.290630462,0.374785019) (0.287317780,0.378570726) (0.284040711,0.382356433) (0.280798721,0.386142140) (0.277591289,0.389927848) (0.274417908,0.393713555) (0.271278082,0.397499262) (0.268171327,0.401284969) (0.265097172,0.405070677) (0.262055155,0.408856384) (0.259044824,0.412642091) (0.256065739,0.416427798) (0.253117471,0.420213506) (0.250199597,0.423999213) (0.247311707,0.427784920) (0.244453398,0.431570628) (0.241624276,0.435356335) (0.238823957,0.439142042) (0.236052064,0.442927749) (0.233308227,0.446713457) (0.230592086,0.450499164) (0.227903288,0.454284871) (0.225241487,0.458070578) (0.222606343,0.461856286) (0.219997525,0.465641993) (0.217414708,0.469427700) (0.214857573,0.473213407) (0.212325807,0.476999115) (0.209819106,0.480784822) (0.207337169,0.484570529) (0.204879701,0.488356236) (0.202446415,0.492141944) (0.200037027,0.495927651) (0.197651261,0.499713358) (0.195288844,0.503499065) (0.192949509,0.507284773) (0.190632994,0.511070480) (0.188339042,0.514856187) (0.186067400,0.518641894) (0.183817822,0.522427602) (0.181590063,0.526213309) (0.179383885,0.529999016) (0.177199054,0.533784724) (0.175035338,0.537570431) (0.172892513,0.541356138) (0.170770355,0.545141845) (0.168668646,0.548927553) (0.166587172,0.552713260) (0.164525721,0.556498967) (0.162484088,0.560284674) (0.160462067,0.564070382) (0.158459458,0.567856089) (0.156476065,0.571641796) (0.154511695,0.575427503) (0.152566156,0.579213211) (0.150639262,0.582998918) (0.148730829,0.586784625) (0.146840674,0.590570332) (0.144968621,0.594356040) (0.143114494,0.598141747) (0.141278120,0.601927454) (0.139459330,0.605713161) (0.137657957,0.609498869) (0.135873836,0.613284576) (0.134106805,0.617070283) (0.132356705,0.620855990) (0.130623380,0.624641698) (0.128906674,0.628427405) (0.127206436,0.632213112) (0.125522516,0.635998819) (0.123854767,0.639784527) (0.122203043,0.643570234) (0.120567202,0.647355941) (0.118947101,0.651141649) (0.117342604,0.654927356) (0.115753572,0.658713063) (0.114179871,0.662498770) (0.112621368,0.666284478) (0.111077932,0.670070185) (0.109549434,0.673855892) (0.108035748,0.677641599) (0.106536747,0.681427307) (0.105052309,0.685213014) (0.103582311,0.688998721) (0.102126633,0.692784428) (0.100685157,0.696570136) (0.099257767,0.700355843) (0.097844346,0.704141550) (0.096444782,0.707927257) (0.095058963,0.711712965) (0.093686778,0.715498672) (0.092328117,0.719284379) (0.090982875,0.723070086) (0.089650944,0.726855794) (0.088332220,0.730641501) (0.087026600,0.734427208) (0.085733982,0.738212915) (0.084454265,0.741998623) (0.083187351,0.745784330) (0.081933141,0.749570037) (0.080691539,0.753355745) (0.079462449,0.757141452) (0.078245778,0.760927159) (0.077041432,0.764712866) (0.075849320,0.768498574) (0.074669351,0.772284281) (0.073501436,0.776069988) (0.072345487,0.779855695) (0.071201415,0.783641403) (0.070069136,0.787427110) (0.068948565,0.791212817) (0.067839616,0.794998524) (0.066742208,0.798784232) (0.065656259,0.802569939) (0.064581687,0.806355646) (0.063518413,0.810141353) (0.062466358,0.813927061) (0.061425443,0.817712768) (0.060395593,0.821498475) (0.059376730,0.825284182) (0.058368779,0.829069890) (0.057371667,0.832855597) (0.056385319,0.836641304) (0.055409664,0.840427011) (0.054444629,0.844212719) (0.053490143,0.847998426) (0.052546137,0.851784133) (0.051612542,0.855569841) (0.050689288,0.859355548) (0.049776308,0.863141255) (0.048873536,0.866926962) (0.047980905,0.870712670) (0.047098350,0.874498377) (0.046225806,0.878284084) (0.045363209,0.882069791) (0.044510497,0.885855499) (0.043667606,0.889641206) (0.042834475,0.893426913) (0.042011043,0.897212620) (0.041197249,0.900998328) (0.040393035,0.904784035) (0.039598340,0.908569742) (0.038813106,0.912355449) (0.038037275,0.916141157) (0.037270791,0.919926864) (0.036513596,0.923712571) (0.035765635,0.927498278) (0.035026853,0.931283986) (0.034297194,0.935069693) (0.033576605,0.938855400) (0.032865031,0.942641107) (0.032162421,0.946426815) (0.031468721,0.950212522) (0.030783879,0.953998229) (0.030107845,0.957783937) (0.029440567,0.961569644) (0.028781995,0.965355351) (0.028132079,0.969141058) (0.027490771,0.972926766) (0.026858020,0.976712473) (0.026233780,0.980498180) (0.025618001,0.984283887) (0.025010638,0.988069595) (0.024411643,0.991855302) (0.023820970,0.995641009) (0.023238572,0.999426716) (0.022664405,1.003212424) (0.022098424,1.006998131) (0.021540584,1.010783838) (0.020990842,1.014569545) (0.020449152,1.018355253) (0.019915473,1.022140960) (0.019389762,1.025926667) (0.018871976,1.029712374) (0.018362073,1.033498082) (0.017860012,1.037283789) (0.017365751,1.041069496) (0.016879251,1.044855203) (0.016400470,1.048640911) (0.015929369,1.052426618) (0.015465909,1.056212325) (0.015010049,1.059998032) (0.014561752,1.063783740) (0.014120979,1.067569447) (0.013687691,1.071355154) (0.013261852,1.075140862) (0.012843423,1.078926569) (0.012432368,1.082712276) (0.012028650,1.086497983) (0.011632234,1.090283691) (0.011243082,1.094069398) (0.010861159,1.097855105) (0.010486430,1.101640812) (0.010118861,1.105426520) (0.009758415,1.109212227) (0.009405060,1.112997934) (0.009058761,1.116783641) (0.008719484,1.120569349) (0.008387196,1.124355056) (0.008061864,1.128140763) (0.007743455,1.131926470) (0.007431936,1.135712178) (0.007127276,1.139497885) (0.006829442,1.143283592) (0.006538403,1.147069299) (0.006254128,1.150855007) (0.005976584,1.154640714) (0.005705743,1.158426421) (0.005441572,1.162212128) (0.005184042,1.165997836) (0.004933123,1.169783543) (0.004688785,1.173569250) (0.004450999,1.177354958) (0.004219735,1.181140665) (0.003994964,1.184926372) (0.003776659,1.188712079) (0.003564789,1.192497787) (0.003359328,1.196283494) (0.003160247,1.200069201) (0.002967518,1.203854908) (0.002781114,1.207640616) (0.002601008,1.211426323) (0.002427172,1.215212030) (0.002259580,1.218997737) (0.002098206,1.222783445) (0.001943022,1.226569152) (0.001794004,1.230354859) (0.001651124,1.234140566) (0.001514358,1.237926274) (0.001383679,1.241711981) (0.001259063,1.245497688) (0.001140484,1.249283395) (0.001027918,1.253069103) (0.000921341,1.256854810) (0.000820726,1.260640517) (0.000726052,1.264426224) (0.000637292,1.268211932) (0.000554424,1.271997639) (0.000477425,1.275783346) (0.000406269,1.279569054) (0.000340935,1.283354761) (0.000281399,1.287140468) (0.000227639,1.290926175) (0.000179631,1.294711883) (0.000137353,1.298497590) (0.000100783,1.302283297) (0.000069899,1.306069004) (0.000044678,1.309854712) (0.000025099,1.313640419) (0.000011141,1.317426126) (0.000002782,1.321211833) (0.000000000,1.324997541)};
\addplot[red!75!black,dashed,thick] coordinates {(0.200000000,0.495986121) (0.495986121,0.495986121) (0.495986121,0.200000000)};
\addplot[black,densely dotted,thick] coordinates {(0.000000000,0.404789190) (0.329465857,0.404789190) (0.329465857,0.000000000)};
\addplot[black,densely dotted,thick] coordinates {(0.000000000,0.329465857) (0.404789190,0.329465857) (0.404789190,0.000000000)};
\addplot[gray!60,thin] coordinates {(0.000000000,0.000000000) (1.400000000,1.400000000)};
\end{axis}
\end{tikzpicture}
\hfill
\begin{tikzpicture}
\begin{axis}[width=.38\textwidth,height=.38\textwidth,scale only axis,axis lines=left,grid=major,grid style={gray!15},tick label style={font=\scriptsize},label style={font=\small},scaled ticks=false,ticklabel style={/pgf/number format/fixed},clip=true,xmin=0,xmax=0.55,ymin=0,ymax=0.55,xlabel={$a$},ylabel={$b$},title={(b) $K=L=0.28$},xtick={0,.1,.2,.3,.4,.5},ytick={0,.1,.2,.3,.4,.5}]
\addplot[draw=blue!70!black,fill=blue!12,thin] coordinates {(0.000000000,0.000000000) (0.000000000,1.324997541) (0.007361097,1.136584878) (0.014722195,1.062421524) (0.022083292,1.007100100) (0.029444390,0.961547814) (0.036805487,0.922247727) (0.044166585,0.887394817) (0.051527682,0.855916041) (0.058888780,0.827111749) (0.066249877,0.800495608) (0.073610974,0.775713269) (0.080972072,0.752497030) (0.088333169,0.730638762) (0.095694267,0.709972832) (0.103055364,0.690364846) (0.110416462,0.671703962) (0.117777559,0.653897480) (0.125138657,0.636866930) (0.132499754,0.620545184) (0.139860852,0.604874280) (0.147221949,0.589803755) (0.154583046,0.575289356) (0.161944144,0.561292008) (0.169305241,0.547777011) (0.176666339,0.534713375) (0.184027436,0.522073288) (0.191388534,0.509831678) (0.198749631,0.497965847) (0.206110729,0.486455165) (0.213471826,0.475280819) (0.220832923,0.464425593) (0.228194021,0.453873684) (0.235555118,0.443610545) (0.242916216,0.433622749) (0.250277313,0.423897874) (0.257638411,0.414424395) (0.258423233,0.413428735) (0.258423333,0.413428609) (0.258424233,0.413427468) (0.258433233,0.413416067) (0.259709812,0.411802411) (0.260996390,0.410183349) (0.262282969,0.408571497) (0.263569547,0.406966800) (0.264856126,0.405369207) (0.264999508,0.405191602) (0.266142704,0.403778666) (0.267429283,0.402195126) (0.268715861,0.400618536) (0.270002440,0.399048846) (0.271289018,0.397486006) (0.272360606,0.396189516) (0.272575597,0.395929969) (0.273862175,0.394380685) (0.275148754,0.392838108) (0.276435332,0.391302189) (0.277721911,0.389772883) (0.279008489,0.388250143) (0.279721703,0.387408826) (0.280000000,0.387081084) (0.280295068,0.387081084) (0.281581646,0.387081084) (0.282868225,0.387081084) (0.284154803,0.387081084) (0.285441382,0.387081084) (0.286727960,0.387081084) (0.287082800,0.387081084) (0.288014539,0.387081084) (0.289301117,0.387081084) (0.290587696,0.387081084) (0.291874274,0.387081084) (0.293160853,0.387081084) (0.294443898,0.387081084) (0.294447431,0.387081084) (0.295734010,0.387081084) (0.297020588,0.386063178) (0.298307167,0.384315962) (0.299593745,0.382591021) (0.300880324,0.380887452) (0.301804995,0.379675816) (0.302166902,0.379204419) (0.303453481,0.377541152) (0.304740059,0.375896937) (0.306026638,0.374271113) (0.307313216,0.372663064) (0.308599795,0.371072218) (0.309166093,0.370377308) (0.309886374,0.369498042) (0.311172952,0.367940038) (0.312459531,0.366397741) (0.313746109,0.364870716) (0.315032688,0.363358554) (0.316319266,0.361860873) (0.316527190,0.361620168) (0.317605845,0.360377316) (0.318892423,0.358907544) (0.320179002,0.357451241) (0.321465580,0.356008109) (0.322752159,0.354577867) (0.323888288,0.353325381) (0.324038737,0.353160252) (0.325325316,0.351755014) (0.326611894,0.350361919) (0.327898473,0.348980745) (0.329185051,0.347611284) (0.330471630,0.346253339) (0.331249385,0.345437948) (0.331362388,0.345319819) (0.331758208,0.344906724) (0.333044787,0.343571264) (0.334331365,0.342246794) (0.335617944,0.340933157) (0.336904522,0.339630208) (0.338191101,0.338337806) (0.338610483,0.337918783) (0.339477679,0.337055821) (0.340764258,0.335784131) (0.342050836,0.334522619) (0.343337415,0.333271176) (0.344623993,0.332029701) (0.345910572,0.330798097) (0.345971580,0.330739939) (0.347197150,0.329576274) (0.348483729,0.328364148) (0.349770307,0.327161640) (0.351056886,0.325968676) (0.352343464,0.324785189) (0.353332677,0.323881643) (0.353630043,0.323611114) (0.354916621,0.322446394) (0.356203200,0.321290974) (0.357489778,0.320144804) (0.358776357,0.319007839) (0.360062935,0.317880038) (0.360693775,0.317330388) (0.361349514,0.316761364) (0.362636092,0.315651783) (0.363922671,0.314551265) (0.365209249,0.313459786) (0.366495828,0.312377322) (0.367782407,0.311303856) (0.368054872,0.311077674) (0.369068985,0.310239370) (0.370355564,0.309183855) (0.371642142,0.308137301) (0.372928721,0.307099704) (0.374215299,0.306071059) (0.375415970,0.305119178) (0.375501878,0.305051370) (0.376788456,0.304040639) (0.378075035,0.303038874) (0.379361613,0.302046085) (0.380648192,0.301062285) (0.381934770,0.300087490) (0.382777067,0.299454197) (0.383221349,0.299121719) (0.384507927,0.298164993) (0.385794506,0.297217337) (0.387081084,0.296278780) (0.387081094,0.279999992) (0.390138165,0.277414101) (0.397499262,0.271278082) (0.404860360,0.265267111) (0.412221457,0.259377754) (0.419582555,0.253606727) (0.426943652,0.247950890) (0.434304749,0.242407238) (0.441665847,0.236972893) (0.449026944,0.231645095) (0.456388042,0.226421198) (0.463749139,0.221298664) (0.471110237,0.216275051) (0.478471334,0.211348017) (0.485832432,0.206515308) (0.493193529,0.201774755) (0.500554626,0.197124271) (0.507915724,0.192561845) (0.515276821,0.188085540) (0.522637919,0.183693487) (0.529999016,0.179383885) (0.537360114,0.175154995) (0.544721211,0.171005136) (0.552082309,0.166932688) (0.559443406,0.162936082) (0.566804503,0.159013802) (0.574165601,0.155164383) (0.581526698,0.151386406) (0.588887796,0.147678498) (0.596248893,0.144039328) (0.603609991,0.140467608) (0.610971088,0.136962091) (0.618332186,0.133521565) (0.625693283,0.130144856) (0.633054381,0.126830827) (0.640415478,0.123578371) (0.647776575,0.120386416) (0.655137673,0.117253920) (0.662498770,0.114179871) (0.669859868,0.111163285) (0.677220965,0.108203208) (0.684582063,0.105298709) (0.691943160,0.102448886) (0.699304258,0.099652858) (0.706665355,0.096909771) (0.714026452,0.094218792) (0.721387550,0.091579111) (0.728748647,0.088989938) (0.736109745,0.086450504) (0.743470842,0.083960061) (0.750831940,0.081517878) (0.758193037,0.079123243) (0.765554135,0.076775464) (0.772915232,0.074473864) (0.780276329,0.072217783) (0.787637427,0.070006576) (0.794998524,0.067839616) (0.802359622,0.065716290) (0.809720719,0.063635999) (0.817081817,0.061598159) (0.824442914,0.059602198) (0.831804012,0.057647560) (0.839165109,0.055733699) (0.846526207,0.053860083) (0.853887304,0.052026192) (0.861248401,0.050231518) (0.868609499,0.048475563) (0.875970596,0.046757841) (0.883331694,0.045077876) (0.890692791,0.043435204) (0.898053889,0.041829369) (0.905414986,0.040259926) (0.912776084,0.038726439) (0.920137181,0.037228481) (0.927498278,0.035765635) (0.934859376,0.034337492) (0.942220473,0.032943652) (0.949581571,0.031583721) (0.956942668,0.030257316) (0.964303766,0.028964061) (0.971664863,0.027703587) (0.979025961,0.026475532) (0.986387058,0.025279541) (0.993748155,0.024115269) (1.001109253,0.022982374) (1.008470350,0.021880522) (1.015831448,0.020809386) (1.023192545,0.019768645) (1.030553643,0.018757984) (1.037914740,0.017777094) (1.045275838,0.016825673) (1.052636935,0.015903422) (1.059998032,0.015010049) (1.067359130,0.014145269) (1.074720227,0.013308801) (1.082081325,0.012500367) (1.089442422,0.011719697) (1.096803520,0.010966526) (1.104164617,0.010240591) (1.111525715,0.009541636) (1.118886812,0.008869409) (1.126247910,0.008223663) (1.133609007,0.007604153) (1.140970104,0.007010642) (1.148331202,0.006442895) (1.155692299,0.005900680) (1.163053397,0.005383770) (1.170414494,0.004891944) (1.177775592,0.004424981) (1.185136689,0.003982667) (1.192497787,0.003564789) (1.199858884,0.003171140) (1.207219981,0.002801514) (1.214581079,0.002455710) (1.221942176,0.002133531) (1.229303274,0.001834781) (1.236664371,0.001559269) (1.244025469,0.001306806) (1.251386566,0.001077207) (1.258747664,0.000870290) (1.266108761,0.000685874) (1.273469858,0.000523784) (1.280830956,0.000383846) (1.288192053,0.000265887) (1.295553151,0.000169741) (1.302914248,0.000095241) (1.310275346,0.000042224) (1.317636443,0.000010530) (1.324997541,0.000000000) (1.324997541,0.000000000) (0.000000000,0.000000000)} --cycle;
\addplot[black,thin] coordinates {(1.324997541,0.000000000) (1.188552913,0.003785707) (1.134006840,0.007571415) (1.092952676,0.011357122) (1.058888838,0.015142829) (1.029296013,0.018928536) (1.002881666,0.022714244) (0.978877686,0.026499951) (0.956783427,0.030285658) (0.936251012,0.034071365) (0.917027059,0.037857073) (0.898920219,0.041642780) (0.881781775,0.045428487) (0.865493420,0.049214194) (0.849959193,0.052999902) (0.835099982,0.056785609) (0.820849644,0.060571316) (0.807152213,0.064357023) (0.793959825,0.068142731) (0.781231170,0.071928438) (0.768930295,0.075714145) (0.757025678,0.079499852) (0.745489502,0.083285560) (0.734297067,0.087071267) (0.723426322,0.090856974) (0.712857486,0.094642681) (0.702572730,0.098428389) (0.692555918,0.102214096) (0.682792390,0.105999803) (0.673268779,0.109785511) (0.663972857,0.113571218) (0.654893403,0.117356925) (0.646020089,0.121142632) (0.637343385,0.124928340) (0.628854475,0.128714047) (0.620545184,0.132499754) (0.612407913,0.136285461) (0.604435585,0.140071169) (0.596621597,0.143856876) (0.588959773,0.147642583) (0.581444330,0.151428290) (0.574069843,0.155213998) (0.566831211,0.158999705) (0.559723638,0.162785412) (0.552742598,0.166571119) (0.545883824,0.170356827) (0.539143280,0.174142534) (0.532517150,0.177928241) (0.526001815,0.181713948) (0.519593846,0.185499656) (0.513289985,0.189285363) (0.507087134,0.193071070) (0.500982348,0.196856777) (0.494972820,0.200642485) (0.489055873,0.204428192) (0.483228956,0.208213899) (0.477489630,0.211999606) (0.471835563,0.215785314) (0.466264526,0.219571021) (0.460774384,0.223356728) (0.455363093,0.227142436) (0.450028691,0.230928143) (0.444769296,0.234713850) (0.439583102,0.238499557) (0.434468374,0.242285265) (0.429423443,0.246070972) (0.424446703,0.249856679) (0.419536610,0.253642386) (0.414691676,0.257428094) (0.409910468,0.261213801) (0.405191602,0.264999508) (0.400533745,0.268785215) (0.395935610,0.272570923) (0.391395953,0.276356630) (0.386913576,0.280142337) (0.382487315,0.283928044) (0.378116050,0.287713752) (0.373798695,0.291499459) (0.369534198,0.295285166) (0.365321542,0.299070873) (0.361159741,0.302856581) (0.357047840,0.306642288) (0.352984914,0.310427995) (0.348970063,0.314213702) (0.345002418,0.317999410) (0.341081132,0.321785117) (0.337205385,0.325570824) (0.333374378,0.329356532) (0.329587339,0.333142239) (0.325843514,0.336927946) (0.322142171,0.340713653) (0.318482599,0.344499361) (0.314864106,0.348285068) (0.311286017,0.352070775) (0.307747679,0.355856482) (0.304248451,0.359642190) (0.300787714,0.363427897) (0.297364860,0.367213604) (0.293979301,0.370999311) (0.290630462,0.374785019) (0.287317780,0.378570726) (0.284040711,0.382356433) (0.280798721,0.386142140) (0.277591289,0.389927848) (0.274417908,0.393713555) (0.271278082,0.397499262) (0.268171327,0.401284969) (0.265097172,0.405070677) (0.262055155,0.408856384) (0.259044824,0.412642091) (0.256065739,0.416427798) (0.253117471,0.420213506) (0.250199597,0.423999213) (0.247311707,0.427784920) (0.244453398,0.431570628) (0.241624276,0.435356335) (0.238823957,0.439142042) (0.236052064,0.442927749) (0.233308227,0.446713457) (0.230592086,0.450499164) (0.227903288,0.454284871) (0.225241487,0.458070578) (0.222606343,0.461856286) (0.219997525,0.465641993) (0.217414708,0.469427700) (0.214857573,0.473213407) (0.212325807,0.476999115) (0.209819106,0.480784822) (0.207337169,0.484570529) (0.204879701,0.488356236) (0.202446415,0.492141944) (0.200037027,0.495927651) (0.197651261,0.499713358) (0.195288844,0.503499065) (0.192949509,0.507284773) (0.190632994,0.511070480) (0.188339042,0.514856187) (0.186067400,0.518641894) (0.183817822,0.522427602) (0.181590063,0.526213309) (0.179383885,0.529999016) (0.177199054,0.533784724) (0.175035338,0.537570431) (0.172892513,0.541356138) (0.170770355,0.545141845) (0.168668646,0.548927553) (0.166587172,0.552713260) (0.164525721,0.556498967) (0.162484088,0.560284674) (0.160462067,0.564070382) (0.158459458,0.567856089) (0.156476065,0.571641796) (0.154511695,0.575427503) (0.152566156,0.579213211) (0.150639262,0.582998918) (0.148730829,0.586784625) (0.146840674,0.590570332) (0.144968621,0.594356040) (0.143114494,0.598141747) (0.141278120,0.601927454) (0.139459330,0.605713161) (0.137657957,0.609498869) (0.135873836,0.613284576) (0.134106805,0.617070283) (0.132356705,0.620855990) (0.130623380,0.624641698) (0.128906674,0.628427405) (0.127206436,0.632213112) (0.125522516,0.635998819) (0.123854767,0.639784527) (0.122203043,0.643570234) (0.120567202,0.647355941) (0.118947101,0.651141649) (0.117342604,0.654927356) (0.115753572,0.658713063) (0.114179871,0.662498770) (0.112621368,0.666284478) (0.111077932,0.670070185) (0.109549434,0.673855892) (0.108035748,0.677641599) (0.106536747,0.681427307) (0.105052309,0.685213014) (0.103582311,0.688998721) (0.102126633,0.692784428) (0.100685157,0.696570136) (0.099257767,0.700355843) (0.097844346,0.704141550) (0.096444782,0.707927257) (0.095058963,0.711712965) (0.093686778,0.715498672) (0.092328117,0.719284379) (0.090982875,0.723070086) (0.089650944,0.726855794) (0.088332220,0.730641501) (0.087026600,0.734427208) (0.085733982,0.738212915) (0.084454265,0.741998623) (0.083187351,0.745784330) (0.081933141,0.749570037) (0.080691539,0.753355745) (0.079462449,0.757141452) (0.078245778,0.760927159) (0.077041432,0.764712866) (0.075849320,0.768498574) (0.074669351,0.772284281) (0.073501436,0.776069988) (0.072345487,0.779855695) (0.071201415,0.783641403) (0.070069136,0.787427110) (0.068948565,0.791212817) (0.067839616,0.794998524) (0.066742208,0.798784232) (0.065656259,0.802569939) (0.064581687,0.806355646) (0.063518413,0.810141353) (0.062466358,0.813927061) (0.061425443,0.817712768) (0.060395593,0.821498475) (0.059376730,0.825284182) (0.058368779,0.829069890) (0.057371667,0.832855597) (0.056385319,0.836641304) (0.055409664,0.840427011) (0.054444629,0.844212719) (0.053490143,0.847998426) (0.052546137,0.851784133) (0.051612542,0.855569841) (0.050689288,0.859355548) (0.049776308,0.863141255) (0.048873536,0.866926962) (0.047980905,0.870712670) (0.047098350,0.874498377) (0.046225806,0.878284084) (0.045363209,0.882069791) (0.044510497,0.885855499) (0.043667606,0.889641206) (0.042834475,0.893426913) (0.042011043,0.897212620) (0.041197249,0.900998328) (0.040393035,0.904784035) (0.039598340,0.908569742) (0.038813106,0.912355449) (0.038037275,0.916141157) (0.037270791,0.919926864) (0.036513596,0.923712571) (0.035765635,0.927498278) (0.035026853,0.931283986) (0.034297194,0.935069693) (0.033576605,0.938855400) (0.032865031,0.942641107) (0.032162421,0.946426815) (0.031468721,0.950212522) (0.030783879,0.953998229) (0.030107845,0.957783937) (0.029440567,0.961569644) (0.028781995,0.965355351) (0.028132079,0.969141058) (0.027490771,0.972926766) (0.026858020,0.976712473) (0.026233780,0.980498180) (0.025618001,0.984283887) (0.025010638,0.988069595) (0.024411643,0.991855302) (0.023820970,0.995641009) (0.023238572,0.999426716) (0.022664405,1.003212424) (0.022098424,1.006998131) (0.021540584,1.010783838) (0.020990842,1.014569545) (0.020449152,1.018355253) (0.019915473,1.022140960) (0.019389762,1.025926667) (0.018871976,1.029712374) (0.018362073,1.033498082) (0.017860012,1.037283789) (0.017365751,1.041069496) (0.016879251,1.044855203) (0.016400470,1.048640911) (0.015929369,1.052426618) (0.015465909,1.056212325) (0.015010049,1.059998032) (0.014561752,1.063783740) (0.014120979,1.067569447) (0.013687691,1.071355154) (0.013261852,1.075140862) (0.012843423,1.078926569) (0.012432368,1.082712276) (0.012028650,1.086497983) (0.011632234,1.090283691) (0.011243082,1.094069398) (0.010861159,1.097855105) (0.010486430,1.101640812) (0.010118861,1.105426520) (0.009758415,1.109212227) (0.009405060,1.112997934) (0.009058761,1.116783641) (0.008719484,1.120569349) (0.008387196,1.124355056) (0.008061864,1.128140763) (0.007743455,1.131926470) (0.007431936,1.135712178) (0.007127276,1.139497885) (0.006829442,1.143283592) (0.006538403,1.147069299) (0.006254128,1.150855007) (0.005976584,1.154640714) (0.005705743,1.158426421) (0.005441572,1.162212128) (0.005184042,1.165997836) (0.004933123,1.169783543) (0.004688785,1.173569250) (0.004450999,1.177354958) (0.004219735,1.181140665) (0.003994964,1.184926372) (0.003776659,1.188712079) (0.003564789,1.192497787) (0.003359328,1.196283494) (0.003160247,1.200069201) (0.002967518,1.203854908) (0.002781114,1.207640616) (0.002601008,1.211426323) (0.002427172,1.215212030) (0.002259580,1.218997737) (0.002098206,1.222783445) (0.001943022,1.226569152) (0.001794004,1.230354859) (0.001651124,1.234140566) (0.001514358,1.237926274) (0.001383679,1.241711981) (0.001259063,1.245497688) (0.001140484,1.249283395) (0.001027918,1.253069103) (0.000921341,1.256854810) (0.000820726,1.260640517) (0.000726052,1.264426224) (0.000637292,1.268211932) (0.000554424,1.271997639) (0.000477425,1.275783346) (0.000406269,1.279569054) (0.000340935,1.283354761) (0.000281399,1.287140468) (0.000227639,1.290926175) (0.000179631,1.294711883) (0.000137353,1.298497590) (0.000100783,1.302283297) (0.000069899,1.306069004) (0.000044678,1.309854712) (0.000025099,1.313640419) (0.000011141,1.317426126) (0.000002782,1.321211833) (0.000000000,1.324997541)};
\addplot[red!75!black,dashed,thick] coordinates {(0.280000000,0.387081084) (0.387081084,0.387081084) (0.387081084,0.280000000)};
\addplot[black,densely dotted,thick] coordinates {(0.000000000,0.268279800) (0.272233321,0.268279800) (0.272233321,0.000000000)};
\addplot[black,densely dotted,thick] coordinates {(0.000000000,0.272233321) (0.268279800,0.272233321) (0.268279800,0.000000000)};
\addplot[gray!60,thin] coordinates {(0.000000000,0.000000000) (0.550000000,0.550000000)};
\addplot[only marks,mark=x,mark size=2.5pt,red!75!black] coordinates {(.35,.35)};
\end{axis}
\end{tikzpicture}
\caption{The qubit pair \eqref{eq:qubit} with $p=0.95$, $q=0.05$: $\cZ(K,L)$ (blue shading) for (a) $K=L=0.2$, below the fidelity threshold, and (b) $K=L=0.28$, above it. Panel (b) shows an enlarged range of exponents. Thin black: boundary of $\cH$. Dashed red: the part of the boundary of $\cC^{\Pet}(K,L)=\cH\cup(0,H_K]^2$ outside $\cH$. Dotted black: the two rectangles $(0,\protect\overleftarrow{H}_K]\times(0,\widetilde H_K]$ and $(0,\widetilde H_K]\times(0,\protect\overleftarrow{H}_K]$ that, together with $\cH$, form the region achievable by \cite[Proposition~6]{JiRegula25}; in (b) these rectangles are contained in $\cH$. The cross in (b) marks $(0.35,0.35)$, which satisfies (P1) and (P2) but violates (F). The thin gray line is the diagonal $a=b$.}
\label{fig:regions}
\end{figure}
Let $d=2$, $0<q<\frac12$, $p=1-q$, and
\begin{equation}\label{eq:qubit}
 \rho=\begin{pmatrix}p&0\\0&q\end{pmatrix},\qquad
 \sigma=V\rho\,V^*,\qquad V=\frac1{\sqrt2}\begin{pmatrix}1&1\\1&-1\end{pmatrix}.
\end{equation}
Since $V=V^*=V^{-1}$, this pair is symmetric in the sense of Corollary~\ref{cor:symmetric}. A direct computation gives
\begin{equation}\label{eq:qubit-psi}
\begin{aligned}
 \psi(\alpha,z)&=-\log\bigl(\lambda_+^{\,z}+\lambda_-^{\,z}\bigr),\qquad
 \lambda_\pm=\frac{\mathsf t\pm\sqrt{\mathsf t^2-4\,\mathsf d}}{2},\\
 \mathsf t&=\tfrac12\bigl(p^{\frac{\alpha}{z}}+q^{\frac{\alpha}{z}}\bigr)\bigl(p^{\frac{1-\alpha}{z}}+q^{\frac{1-\alpha}{z}}\bigr),\qquad
 \mathsf d=(pq)^{\frac1z}.
\end{aligned}
\end{equation}
For this family, $r_0<r_{\rm c}$ for every $p\in(\frac12,1)$, because $\rho$ and $\sigma$ do not commute; see \eqref{eq:sym-threshold-equality}. In the following we take $p=0.95$ and $q=0.05$. Then
\begin{align*}
 C&=\log\frac{20}{10+\sqrt{19}}\approx0.33136239,&
 r_0&=\tfrac12\log\frac{200}{119}\approx0.25959694,\\
 \Dsand_{1/2}(\rho\|\sigma)&=\log\frac{200}{119}\approx0.51919387,&
 r_{\rm c}&\approx0.30056351.
\end{align*}
To specify the comparison with \cite[Proposition~6]{JiRegula25}, define, for this symmetric pair and $r>0$,
\begin{equation}\label{eq:JR-quantities-main}
 \widetilde H_r\eqdef\sup_{0<s<1}\frac{\psi(s,s)-(1-s)r}{s},\qquad
 \overleftarrow H_r\eqdef\sup_{0<s<1}\frac{\psi(s,1-s)-(1-s)r}{s}.
\end{equation}
These are the Hoeffding-type quantities built from the sandwiched and reverse sandwiched divergences, respectively. The equal-error exponent guaranteed by that proposition is
$\max\{C,\min\{\widetilde H_r,\overleftarrow H_r\}\}$. The corresponding achievable region contains $\cH$ and the two rectangles displayed in Fig.~\ref{fig:regions}.

Table~\ref{tab:example} and Fig.~\ref{fig:equal} compare the optimal equal-error exponent $e^\ast(r)$ of Corollary~\ref{cor:symmetric} with the Petz-type converse $\max\{C,H_r\}$ and the exponent achievable by \cite[Proposition~6]{JiRegula25}. For $r\le r_0$ the Petz-type converse is attained (Remark~\ref{rem:petz}). At $r_0$ the optimal exponent has a downward jump:
\[
 e^\ast(r_0)=H_{r_0}\approx0.41194480,
 \qquad \lim_{r\downarrow r_0}e^\ast(r)
 =r_0+\tfrac12\psi_{1/2}'(\tfrac12)\approx0.40580607.
\]
The jump is approximately $0.00613873$. 
For this pair, (F) is binding throughout $r_0<r<r_{\rm c}$. For example,
\[
 H_{0.28}\approx0.38708108,
 \qquad e^\ast(0.28)\approx0.33826444,
 \qquad z_{0.28}\approx0.77795.
\]
Here the optimal witness $(\frac12,z_{0.28})$ lies in the interior of $\cD$. For $r\ge r_{\rm c}$ the optimal equal-error exponent is $C$. Take $(a,b,K,L)=(0.35,0.35,0.28,0.28)$, one can verify that (P1) and (P2) hold with $s=\frac12$ but \begin{equation*}
 \psi(\alpha,z)\le f(z)\le \ell(z)
 <(2z-1)0.35+2(1-z)0.28
 \qquad ((\alpha,z)\in\cD).
\end{equation*}
which implies that (F) fails throughout $\cD$.

Fig.~\ref{fig:regions} shows the regions $\cZ(K,L)$ for $K=L=0.2$ and $K=L=0.28$. In the first case $K+L<\Dsand_{1/2}(\rho\|\sigma)$ and
$\cZ(K,L)=\cC^{\Pet}(K,L)=\cH\cup(0,H_{0.2}]^2$, with $H_{0.2}\approx0.49598612$. In the second case the fidelity threshold is exceeded and (F) cuts away part of the Petz region; the point $(0.35,0.35)$ excluded above is marked in the figure. The boundaries are computed directly from (P1), (P2), and (F).

\section{Sector Geometry and the Pair Characterization}\label{sec:sectors}

This section proves Theorem~\ref{thm:pair}. The first step uses the block decomposition of the tensor powers. The remaining steps describe the geometry of the resulting limiting sector set.

\subsection{Block decomposition}

Let $\fA_n\eqdef\spn\{A^{\otimes n}:A\in M_d\}\subseteq M_d^{\otimes n}$, and put $
 N_n\eqdef(n+1)^{d^2-1}.$

\begin{lemma}[Block structure \cite{Harrow13church,Christandl06thesis,GoodmanWallach09,ChristandlMitchison06}]\label{lem:blocks}
For every $n\ge1$ there are a finite set $\Lambda_n$, positive integers $D_\lambda,m_\lambda$ $(\lambda\in\Lambda_n)$ and a unitary $W_n:(\bbC^d)^{\otimes n}\to\bigoplus_{\lambda\in\Lambda_n}\bbC^{D_\lambda}\otimes\bbC^{m_\lambda}$ such that
\begin{equation}\label{eq:block-form}
 W_n\,\fA_n\,W_n^*=\bigoplus_{\lambda\in\Lambda_n}M_{D_\lambda}\otimes\one_{m_\lambda}
\end{equation}
and
\begin{equation}\label{eq:SW-bounds}
 |\Lambda_n|\le(n+1)^{d},\qquad D_\lambda\le(n+1)^{\frac{d(d-1)}{2}}\quad(\lambda\in\Lambda_n).
\end{equation}
\end{lemma}

Lemma~\ref{lem:blocks} is standard; we indicate where its parts can be found. By \cite[Corollary~4]{Harrow13church}, $\fA_n$ is the commutant of the representation of the symmetric group $S_n$ on $(\bbC^d)^{\otimes n}$ that permutes the tensor factors. By Schur--Weyl duality \cite[Theorems~1.6, 1.8 and~1.10]{Christandl06thesis} (see also \cite[Section~9.1]{GoodmanWallach09}), $(\bbC^d)^{\otimes n}\cong\bigoplus_\lambda U_\lambda\otimes V_\lambda$, where $\lambda$ runs over the Young diagrams with $n$ boxes and at most $d$ rows, $U_\lambda$ and $V_\lambda$ are irreducible representations of the unitary group $U(d)$ and of $S_n$, respectively, and this commutant is $\bigoplus_\lambda\operatorname{End}(U_\lambda)\otimes\one_{V_\lambda}$. This is \eqref{eq:block-form}, with $\Lambda_n$ the set of these Young diagrams, $D_\lambda=\dim U_\lambda$ and $m_\lambda=\dim V_\lambda$. A Young diagram with at most $d$ rows and $n$ boxes is determined by its $d$ row lengths, each of which lies in $\{0,1,\dots,n\}$; this gives the first bound in \eqref{eq:SW-bounds}. The second bound is \cite[eq.~(7)]{ChristandlMitchison06}; see also \cite[eq.~(1.22)]{Christandl06thesis}. For further background see, e.g., \cite{KeylWerner01,Harrow05,Hayashi17book}; the decomposition has been used in quantum hypothesis testing in \cite{Hayashi02,HMH09}.

We use Lemma~\ref{lem:blocks} only through the following two consequences. First, it follows from \eqref{eq:SW-bounds} and the definition of $N_n$ that
\begin{equation}\label{eq:block-bounds}
 |\Lambda_n|\le N_n,\qquad D_\lambda\le N_n^{1/2}\ \ (\lambda\in\Lambda_n),\qquad\sum_{\lambda\in\Lambda_n}D_\lambda\le|\Lambda_n|\max_{\lambda\in\Lambda_n}D_\lambda\le N_n .
\end{equation}
Second, $\rhon$ and $\sigman$ are positive definite elements of $\fA_n$. Hence, by \eqref{eq:block-form}, there are positive definite matrices $\rho_\lambda,\sigma_\lambda\in M_{D_\lambda}$ with $W_n\rhon W_n^*=\bigoplus_\lambda\rho_\lambda\otimes\one_{m_\lambda}$ and $W_n\sigman W_n^*=\bigoplus_\lambda\sigma_\lambda\otimes\one_{m_\lambda}$. Since real powers of a block-diagonal positive definite matrix are taken block by block and $(\rho_\lambda\otimes\one_{m_\lambda})^t=\rho_\lambda^{\,t}\otimes\one_{m_\lambda}$, we obtain, for all $t\in\bbR$,
\begin{equation}\label{eq:block-states}
 W_n(\rhon)^{t}W_n^*=\bigoplus_{\lambda\in\Lambda_n}\rho_\lambda^{\,t}\otimes\one_{m_\lambda},\qquad
 W_n(\sigman)^{t}W_n^*=\bigoplus_{\lambda\in\Lambda_n}\sigma_\lambda^{\,t}\otimes\one_{m_\lambda}.
\end{equation}
\emph{From now on we identify $(\bbC^d)^{\otimes n}$ with $\bigoplus_\lambda\bbC^{D_\lambda}\otimes\bbC^{m_\lambda}$ and omit $W_n$.}

\subsection{Sectors}

\begin{definition}[Sectors and their exponents]\label{def:sector}
A \emph{sector} at blocklength $n$ is a triple $\theta=(\lambda,P,Q)$ consisting of a block $\lambda\in\Lambda_n$, the orthogonal projection $P$ onto an eigenspace of $\rho_\lambda$ and the orthogonal projection $Q$ onto an eigenspace of $\sigma_\lambda$. With $\varrho_P$ and $\varsigma_Q$ the corresponding eigenvalues ($\rho_\lambda P=\varrho_PP$, $\sigma_\lambda Q=\varsigma_QQ$), the three \emph{exponents} of the sector are
\begin{equation}\label{eq:sector-exp}
\begin{aligned}
 \drho(\theta)&\eqdef-\frac1n\log\bigl(m_\lambda\varrho_P\bigr),\qquad
 \dsig(\theta)\eqdef-\frac1n\log\bigl(m_\lambda\varsigma_Q\bigr),\\
 \chi(\theta)&\eqdef-\frac1n\log\norm{PQ}_\infty\in[0,\infty].
\end{aligned}
\end{equation}
Let $\Theta_n$ denote the set of sectors at blocklength $n$ with $PQ\ne0$, and $\bv(\theta)\eqdef(\drho(\theta),\allowbreak\dsig(\theta),\allowbreak\chi(\theta))\in\bbR^3$ for $\theta\in\Theta_n$.
\end{definition}

Each $\Theta_n$ is finite, and it is non-empty: in every block, $\sum_PP=\one=\sum_QQ$ (sums over the eigenprojections of $\rho_\lambda$ and of $\sigma_\lambda$), hence $\sum_{P,Q}PQ=\one\ne0$, so that $PQ\ne0$ for at least one pair $(P,Q)$.

Thus $e^{-n\drho(\theta)}$ is the weight that $\rhon$ gives to a single eigenvector in $\ran P$, counted with the multiplicity $m_\lambda$ of the block, and $e^{-n\chi(\theta)}$ is the largest overlap between unit vectors of the two eigenspaces: for $\theta=(\lambda,P,Q)\in\Theta_n$,
\begin{equation}\label{eq:sector-overlap}
 \abs{\ip xy}=\abs{\bra xPQ\ket y}\le\norm{PQ}_\infty=e^{-n\chi(\theta)}\qquad\text{for all unit vectors } x\in\ran P,\ y\in\ran Q,
\end{equation}
and equality holds for some pair of unit vectors $x\in\ran P$, $y\in\ran Q$. (Let $u,v$ be unit vectors with $\abs{\bra uPQ\ket v}=\norm{PQ}_\infty>0$; then $Pu\ne0$ and $Qv\ne0$ because $\ip{Pu}{Qv}=\bra uPQ\ket v\ne0$. Put $x\eqdef Pu/\norm{Pu}$ and $y\eqdef Qv/\norm{Qv}$; then $\abs{\ip xy}=\abs{\ip{Pu}{Qv}}/(\norm{Pu}\norm{Qv})\ge\abs{\ip{Pu}{Qv}}=\norm{PQ}_\infty$, since $\norm{Pu},\norm{Qv}\le1$.) 

Let
$ \lmax\eqdef\max\bigl\{\log\norm{\rho^{-1}}_\infty,\ \log\norm{\sigma^{-1}}_\infty\bigr\}.$
Then every sector satisfies
\begin{equation}\label{eq:sector-bounds}
 0\le\drho(\theta)\le\lmax,\qquad 0\le\dsig(\theta)\le\lmax,\qquad\chi(\theta)\ge0 .
\end{equation}
Indeed,
\begin{align*}
 e^{-n\drho(\theta)}&=m_\lambda\varrho_P\overset{(a)}{\le}m_\lambda\varrho_P\tr P=\tr\bigl[(\varrho_PP)\otimes\one_{m_\lambda}\bigr]\overset{(b)}{\le}\tr\bigl[\rho_\lambda\otimes\one_{m_\lambda}\bigr]\overset{(c)}{\le}\tr\rhon=1,\\
 e^{-n\drho(\theta)}&=m_\lambda\varrho_P\overset{(d)}{\ge}\varrho_P\overset{(e)}{\ge}\norm{\rho^{-1}}_\infty^{-n}\ge e^{-n\lmax},\qquad
 e^{-n\chi(\theta)}=\norm{PQ}_\infty\le\norm{P}_\infty\norm{Q}_\infty=1,
\end{align*}
where $(a)$ follows from $\tr P\ge1$, $(b)$ follows from $\varrho_PP\le\rho_\lambda$, $(c)$ follows from \eqref{eq:block-states} with $t=1$, $(d)$ follows from $m_\lambda\ge1$, and $(e)$ holds because $\varrho_P$ is an eigenvalue of $\rhon$ by \eqref{eq:block-states}, and the smallest eigenvalue of $\rhon$ is $\norm{\rho^{-1}}_\infty^{-n}$. The bounds for $\dsig$ follow in the same way from the second identity of \eqref{eq:block-states}.

\subsection{The \texorpdfstring{$\alpha$--$z$}{alpha-z} quantity as a support function}

\begin{lemma}[Sector representation]\label{lem:support}
For all $\alpha\in\bbR$, $z>0$ and $n\ge1$,
\begin{equation}\label{eq:support-n}
 \psi_n(\alpha,z)-\kappa_z\frac{\log(n+1)}{n}\ \le\ \psi(\alpha,z)\ \le\ \psi_n(\alpha,z),
\end{equation}
where
\begin{equation}\label{eq:mn}
 \psi_n(\alpha,z)\eqdef\min_{\theta\in\Theta_n}\bigl[\alpha\,\drho(\theta)+(1-\alpha)\,\dsig(\theta)+2z\,\chi(\theta)\bigr],\qquad
 \kappa_z\eqdef(d^2-1)\bigl(\tfrac32+2z\bigr).
\end{equation}
In particular, for every $n\ge1$ and every sector $\theta\in\Theta_n$,
\begin{equation}\label{eq:support-each}
 \alpha\,\drho(\theta)+(1-\alpha)\,\dsig(\theta)+2z\,\chi(\theta)\ \ge\ \psi(\alpha,z).
\end{equation}
\end{lemma}

\begin{proof}
Let $R\eqdef\rho^{\frac{\alpha}{2z}}\sigma^{\frac{1-\alpha}{2z}}$ as in \eqref{eq:Qnorm}, and for $\lambda\in\Lambda_n$ let $R_\lambda\eqdef\rho_\lambda^{\frac{\alpha}{2z}}\sigma_\lambda^{\frac{1-\alpha}{2z}}$. By \eqref{eq:block-states},
\begin{equation}\label{eq:R-blocks}
 R^{\otimes n}=(\rhon)^{\frac{\alpha}{2z}}(\sigman)^{\frac{1-\alpha}{2z}}=\bigoplus_{\lambda\in\Lambda_n}R_\lambda\otimes\one_{m_\lambda}.
\end{equation}
Therefore
\begin{equation}\label{eq:Qn-blocks}
 e^{-n\psi(\alpha,z)}\overset{(a)}{=}\bigl(\norm{R}_{2z}^{2z}\bigr)^n\overset{(b)}{=}\norm{R^{\otimes n}}_{2z}^{2z}\overset{(c)}{=}\sum_{\lambda\in\Lambda_n}m_\lambda\norm{R_\lambda}_{2z}^{2z},
\end{equation}
where $(a)$ follows from \eqref{eq:Qaz} and \eqref{eq:Qnorm}; $(b)$ holds because the singular values of $R^{\otimes n}$ are the $n$-fold products of the singular values of $R$; and $(c)$ follows from \eqref{eq:R-blocks}, because the singular values of $\bigoplus_\lambda R_\lambda\otimes\one_{m_\lambda}$ are those of the $R_\lambda$, each repeated $m_\lambda$ times. Let $\rho_\lambda=\sum_P\varrho_PP$ and $\sigma_\lambda=\sum_Q\varsigma_QQ$ be the spectral decompositions. Then $\rho_\lambda^{\,t}=\sum_P\varrho_P^{\,t}P$, $\sigma_\lambda^{\,t}=\sum_Q\varsigma_Q^{\,t}Q$, and hence
\begin{equation}\label{eq:Z-expansion}
 R_\lambda=\sum_{P,Q}\varrho_P^{\frac{\alpha}{2z}}\varsigma_Q^{\frac{1-\alpha}{2z}}\,PQ,\qquad
 PR_\lambda Q=\varrho_P^{\frac{\alpha}{2z}}\varsigma_Q^{\frac{1-\alpha}{2z}}\,PQ .
\end{equation}

\textbf{Right inequality of \eqref{eq:support-n}.} Let $\theta=(\lambda,P,Q)\in\Theta_n$. Then
\begin{align}
 e^{-n\psi(\alpha,z)}
 &\overset{(a)}{\ge}m_\lambda\norm{R_\lambda}_{2z}^{2z}
 \overset{(b)}{\ge}m_\lambda\norm{R_\lambda}_\infty^{2z}
 \overset{(c)}{\ge}m_\lambda\norm{PR_\lambda Q}_\infty^{2z}
 \overset{(d)}{=}m_\lambda\,\varrho_P^{\alpha}\varsigma_Q^{1-\alpha}\norm{PQ}_\infty^{2z}\notag\\
 &\overset{(e)}{=}\bigl(m_\lambda\varrho_P\bigr)^{\alpha}\bigl(m_\lambda\varsigma_Q\bigr)^{1-\alpha}\norm{PQ}_\infty^{2z}
 \overset{(f)}{=}\exp\Bigl(-n\bigl[\alpha\,\drho(\theta)+(1-\alpha)\,\dsig(\theta)+2z\,\chi(\theta)\bigr]\Bigr),\notag%\label{eq:upper-chain}
\end{align}
where $(a)$ follows by dropping all other terms of \eqref{eq:Qn-blocks}; $(b)$ holds because $\norm{R_\lambda}_{2z}^{2z}$ is the sum of the $2z$-th powers of the singular values of $R_\lambda$, the largest of which is $\norm{R_\lambda}_\infty$; $(c)$ follows from $\norm{PR_\lambda Q}_\infty\le\norm P_\infty\norm{R_\lambda}_\infty\norm Q_\infty$; $(d)$ follows from \eqref{eq:Z-expansion}; $(e)$ follows from $m_\lambda=m_\lambda^{\alpha}m_\lambda^{1-\alpha}$; and $(f)$ follows from \eqref{eq:sector-exp}. Taking logarithms gives \eqref{eq:support-each}, and minimizing over $\theta\in\Theta_n$ gives the right inequality of \eqref{eq:support-n}.

\textbf{Left inequality of \eqref{eq:support-n}.} Fix $\lambda$. The matrix $R_\lambda\in M_{D_\lambda}$ has $D_\lambda$ singular values, and each of $\rho_\lambda$, $\sigma_\lambda$ has at most $D_\lambda$ distinct eigenvalues. Hence
\begin{align}
 \norm{R_\lambda}_{2z}^{2z}
 &\overset{(a)}{\le}D_\lambda\norm{R_\lambda}_\infty^{2z}
 \overset{(b)}{\le}D_\lambda\Bigl(\sum_{P,Q}\varrho_P^{\frac{\alpha}{2z}}\varsigma_Q^{\frac{1-\alpha}{2z}}\norm{PQ}_\infty\Bigr)^{2z}
 \overset{(c)}{\le}D_\lambda^{1+4z}\max_{P,Q}\,\varrho_P^{\alpha}\varsigma_Q^{1-\alpha}\norm{PQ}_\infty^{2z},\label{eq:lower-chain-1}
\end{align}
where $(a)$ follows by bounding each singular value by the largest one, $(b)$ follows from \eqref{eq:Z-expansion} and the triangle inequality for $\norm\cdot_\infty$, and $(c)$ follows by bounding each of the at most $D_\lambda^2$ terms by the largest one. Since $R_\lambda$ is invertible, $R_\lambda\ne0$, so by \eqref{eq:Z-expansion} the maximum in \eqref{eq:lower-chain-1} is positive and is attained at a pair with $PQ\ne0$. Inserting \eqref{eq:lower-chain-1} into \eqref{eq:Qn-blocks},
\begin{align}
 e^{-n\psi(\alpha,z)}
 &\overset{(d)}{\le}\sum_{\lambda\in\Lambda_n}D_\lambda^{1+4z}\max_{\theta=(\lambda,P,Q)\in\Theta_n}\exp\Bigl(-n\bigl[\alpha\,\drho(\theta)+(1-\alpha)\,\dsig(\theta)+2z\,\chi(\theta)\bigr]\Bigr)\notag\\
 &\overset{(e)}{\le}\Bigl(\sum_{\lambda\in\Lambda_n}D_\lambda^{1+4z}\Bigr)e^{-n\psi_n(\alpha,z)}
 \overset{(f)}{\le}N_n^{\frac32+2z}\,e^{-n\psi_n(\alpha,z)},\label{eq:lower-chain-2}
\end{align}
where $(d)$ follows from $m_\lambda\varrho_P^{\alpha}\varsigma_Q^{1-\alpha}\norm{PQ}_\infty^{2z}=\exp(-n[\alpha\drho(\theta)+(1-\alpha)\dsig(\theta)+2z\chi(\theta)])$; $(e)$ follows from the definition \eqref{eq:mn} of $\psi_n$; and $(f)$ follows from $\sum_\lambda D_\lambda^{1+4z}\le|\Lambda_n|\max_\lambda D_\lambda^{1+4z}\le N_n\cdot N_n^{(1+4z)/2}$. Taking logarithms and recalling $N_n=(n+1)^{d^2-1}$, we obtain the left inequality of \eqref{eq:support-n} as desired.
\end{proof}

\begin{definition}[Limit set]\label{def:S}
$\cS\subset\bbR^3$ is the set of all $\bv$ for which there is a strictly increasing sequence of positive integers $\{n_k\}_{k=1}^{\infty}$ and a sequence of sectors $\theta_k\in\Theta_{n_k}$ with $\bv(\theta_k)\to\bv$ as $k\to\infty$.
\end{definition}

\begin{lemma}[Limit set and support representation]\label{lem:limit}
The limit set $\cS$ has the following properties:
\begin{enumerate}[label=(\alph*),leftmargin=*]
\item if $\theta_k\in\Theta_{n_k}$, where $(n_k)_k$ is a strictly increasing sequence of positive integers, and $\sup_k\chi(\theta_k)<\infty$, then $(\bv(\theta_k))_k$ has a subsequence converging to a point of $\cS$;
\item the set $\cS$ is closed and $\cS\subseteq[0,\lmax]^2\times[0,\infty)$;
\item for all $(\alpha,z)\in\bbR\times(0,\infty)$ and all $(\drho,\dsig,\chi)\in\cS$: $\ \alpha\,\drho+(1-\alpha)\,\dsig+2z\,\chi\ge\psi(\alpha,z)$;
\item for every $(\alpha,z)\in\bbR\times(0,\infty)$ there is a point of $\cS$ at which equality holds in (c). Consequently
\begin{equation}\label{eq:support}
 \psi(\alpha,z)=\min_{(\drho,\dsig,\chi)\in\cS}\bigl[\alpha\,\drho+(1-\alpha)\,\dsig+2z\,\chi\bigr].
\end{equation}
\end{enumerate}
\end{lemma}

\begin{proof}
(a) By \eqref{eq:sector-bounds}, the sequence $(\bv(\theta_k))_k$ lies in the compact set $[0,\lmax]^2\times[0,\sup_k\chi(\theta_k)]$. Hence it has a convergent subsequence and its limit belongs to $\cS$.

(b) The inclusion follows from \eqref{eq:sector-bounds}. 
Let $\bv^{(j)}\in\cS$ with $\bv^{(j)}\to\bv$. Set $n_0=0$. By Definition~\ref{def:S} we can choose inductively integers $n_j>n_{j-1}$ and sectors $\theta_j\in\Theta_{n_j}$ with $\norm{\bv(\theta_j)-\bv^{(j)}}\le1/j$. Then $\norm{\bv(\theta_j)-\bv}\le1/j+\norm{\bv^{(j)}-\bv}\to0$, so $\bv\in\cS$.

(c) Let $\bv(\theta_k)\to(\drho,\dsig,\chi)$ with $\theta_k\in\Theta_{n_k}$. By \eqref{eq:support-each}, $\alpha\drho(\theta_k)+(1-\alpha)\dsig(\theta_k)+2z\chi(\theta_k)\ge\psi(\alpha,z)$ for every $k$; letting $k\to\infty$ gives the claim.

(d) For each $n$ let $\theta_n\in\Theta_n$ attain the minimum in \eqref{eq:mn}. By \eqref{eq:sector-bounds} and \eqref{eq:support-n},
\begin{align}
 &\psi(\alpha,z)\le\psi_n(\alpha,z)\le\psi(\alpha,z)+\kappa_z\frac{\log(n+1)}{n},\label{eq:limit-1}\\
 &2z\,\chi(\theta_n)=\psi_n(\alpha,z)-\alpha\drho(\theta_n)-(1-\alpha)\dsig(\theta_n)\le\psi(\alpha,z)+\kappa_z\log2+\bigl(\abs\alpha+\abs{1-\alpha}\bigr)\lmax .\notag
\end{align}
Thus $\sup_n\chi(\theta_n)<\infty$, and by (a) there exists a subsequence $\{n_k\}_{k=1}^{\infty}$ such that $\bv(\theta_{n_k})\to(\drho,\dsig,\chi)\in\cS$. By \eqref{eq:limit-1},
\begin{equation*}
 \alpha\drho+(1-\alpha)\dsig+2z\chi=\lim_{k\to\infty}\psi_{n_k}(\alpha,z)=\psi(\alpha,z).
\end{equation*}
Together with (c) this proves \eqref{eq:support}.
\end{proof}

In the language of convex analysis, \eqref{eq:support} says that $\log Q_{\alpha,z}(\rho\|\sigma)=-\psi(\alpha,z)=h_{\cS}\bigl(-(\alpha,1-\alpha,2z)\bigr)$, where $h_{\cS}(\mathbf u)\eqdef\sup_{\bv\in\cS}\ip{\mathbf u}{\bv}$ is the support function of $\cS$.

\begin{proof}[Proof of Proposition~\ref{prop:psi}(a), (b), (d)]
(a) Fix $(\alpha_0,z_0)\in\bbR\times(0,\infty)$. The matrix
\begin{equation*}
 A(\alpha,z)\eqdef\rho^{\frac{\alpha}{2z}}\sigma^{\frac{1-\alpha}{z}}\rho^{\frac{\alpha}{2z}}=e^{\frac{\alpha}{2z}\log\rho}\,e^{\frac{1-\alpha}{z}\log\sigma}\,e^{\frac{\alpha}{2z}\log\rho}
\end{equation*}
has entries that are holomorphic functions of $(\alpha,z)$ in a complex neighborhood $\mathcal U$ of $(\alpha_0,z_0)$, and $A(\alpha_0,z_0)$ is positive definite. Let $\Gamma$ be a simple closed curve in the open right half-plane $\{\operatorname{Re}\zeta>0\}$ that does not meet the spectrum of $A(\alpha_0,z_0)$ and winds once around each of its eigenvalues. Since the eigenvalues of a matrix depend continuously on its entries \cite{Kato95}, we may shrink $\mathcal U$ so that the same holds for $A(\alpha,z)$ for all $(\alpha,z)\in\mathcal U$. By the Cauchy integral formula for matrix functions (Dunford--Taylor integral, \cite{Kato95}),
\begin{equation*}
 \tr A(\alpha,z)^{z}=\frac1{2\pi i}\oint_\Gamma\zeta^{z}\,\tr\bigl(\zeta\one-A(\alpha,z)\bigr)^{-1}d\zeta ,
\end{equation*}
with the principal branch of $\zeta^z=e^{z\log\zeta}$. The integrand is holomorphic in $(\alpha,z)\in\mathcal U$ for every $\zeta\in\Gamma$ and continuous in $(\zeta,\alpha,z)$, so the integral is holomorphic in $\mathcal U$. For real $(\alpha,z)$ it equals $Q_{\alpha,z}(\rho\|\sigma)>0$. Hence $\psi=-\log Q_{\alpha,z}$ is real-analytic near $(\alpha_0,z_0)$.

(b) By \eqref{eq:support}, $\psi$ is the pointwise minimum of the functions $(\alpha,z)\mapsto\alpha\drho+(1-\alpha)\dsig+2z\chi$, $(\drho,\dsig,\chi)\in\cS$, each of which is affine in $(\alpha,z)$. A pointwise minimum of affine functions is concave.

(d) By \eqref{eq:Qaz}, $\psi(0,z)=-\log\tr\bigl(\sigma^{1/z}\bigr)^{z}=-\log\tr\sigma=0$ and $\psi(1,z)=-\log\tr\bigl(\rho^{\frac1{2z}}\rho^{\frac1{2z}}\bigr)^{z}=-\log\tr\rho=0$. By (b), for $\alpha\in[0,1]$, $\psi(\alpha,z)\ge(1-\alpha)\psi(0,z)+\alpha\psi(1,z)=0$.
\end{proof}

The next lemma identifies the points of $\cS$ at which the minimum in \eqref{eq:support} is attained.

\begin{lemma}[Unique and continuous minimizing sectors]\label{lem:exposed}
Let $(\alpha,z)\in\bbR\times(0,\infty)$. The minimum in \eqref{eq:support} is attained at exactly one point of $\cS$, namely at
\begin{equation}\label{eq:vstar}
 \bv^\ast(\alpha,z)\eqdef\bigl(\drho^\ast(\alpha,z),\dsig^\ast(\alpha,z),\chi^\ast(\alpha,z)\bigr)
\end{equation}
with the functions of \eqref{eq:gradient}. In particular $\bv^\ast(\alpha,z)\in\cS$, and $\bv^\ast$ is continuous on $\bbR\times(0,\infty)$.
\end{lemma}

\begin{proof}[Proof of Lemma~\ref{lem:exposed} and of Proposition~\ref{prop:psi}(c)]
By Lemma~\ref{lem:limit}(d) there is a minimizer $\bv'=(\drho',\dsig',\chi')\in\cS$. Define, on $\bbR\times(0,\infty)$,
\begin{equation*}
 h(\alpha',z')\eqdef\alpha'\drho'+(1-\alpha')\dsig'+2z'\chi'-\psi(\alpha',z').
\end{equation*}
By Lemma~\ref{lem:limit}(c), $h\ge0$ on $\bbR\times(0,\infty)$, and $h(\alpha,z)=0$ because $\bv'$ is a minimizer. Thus $h$ attains its minimum over the open set $\bbR\times(0,\infty)$ at $(\alpha,z)$, and $h$ is differentiable by Proposition~\ref{prop:psi}(a). Therefore its gradient vanishes at $(\alpha,z)$:
\begin{align*}
 0&=\partial_{\alpha'}h(\alpha,z)=\drho'-\dsig'-\partial_\alpha\psi(\alpha,z),\\
 0&=\partial_{z'}h(\alpha,z)=2\chi'-\partial_z\psi(\alpha,z),\\
 0&=h(\alpha,z)=\alpha\drho'+(1-\alpha)\dsig'+2z\chi'-\psi(\alpha,z).
\end{align*}
These are the equations \eqref{eq:gradient-eq}, whose unique solution is $\bv^\ast(\alpha,z)$. Hence the minimizer $\bv'$ is unique and equal to $\bv^\ast(\alpha,z)$. Continuity of $\bv^\ast$ follows from \eqref{eq:gradient} and Proposition~\ref{prop:psi}(a). Finally, since $\bv^\ast(\alpha,z)\in\cS$ it follows that $\partial_z\psi(\alpha,z)=2\chi^\ast(\alpha,z)\ge0$, which is Proposition~\ref{prop:psi}(c).
\end{proof}

\subsection{Exact minimax from continuous minimizing sectors}\label{sec:proof-single}

The support representation gives the minimum of every relevant linear functional on $\cS$. To use these minima to test several inequalities simultaneously, we need an exact minimax statement. Convexity of $\cS$ is not assumed. The additional ingredient is Lemma~\ref{lem:exposed}: each functional has a unique minimizing sector, and this sector depends continuously on its direction.

We write $\ip{\bn}{\bv}$ for the Euclidean inner product on $\bbR^3$ and
\[
 \Delta_q\eqdef\bigl\{\bs=(s_1,\ldots,s_q)\in[0,1]^q:\textstyle\sum_{l=1}^q s_l=1\bigr\}
\]
for the probability simplex. The three directions that will occur are $(1,0,2)$, $(0,1,2)$ and $(1,1,2)$.

\begin{lemma}[Exact minimax and separation]\label{lem:separation}
Let $q\ge1$, $\bn_1,\ldots,\bn_q\in\{(1,0,2),\allowbreak(0,1,2),\allowbreak(1,1,2)\}$ and $t_1,\ldots,t_q\in\bbR$. Put
\begin{align*}
 r_l(\bv)&\eqdef\ip{\bn_l}{\bv}-t_l,\qquad
 \cV\eqdef\{\bv\in\bbR^3:r_l(\bv)<0\text{ for every }l\},\\
 \bn_{\bs}&\eqdef\sum_{l=1}^q s_l\bn_l,\qquad
 g(\bs)\eqdef\min_{\bv\in\cS}\sum_{l=1}^q s_l r_l(\bv)
 \quad(\bs\in\Delta_q).
\end{align*}
Then the following statements hold.
\begin{enumerate}[label=(\alph*),leftmargin=*]
\item For every $\bs\in\Delta_q$ there are $c_{\bs}\in[1,2]$ and $(\alpha_{\bs},z_{\bs})\in\cD$ such that
\[
 \bn_{\bs}=c_{\bs}(\alpha_{\bs},1-\alpha_{\bs},2z_{\bs}),\qquad
 g(\bs)=c_{\bs}\psi(\alpha_{\bs},z_{\bs})-\sum_l s_lt_l.
\]
The unique minimizing sector is $\bv_{\bs}\eqdef\bv^\ast(\alpha_{\bs},z_{\bs})$. Both $g$ and $\bs\mapsto\bv_{\bs}$ are continuous on $\Delta_q$.
\item The exact minimax identity
\begin{equation}\label{eq:exact-minimax}
 \min_{\bv\in\cS}\max_{1\le l\le q}r_l(\bv)
 =\max_{\bs\in\Delta_q}\min_{\bv\in\cS}\sum_{l=1}^q s_lr_l(\bv)
 =\max_{\bs\in\Delta_q}g(\bs)
\end{equation}
holds, and both extrema are attained. For every maximizer $\bs^\ast$ of $g$, the sector $\bv_{\bs^\ast}$ attains the minimum on the left and satisfies
\begin{equation}\label{eq:minimax-witness}
 r_l(\bv_{\bs^\ast})\le g(\bs^\ast)\qquad(l=1,\ldots,q).
\end{equation}
\item $\cS\cap\cV=\emptyset$ if and only if $g(\bs)\ge0$ for some $\bs\in\Delta_q$. If $\cS\cap\cV\ne\emptyset$, then $g(\bs^\ast)<0$ and $\bv_{\bs^\ast}\in\cS\cap\cV$ for every maximizer $\bs^\ast$ of $g$.
\end{enumerate}
\end{lemma}

\begin{proof}
(a) Write $\bn_{\bs}=(p_{\bs},q_{\bs},2)$. The allowed normals give $0\le p_{\bs},q_{\bs}\le1$ and $1\le c_{\bs}\eqdef p_{\bs}+q_{\bs}\le2$. Define
\[
 \alpha_{\bs}\eqdef\frac{p_{\bs}}{c_{\bs}},\qquad z_{\bs}\eqdef\frac1{c_{\bs}}.
\]
Then $z_{\bs}\le1$,
\[
 z_{\bs}-\alpha_{\bs}=\frac{1-p_{\bs}}{c_{\bs}}\ge0,\qquad
 \alpha_{\bs}+z_{\bs}-1=\frac{1-q_{\bs}}{c_{\bs}}\ge0,
\]
so $(\alpha_{\bs},z_{\bs})\in\cD$. The support representation \eqref{eq:support} gives the formula for $g$, and Lemma~\ref{lem:exposed} gives its unique minimizing sector. The parameter maps are continuous, and the continuity of $\psi$ and $\bv^\ast$ therefore gives both continuity assertions. In particular, continuity of the minimizing sector follows from the explicit gradient formula \eqref{eq:gradient}, including at the boundary of $\cD$.

(b) Compactness of $\Delta_q$ and continuity of $g$ give a maximizer $\bs^\ast$. Write
\[
 h(\bv,\bs)\eqdef\sum_l s_lr_l(\bv),\qquad M\eqdef g(\bs^\ast).
\]
Fix $l$, let $\mathbf e_l$ be the $l$-th vertex of $\Delta_q$, and put $\bs_\tau=(1-\tau)\bs^\ast+\tau\mathbf e_l$ for $0<\tau\le1$. Maximality of $\bs^\ast$, affinity in $\bs$, and the definition of $g$ give, in this order,
\begin{align*}
 M&\ge g(\bs_\tau)=h(\bv_{\bs_\tau},\bs_\tau)\\
  &=(1-\tau)h(\bv_{\bs_\tau},\bs^\ast)+\tau r_l(\bv_{\bs_\tau})\\
  &\ge(1-\tau)M+\tau r_l(\bv_{\bs_\tau}).
\end{align*}
Thus $r_l(\bv_{\bs_\tau})\le M$. By the continuity in (a), letting $\tau\downarrow0$ gives $r_l(\bv_{\bs^\ast})\le M$. This holds for every $l$, and all limits refer to the same sector $\bv_{\bs^\ast}$. Since
\[
 M=h(\bv_{\bs^\ast},\bs^\ast)
   \le\max_l r_l(\bv_{\bs^\ast})\le M,
\]
the maximum residual at this sector equals $M$. For every other $\bv\in\cS$,
\[
 \max_l r_l(\bv)\ge h(\bv,\bs^\ast)\ge g(\bs^\ast)=M.
\]
This proves \eqref{eq:exact-minimax}, attainment, and \eqref{eq:minimax-witness}.

(c) A sector belongs to $\cV$ exactly when its maximum residual is strictly negative. By (b), such a sector exists if and only if $M<0$, and $\bv_{\bs^\ast}$ then supplies one. Otherwise $M\ge0$, which is equivalent to the existence of a weight $\bs$ with $g(\bs)\ge0$.
\end{proof}

The proof shows the role of the minimizing sectors explicitly. Uniqueness identifies one sector for each weight, and the gradient formula makes this choice continuous. Perturbing a maximizing weight toward each simplex vertex then controls every residual at the same sector. This proves the minimax identity directly on $\cS$, without first replacing $\cS$ by its convex hull.

\subsection{The convex upper image of the sector set}\label{sec:upper-image}

The three normals in Lemma~\ref{lem:separation} suggest the linear coordinates
\begin{equation}\label{eq:sector-coordinates}
 \mathsf T(\drho,\dsig,\chi)
 \eqdef(\drho+2\chi,\ \dsig+2\chi,\ \drho+\dsig+2\chi).
\end{equation}
Let $\mathcal E\eqdef\mathsf T(\cS)$ and define its \emph{upper image}
\begin{equation}\label{eq:upper-image}
 \mathcal U\eqdef\mathcal E+\bbR_+^3
 =\{\mathbf y\in\bbR^3:\mathsf T\bv\le\mathbf y\text{ for some }\bv\in\cS\},
 \qquad\bbR_+^3\eqdef[0,\infty)^3,
\end{equation}
where inequalities between vectors are componentwise. Thus $\mathcal U$ consists of the threshold triples that dominate the three coordinates of at least one point of the limiting sector set.

For $\bs\in\Delta_3$, define
\begin{equation}\label{eq:s-to-az}
 c_{\bs}=1+s_3,\qquad
 \alpha_{\bs}=\frac{s_1+s_3}{1+s_3},\qquad
 z_{\bs}=\frac1{1+s_3}.
\end{equation}
The support representation yields
 $m(\bs)\eqdef\min_{\mathbf u\in\mathcal E}\ip{\bs}{\mathbf u}
 =c_{\bs}\psi(\alpha_{\bs},z_{\bs}).$
The map \eqref{eq:s-to-az} is a bijection from $\Delta_3$ onto $\cD$, with inverse
\begin{equation}\label{eq:az-to-s}
 s_1=\frac{\alpha+z-1}{z},\qquad
 s_2=\frac{z-\alpha}{z},\qquad
 s_3=\frac{1-z}{z}.
\end{equation}
Indeed, the three entries in \eqref{eq:az-to-s} are nonnegative on $\cD$, sum to one, and recover $(\alpha,z)$ through \eqref{eq:s-to-az}.

\begin{proposition}[Convex upper image]\label{prop:upper-image}
The upper image has the exact descriptions
\begin{align}
 \mathcal U
 &=\bigcap_{\bs\in\Delta_3}\{\mathbf y\in\bbR^3:\ip{\bs}{\mathbf y}\ge m(\bs)\}\label{eq:upper-halfspaces}\\
 &=\bigcap_{(\alpha,z)\in\cD}
 \bigl\{\mathbf y\in\bbR^3:
 (\alpha+z-1)y_1+(z-\alpha)y_2+(1-z)y_3\ge\psi(\alpha,z)\bigr\}.
 \notag
\end{align}
In particular, $\mathcal U$ is closed and convex. Moreover,
\begin{equation}\label{eq:upper-exposed}
 \mathcal U=\mathsf T\bigl(\bv^\ast(\cD)\bigr)+\bbR_+^3.
\end{equation}
Consequently, for every $\bv\in\cS$ there is $(\alpha,z)\in\cD$ such that
\begin{equation}\label{eq:exposed-domination}
 \mathsf T\bv^\ast(\alpha,z)\le\mathsf T\bv.
\end{equation}
\end{proposition}

\begin{proof}
For any $\mathbf y\in\bbR^3$, apply Lemma~\ref{lem:separation}(b) to the three normals in \eqref{eq:sector-coordinates} with thresholds $y_1,y_2,y_3$. It gives
\begin{equation}\label{eq:upper-minimax}
 \delta(\mathbf y)\eqdef\min_{\bv\in\cS}\max_{1\le i\le3}
       \bigl[(\mathsf T\bv)_i-y_i\bigr]
 =\max_{\bs\in\Delta_3}\bigl[m(\bs)-\ip{\bs}{\mathbf y}\bigr].
\end{equation}
The minimum on the left is attained. Hence $\mathbf y\in\mathcal U$ if and only if $\delta(\mathbf y)\le0$, which, by the right-hand side, is equivalent to all the inequalities in the first line of \eqref{eq:upper-halfspaces}. The second line follows by dividing those inequalities by $c_{\bs}$ and using \eqref{eq:az-to-s}. An intersection of closed half-spaces is closed and convex.

If $\mathbf y\in\mathcal U$, choose a maximizing weight $\bs^\ast$ in \eqref{eq:upper-minimax}. By \eqref{eq:minimax-witness}, its exposed sector satisfies
\[
 (\mathsf T\bv_{\bs^\ast})_i-y_i\le\delta(\mathbf y)\le0
 \qquad(i=1,2,3).
\]
This proves the inclusion of $\mathcal U$ in the right-hand side of \eqref{eq:upper-exposed}. The reverse inclusion follows from $\bv^\ast(\cD)\subseteq\cS$. Finally, take $\mathbf y=\mathsf T\bv$ to obtain \eqref{eq:exposed-domination}.
\end{proof}

The convex object is therefore $\mathcal U$; convexity of $\cS$ or $\mathcal E$ is not needed. Equation~\eqref{eq:upper-exposed} is stronger than a support-function description of a convex hull: every point of $\mathcal U$ dominates an actual exposed sector. This will let us detect every violation of the pair inequality on the two-parameter family $\bv^\ast(\cD)$.

\subsection{Forbidden polyhedra and the pair characterization}\label{sec:proof-pair}

Fix real $a,b,K,L$ and define
\begin{align}
 \cV_1&\eqdef\bigl\{(\drho,\dsig,\chi):\drho+2\chi<\min\{a,K\},\ \dsig+2\chi<b\bigr\},\notag\\
 \cV_2&\eqdef\bigl\{(\drho,\dsig,\chi):\drho+2\chi<a,\ \dsig+2\chi<\min\{b,L\}\bigr\},\label{eq:B-sets}\\
 \cV_3&\eqdef\bigl\{(\drho,\dsig,\chi):\drho+2\chi<a,\ \dsig+2\chi<b,\ \drho+\dsig+2\chi<K+L\bigr\}.\notag
\end{align}
These are subsets of $\bbR^3$. In the coordinates \eqref{eq:sector-coordinates}, the first two are open boxes in the first two coordinates, with the third unrestricted, and the third is an open box in all three coordinates. Each is downward closed: decreasing any of the three transformed coordinates preserves membership.

\begin{lemma}[Polyhedral form of the pair inequality]\label{lem:pairfn}
For $a,b,K,L,\drho,\dsig\in\bbR$ and $\chi\ge0$,
\begin{equation}\label{eq:pair-violation}
 \Phi_{a,b,K,L}(\drho,\dsig)>2\chi
 \quad\Longleftrightarrow\quad
 (\drho,\dsig,\chi)\in\cV_1\cup\cV_2\cup\cV_3.
\end{equation}
\end{lemma}

\begin{proof}
Since $2\chi\ge0$, $x_+>2\chi$ is equivalent to $x>2\chi$. Also
$
 x_++y_+=\max\{x+y,x,y,0\}.
$
Thus the minimum in \eqref{eq:pairfn} exceeds $2\chi$ exactly when
$
 a-\drho>2\chi,\qquad b-\dsig>2\chi,
$
and at least one of $K-\drho>2\chi$, $L-\dsig>2\chi$, or $K+L-\drho-\dsig>2\chi$ holds. These three alternatives give $\cV_1$, $\cV_2$, and $\cV_3$, respectively.
\end{proof}

\begin{proposition}[Equivalent sector, support, and pair conditions]\label{prop:single}
For $(a,b,K,L)\in(0,\infty)^4$ the following are equivalent:
\begin{enumerate}[label=(\arabic*),leftmargin=*]
\item the all-sector condition
\begin{equation}\label{eq:sector}
 \Phi_{a,b,K,L}(\drho,\dsig)\le2\chi
 \qquad\text{for every }(\drho,\dsig,\chi)\in\cS;
\end{equation}
\item the three support conditions \eqref{eq:P1}, \eqref{eq:P2}, and \eqref{eq:F}, i.e., $(a,b,K,L)\in\cZ$;
\item the exposed-sector condition \eqref{eq:pair}.
\end{enumerate}
\end{proposition}

\begin{proof}
By Lemmas~\ref{lem:limit}(b) and~\ref{lem:pairfn},
\begin{equation}\label{eq:sector-B}
 \text{\eqref{eq:sector}}
 \quad\Longleftrightarrow\quad
 \cS\cap\cV_j=\emptyset\quad(j=1,2,3).
\end{equation}

\emph{(1)$\iff$(2): supporting half-spaces.}
For $\cV_1$, use Lemma~\ref{lem:separation}(c) with $q=2$, normals $(1,0,2),(0,1,2)$, and thresholds $\min\{a,K\},b$. For $\bs=(s,1-s)$ its function $g$ is
\[
 g(\bs)=\psi(s,1)-s\min\{a,K\}-(1-s)b.
\]
The condition $g(\bs)\ge0$ for some $s\in[0,1]$ is exactly (P1). Using thresholds $a,\min\{b,L\}$ gives (P2) in the same way.

For $\cV_3$, take all three normals and thresholds $(a,b,K+L)$. Equations~\eqref{eq:s-to-az}--\eqref{eq:az-to-s} give
\begin{equation}\label{eq:g-F}
 \begin{aligned}
 \frac{g(\bs)}{c_{\bs}}
 &=\psi(\alpha_{\bs},z_{\bs})
   -(\alpha_{\bs}+z_{\bs}-1)a\\
 &\quad-(z_{\bs}-\alpha_{\bs})b-(1-z_{\bs})(K+L).
 \end{aligned}
\end{equation}
Since $c_{\bs}>0$ and \eqref{eq:s-to-az} maps $\Delta_3$ onto $\cD$, the existence of $\bs$ with $g(\bs)\ge0$ is equivalent to (F). Lemma~\ref{lem:separation}(c) and \eqref{eq:sector-B} prove (1)$\iff$(2).

\emph{(1)$\iff$(3): exposed sectors suffice.}
The forward implication follows from $\bv^\ast(\cD)\subseteq\cS$. Conversely, if (1) fails, some $\bv\in\cS$ belongs to one of the forbidden sets $\cV_j$. By Proposition~\ref{prop:upper-image}, there is $(\alpha,z)\in\cD$ with $\mathsf T\bv^\ast(\alpha,z)\le\mathsf T\bv$. Since $\cV_j$ is downward closed in these coordinates, $\bv^\ast(\alpha,z)$ also belongs to $\cV_j$. Lemma~\ref{lem:pairfn} then shows that (3) fails.
\end{proof}

\begin{proof}[Proof of Theorem~\ref{thm:pair}]
This is the equivalence (2)$\iff$(3) in Proposition~\ref{prop:single}.
\end{proof}

The three descriptions now have distinct geometric roles. The all-sector condition excludes three forbidden boxes; (P1), (P2), and (F) provide supporting half-spaces that exclude them; and the pair condition checks only the exposed sectors that generate the upper image. The edge $z=1$ corresponds to weights with $s_3=0$, which suffice for the first two boxes. The third box allows weights throughout $\Delta_3$, corresponding to the full triangle $\cD$. If a condition fails, maximizing its function $g$ gives an exposed sector witnessing the failure by Lemma~\ref{lem:separation}(c).

\section{Two One-Shot Lemmas for a Pair of Orthonormal Bases}\label{sec:oneshot}

This section is independent of the i.i.d.\ structure. The first lemma is a form of the monotonicity of the fidelity under measurements and gives the converse; the second lemma is new and gives the achievability. For a statement $S$ we write $[S]\eqdef1$ if $S$ is true and $[S]\eqdef0$ otherwise.

\subsection{Pairwise necessity}

For a unit vector $w\in\bbC^D$, an integer $m\ge1$ and a matrix $0\le E\le\one$ on $\bbC^D\otimes\bbC^m$, let
\begin{equation}\label{eq:ebar}
 \bar e_w(E)\eqdef\frac1m\tr\bigl[(\ketbra{w}\otimes\one_m)\,E\bigr]\in[0,1].
\end{equation}

\begin{lemma}[Pairwise necessity]\label{lem:nec}
Let $m\ge1$, let $x,y\in\bbC^D$ be unit vectors, and let $(E_0,E_1,E_2)$ be a POVM on $\bbC^D\otimes\bbC^m$. Then
\begin{equation}\label{eq:nec}
 \abs{\ip{x}{y}}\ \le\ \sum_{k=0}^{2}\sqrt{\bar e_x(E_k)\,\bar e_y(E_k)} .
\end{equation}
\end{lemma}

\begin{proof}
Let $\tr_2$ denote the partial trace over $\bbC^m$ and put $E_k'\eqdef\frac1m\tr_2E_k$, $k=0,1,2$. Then $E_k'\ge0$, $\sum_kE_k'=\frac1m\tr_2\one_{Dm}=\one_D$, and, by the defining property of the partial trace and \eqref{eq:ebar},
\begin{equation}\label{eq:nec-avg}
 \bra{w}E_k'\ket{w}=\frac1m\tr\bigl[(\ketbra w\otimes\one_m)\,E_k\bigr]=\bar e_w(E_k)\qquad\text{for every unit vector } w\in\bbC^D .
\end{equation}
Therefore
\begin{equation}\label{eq:nec-chain}
\begin{aligned}
 \abs{\ip{x}{y}}
 &\overset{(a)}{=}\Bigl|\sum_{k=0}^2\bra{x}E_k'\ket{y}\Bigr|
 \overset{(b)}{\le}\sum_{k=0}^2\bigl|\ip{E_k'^{1/2}x}{E_k'^{1/2}y}\bigr|\\
 &\overset{(c)}{\le}\sum_{k=0}^2\sqrt{\bra{x}E_k'\ket{x}\,\bra{y}E_k'\ket{y}}
 \overset{(d)}{=}\sum_{k=0}^2\sqrt{\bar e_x(E_k)\,\bar e_y(E_k)},
\end{aligned}
\end{equation}
where $(a)$ follows from $\sum_kE_k'=\one_D$, $(b)$ follows from the triangle inequality, $(c)$ follows from the Cauchy--Schwarz inequality, and $(d)$ follows from \eqref{eq:nec-avg}.
\end{proof}

Inequality \eqref{eq:nec} states that the fidelity $\abs{\ip{x}{y}}$ of the states $\ketbra{x}\otimes\frac1m\one_m$ and $\ketbra{y}\otimes\frac1m\one_m$ does not exceed the Bhattacharyya coefficient of the outcome distributions of any measurement \cite{FuchsCaves95}. For zero-error (unambiguous) discrimination of two mixed states, bounds of this type on the probability of the inconclusive outcome are due to \cite{RST03,RaynalLutkenhaus05}; see also \cite{Zhuang20}.

\subsection{Pairwise sufficiency}

In the next lemma we think of the vectors $x_i$ as eigenvectors of the null hypothesis and of the vectors $y_j$ as eigenvectors of the alternative. For every $x_i$ we prescribe a budget $\varepsilon_i$ for the probability of the wrong decision (outcome $1$) and a budget $\mu_i$ for the probability of rejection (outcome $2$); likewise $\delta_j$ and $\nu_j$ are budgets for the wrong decision (outcome $0$) and for rejection on $y_j$.

\begin{lemma}[Pairwise sufficiency]\label{lem:suff}
Let $\{x_i\}_{i=1}^D$ and $\{y_j\}_{j=1}^D$ be orthonormal bases of $\bbC^D$, let $U_{ij}\eqdef\ip{x_i}{y_j}$, and let $\varepsilon_i,\mu_i,\delta_j,\nu_j$ $(i,j=1,\dots,D)$ be real numbers with
\begin{equation}\label{eq:ordering}
 0<\varepsilon_i\le\mu_i\le1,\qquad0<\delta_j\le\nu_j\le1 .
\end{equation}
Put
\begin{equation}\label{eq:index-sets}
\begin{aligned}
 I_c&\eqdef\{i:\varepsilon_i<1\}, &\qquad I_0&\eqdef\{i:\mu_i<1\},\\
 J_c&\eqdef\{j:\delta_j<1\}, &\qquad J_0&\eqdef\{j:\nu_j<1\};
\end{aligned}
\end{equation}
by \eqref{eq:ordering}, $I_0\subseteq I_c$ and $J_0\subseteq J_c$. (The index $c$ refers to an active constraint on the wrong decision, and the index $0$ to an active rejection constraint.)
Let $0<\gamma\le\gamma_D\eqdef\dfrac{1}{400\,D^4}$, and suppose that
\begin{equation}\label{eq:hyp}
 \abs{U_{ij}}^2\le\gamma\,\max\bigl\{\varepsilon_i,\ \delta_j,\ \mu_i\nu_j\bigr\}\qquad\text{for all } (i,j)\in(I_0\times J_c)\cup(I_c\times J_0).
\end{equation}
Then there is a POVM $(E_0,E_1,E_2)$ on $\bbC^D$ such that, for all $i,j\in\{1,\dots,D\}$,
\begin{equation}\label{eq:budgets}
 \bra{x_i}E_1\ket{x_i}\le\varepsilon_i,\qquad
 \bra{x_i}E_2\ket{x_i}\le\mu_i,\qquad
 \bra{y_j}E_0\ket{y_j}\le\delta_j,\qquad
 \bra{y_j}E_2\ket{y_j}\le\nu_j .
\end{equation}
\end{lemma}

No condition is imposed on pairs $(i,j)$ with $i\notin I_c$, with $j\notin J_c$, or with $i\in I_c\setminus I_0$ and $j\in J_c\setminus J_0$; the overlaps of such pairs may be arbitrarily large.

\begin{proof}
\emph{Step 1: construction of the measurement.} From $\gamma\le1/(400D^4)$ and $D\ge1$, we have
\begin{equation}\label{eq:consts}
 D^2\sqrt\gamma\le\frac{D^2}{20D^2}=\frac1{20},\qquad D\sqrt\gamma\le\frac{1}{20D}\le\frac1{20},\qquad D\gamma\le D^2\gamma\le\frac{1}{400D^2}\le\frac1{400}.
\end{equation}
To every pair $(i,j)\in I_c\times J_c$ assign a label $\ell(i,j)\in\{0,1,2\}$:
\begin{equation}\label{eq:labels}
 \ell(i,j)\eqdef\begin{cases}
 1,&\text{if }\abs{U_{ij}}^2\le\gamma\,\varepsilon_i,\\
 0,&\text{if }\abs{U_{ij}}^2>\gamma\,\varepsilon_i\ \text{and}\ \abs{U_{ij}}^2\le\gamma\,\delta_j,\\
 2,&\text{otherwise.}
 \end{cases}
\end{equation}
(The label records which budget pays for the overlap of the pair.) Let $A_i\eqdef\{j\in J_c:\ell(i,j)=1\}$ for $i\in I_c$ and $B_j\eqdef\{i\in I_c:\ell(i,j)=0\}$ for $j\in J_c$, and define
\begin{equation}\label{eq:tilde}
 \tilde x_i\eqdef x_i-\sum_{j\in A_i}\ip{y_j}{x_i}\,y_j\quad(i\in I_c),\qquad
 \tilde y_j\eqdef y_j-\sum_{i\in B_j}\ip{x_i}{y_j}\,x_i\quad(j\in J_c).
\end{equation}
Let $\cX\eqdef\spn\{\tilde x_i:i\in I_c\}$ and $\cY\eqdef\spn\{\tilde y_j:j\in J_c\}$, with orthogonal projections $P_\cX$ and $P_\cY$, and
\begin{equation}\label{eq:fg}
 p_i\eqdef P_\cY\,\tilde x_i,\quad f_i\eqdef\tilde x_i-p_i\quad(i\in I_0),\qquad
 q_j\eqdef P_\cX\,\tilde y_j,\quad g_j\eqdef\tilde y_j-q_j\quad(j\in J_0).
\end{equation}
Let $\Fm\eqdef[f_i]_{i\in I_0}$ be the $D\times|I_0|$ matrix with columns $f_i$, $\Gm\eqdef[g_j]_{j\in J_0}$, and
\begin{equation}\label{eq:TxTy}
 T_x\eqdef2D\sqrt\gamma\,\diag(\mu_i)_{i\in I_0},\qquad T_y\eqdef2D\sqrt\gamma\,\diag(\nu_j)_{j\in J_0}.
\end{equation}
The POVM is
\begin{equation}\label{eq:povm}
 E_0\eqdef \Fm\bigl(\Fm^*\Fm+T_x\bigr)^{-1}\Fm^*,\qquad
 E_1\eqdef \Gm\bigl(\Gm^*\Gm+T_y\bigr)^{-1}\Gm^*,\qquad
 E_2\eqdef\one-E_0-E_1 .
\end{equation}
The inverses exist because $T_x,T_y$ are positive definite, and $E_0,E_1\ge0$. \emph{Conventions for empty index sets:} $E_0\eqdef0$ if $I_0=\emptyset$ and $E_1\eqdef0$ if $J_0=\emptyset$; the span of an empty family is $\{0\}$, with projection $0$; statements about the elements of an empty set are void. We show in Steps 2--7 that $E_0+E_1\le\one$ and in Steps 8--9 that \eqref{eq:budgets} holds.

\emph{Step 2: the modified vectors are close to the original ones.} For $i\in I_c$ and $j\in J_c$,
\begin{equation}\label{eq:close}
 \norm{x_i-\tilde x_i}^2\overset{(a)}{=}\sum_{j\in A_i}\abs{U_{ij}}^2\overset{(b)}{\le}|A_i|\,\gamma\,\varepsilon_i\le D\gamma\,\varepsilon_i,\qquad
 \norm{y_j-\tilde y_j}^2\overset{(a')}{=}\sum_{i\in B_j}\abs{U_{ij}}^2\overset{(b')}{\le}D\gamma\,\delta_j,
\end{equation}
where $(a)$ and $(a')$ follow from \eqref{eq:tilde} and the orthonormality of $\{y_j\}$ and of $\{x_i\}$, respectively, and $(b)$ and $(b')$ follow from \eqref{eq:labels}.

\emph{Step 3: Gram matrices.} For $A\subseteq\{1,\dots,D\}$ let $Y_A\eqdef\sum_{j\in A}\ketbra{y_j}$. These projections commute, and $(\one-Y_A)(\one-Y_{A'})=\one-Y_{A\cup A'}$. Since $\tilde x_i=(\one-Y_{A_i})x_i$ by \eqref{eq:tilde}, for $i,i'\in I_c$,
\begin{equation}\label{eq:gram-x}
 \ip{\tilde x_i}{\tilde x_{i'}}=\bra{x_i}(\one-Y_{A_i\cup A_{i'}})\ket{x_{i'}}=[i=i']-\sum_{j\in A_i\cup A_{i'}}U_{ij}\overline{U_{i'j}} .
\end{equation}
For $j\in A_i$ we have $\abs{U_{ij}}\le\sqrt{\gamma\varepsilon_i}\le\sqrt\gamma$ and $\abs{U_{i'j}}\le1$; for $j\in A_{i'}$ we have $\abs{U_{i'j}}\le\sqrt\gamma$ and $\abs{U_{ij}}\le1$. Hence every term of the sum in \eqref{eq:gram-x} has modulus at most $\sqrt\gamma$, and the sum has at most $|J_c|\le D$ terms. So every entry of $\Gamma_X-\one$, where $\Gamma_X\eqdef[\ip{\tilde x_i}{\tilde x_{i'}}]_{i,i'\in I_c}$, has modulus at most $D\sqrt\gamma$. Hence
\begin{equation}\label{eq:gram-bound}
\begin{aligned}
 &\norm{\Gamma_X-\one}_\infty\le\norm{\Gamma_X-\one}_2\le|I_c|\cdot D\sqrt\gamma\le D^2\sqrt\gamma\overset{(a)}{\le}\tfrac1{20},\\
 &\Gamma_X\ge\tfrac{19}{20}\one,\qquad\norm{\Gamma_X^{-1}}_\infty\le\tfrac{20}{19}\le\tfrac43,
\end{aligned}
\end{equation}
where $(a)$ follows from \eqref{eq:consts}, and the second line follows from the first. Exchanging the roles of the two bases (with $X_B\eqdef\sum_{i\in B}\ketbra{x_i}$, $\tilde y_j=(\one-X_{B_j})y_j$ and $\abs{U_{ij}}\le\sqrt{\gamma\delta_j}\le\sqrt\gamma$ for $i\in B_j$), the same computation shows that $\Gamma_Y\eqdef[\ip{\tilde y_j}{\tilde y_{j'}}]_{j,j'\in J_c}$ satisfies
\begin{equation}\label{eq:gram-bound-y}
 \norm{\Gamma_Y-\one}_\infty\le|J_c|\cdot D\sqrt\gamma\le\tfrac1{20},\qquad\Gamma_Y\ge\tfrac{19}{20}\one,\qquad\norm{\Gamma_Y^{-1}}_\infty\le\tfrac43 .
\end{equation}
In particular the families $\{\tilde x_i\}_{i\in I_c}$ and $\{\tilde y_j\}_{j\in J_c}$ are linearly independent.

\emph{Step 4: cross overlaps.} For $(i,j)\in I_c\times J_c$,
\begin{align}
 c_{ij}&\eqdef\ip{\tilde x_i}{\tilde y_j}
 =\ip{x_i}{y_j}-\ip{Y_{A_i}x_i}{y_j}-\ip{x_i}{X_{B_j}y_j}+\ip{Y_{A_i}x_i}{X_{B_j}y_j}\notag\\
 &\overset{(a)}{=}U_{ij}\bigl(1-[j\in A_i]-[i\in B_j]\bigr)+\ip{x_i-\tilde x_i}{y_j-\tilde y_j}
 \overset{(b)}{=}U_{ij}\,[\ell(i,j)=2]+\Delta_{ij},\label{eq:cross}
\end{align}
where $\Delta_{ij}\eqdef\ip{x_i-\tilde x_i}{y_j-\tilde y_j}$; $(a)$ follows from the facts that $\ip{Y_{A_i}x_i}{y_j}=\ip{x_i}{Y_{A_i}y_j}=U_{ij}[j\in A_i]$, $\ip{x_i}{X_{B_j}y_j}=U_{ij}[i\in B_j]$, $Y_{A_i}x_i=x_i-\tilde x_i$ and $X_{B_j}y_j=y_j-\tilde y_j$; and $(b)$ follows from the definitions of $A_i$ and $B_j$: $j\in A_i$ means $\ell(i,j)=1$, and $i\in B_j$ means $\ell(i,j)=0$. Moreover,
\begin{equation}\label{eq:Delta}
 \abs{\Delta_{ij}}\overset{(a)}{\le}\norm{x_i-\tilde x_i}\,\norm{y_j-\tilde y_j}\overset{(b)}{\le}D\gamma\sqrt{\varepsilon_i\delta_j}\overset{(c)}{\le}D\gamma\sqrt{\mu_i\nu_j},
\end{equation}
where $(a)$ follows from the Cauchy--Schwarz inequality, $(b)$ follows from \eqref{eq:close}, and $(c)$ follows from \eqref{eq:ordering}.
Now let $(i,j)\in(I_0\times J_c)\cup(I_c\times J_0)$. If $\ell(i,j)=2$, then $\abs{U_{ij}}^2>\gamma\varepsilon_i$ and $\abs{U_{ij}}^2>\gamma\delta_j$ by \eqref{eq:labels}, so the hypothesis \eqref{eq:hyp} can only hold through the third term of the maximum: $\abs{U_{ij}}^2\le\gamma\mu_i\nu_j$. Therefore
\begin{equation}\label{eq:cross-bound}
 \abs{c_{ij}}^2\overset{(a)}{\le}2\abs{U_{ij}}^2[\ell(i,j)=2]+2\abs{\Delta_{ij}}^2\overset{(b)}{\le}2\gamma\,\mu_i\nu_j+2D^2\gamma^2\mu_i\nu_j\overset{(c)}{\le}3\gamma\,\mu_i\nu_j
\end{equation}
for all $(i,j)\in(I_0\times J_c)\cup(I_c\times J_0)$, where $(a)$ follows from \eqref{eq:cross} and $\abs{u+v}^2\le2\abs u^2+2\abs v^2$, $(b)$ follows from $\abs{U_{ij}}^2\le\gamma\mu_i\nu_j$ (if $\ell(i,j)=2$) and \eqref{eq:Delta}, and $(c)$ follows from \eqref{eq:consts}. This is the only place where \eqref{eq:hyp} is used.

\emph{Step 5: the projections $p_i$ and $q_j$ are small.} Let $\tilde Y\eqdef[\tilde y_j]_{j\in J_c}$ and $\tilde X\eqdef[\tilde x_i]_{i\in I_c}$. By Step 3, $\tilde Y^*\tilde Y=\Gamma_Y$ is invertible, and $P_\cY=\tilde Y\Gamma_Y^{-1}\tilde Y^*$: the right-hand side is Hermitian, idempotent, and has range $\ran\tilde Y=\cY$. Likewise $P_\cX=\tilde X\Gamma_X^{-1}\tilde X^*$. For $i\in I_0$,
\begin{align}
 \norm{p_i}^2&=\bra{\tilde x_i}P_\cY\ket{\tilde x_i}=\bigl(\tilde Y^*\tilde x_i\bigr)^*\Gamma_Y^{-1}\bigl(\tilde Y^*\tilde x_i\bigr)
 \le\norm{\Gamma_Y^{-1}}_\infty\sum_{j\in J_c}\abs{\ip{\tilde y_j}{\tilde x_i}}^2
 \overset{(a)}{\le}\frac43\sum_{j\in J_c}3\gamma\,\mu_i\nu_j\overset{(b)}{\le}4D\gamma\,\mu_i,\label{eq:p-bound}
\end{align}
where $(a)$ follows from \eqref{eq:gram-bound-y} and \eqref{eq:cross-bound}, and $(b)$ follows from $\nu_j\le1$ and $|J_c|\le D$. In the same way, for $j\in J_0$, by \eqref{eq:gram-bound}, \eqref{eq:cross-bound} (applicable since $i\in I_c$, $j\in J_0$), $\mu_i\le1$ and $|I_c|\le D$,
\begin{equation}\label{eq:q-bound}
 \norm{q_j}^2=\bigl(\tilde X^*\tilde y_j\bigr)^*\Gamma_X^{-1}\bigl(\tilde X^*\tilde y_j\bigr)\le\norm{\Gamma_X^{-1}}_\infty\sum_{i\in I_c}\abs{c_{ij}}^2\le\frac43\sum_{i\in I_c}3\gamma\,\mu_i\nu_j\le4D\gamma\,\nu_j .
\end{equation}

\emph{Step 6: the two accepted families are almost orthogonal.} By \eqref{eq:fg}, $f_i\perp\cY$ and $g_j\perp\cX$. For $(i,j)\in I_0\times J_0$, since $\tilde y_j\in\cY$ and $\tilde x_i\in\cX$,
\begin{equation*}
 \ip{p_i}{\tilde y_j}=\ip{\tilde x_i}{P_\cY\tilde y_j}=c_{ij},\qquad
 \ip{\tilde x_i}{q_j}=\ip{P_\cX\tilde x_i}{\tilde y_j}=c_{ij},
\end{equation*}
and therefore
\begin{align}
 w_{ij}\eqdef\ip{f_i}{g_j}&=\ip{\tilde x_i}{\tilde y_j}-\ip{p_i}{\tilde y_j}-\ip{\tilde x_i}{q_j}+\ip{p_i}{q_j}=-c_{ij}+\ip{p_i}{q_j},\notag\\
 \abs{w_{ij}}&\le\abs{c_{ij}}+\norm{p_i}\norm{q_j}
 \overset{(a)}{\le}\bigl(\sqrt3+4D\sqrt\gamma\bigr)\sqrt{\gamma\,\mu_i\nu_j}
 \overset{(b)}{\le}2\sqrt{\gamma\,\mu_i\nu_j},\label{eq:w-bound}
\end{align}
where $(a)$ follows from \eqref{eq:cross-bound}, \eqref{eq:p-bound} and \eqref{eq:q-bound}, and $(b)$ follows from \eqref{eq:consts}, which gives $4D\sqrt\gamma\le\frac15$.

\emph{Step 7: $E_0+E_1\le\one$.} Assume first that $I_0\ne\emptyset$ and $J_0\ne\emptyset$. Let $W\eqdef\Fm^*\Gm=[w_{ij}]_{i\in I_0,j\in J_0}$, $\Xi\eqdef[\Fm\ \,\Gm]$ and $\mathsf M\eqdef(\Fm^*\Fm+T_x)^{-1}\oplus(\Gm^*\Gm+T_y)^{-1}$, a positive definite matrix. Then $E_0+E_1=\Xi\mathsf M\Xi^*$ by \eqref{eq:povm}, and
\begin{align}
 \Xi\mathsf M\Xi^*\le\one
 &\overset{(a)}{\iff}\mathsf M^{1/2}\Xi^*\Xi\mathsf M^{1/2}\le\one
 \overset{(b)}{\iff}\Xi^*\Xi\le\mathsf M^{-1}\notag\\
 &\overset{(c)}{\iff}\begin{pmatrix}\Fm^*\Fm&W\\W^*&\Gm^*\Gm\end{pmatrix}\le\begin{pmatrix}\Fm^*\Fm+T_x&0\\0&\Gm^*\Gm+T_y\end{pmatrix}
 \iff\begin{pmatrix}T_x&-W\\-W^*&T_y\end{pmatrix}\ge0\notag\\
 &\overset{(d)}{\iff}\begin{pmatrix}\one&-\widehat W\\-\widehat W^*&\one\end{pmatrix}\ge0
 \overset{(e)}{\Longleftarrow}\norm{\widehat W}_\infty\le1,
 \qquad\widehat W\eqdef T_x^{-1/2}\,W\,T_y^{-1/2},\label{eq:positivity}
\end{align}
where $(a)$ holds because $(\Xi\mathsf M^{1/2})(\Xi\mathsf M^{1/2})^*$ and $(\Xi\mathsf M^{1/2})^*(\Xi\mathsf M^{1/2})$ have the same non-zero eigenvalues; $(b)$ follows from a congruence with $\mathsf M^{-1/2}$; $(c)$ follows by writing out $\Xi^*\Xi$ and $\mathsf M^{-1}$; $(d)$ follows from a congruence with $T_x^{-1/2}\oplus T_y^{-1/2}$; and $(e)$ holds because, for vectors $\xi,\zeta$, the quadratic form of the last block matrix is $\norm\xi^2+\norm\zeta^2-2\operatorname{Re}\ip{\xi}{\widehat W\zeta}\ge(\norm\xi-\norm\zeta)^2\ge0$ if $\norm{\widehat W}_\infty\le1$. By \eqref{eq:TxTy} and \eqref{eq:w-bound}, the entries of $\widehat W$ satisfy
\begin{equation}\label{eq:What}
 \abs{\widehat W_{ij}}=\frac{\abs{w_{ij}}}{2D\sqrt\gamma\sqrt{\mu_i\nu_j}}\le\frac1D,
 \qquad\text{hence}\qquad\norm{\widehat W}_\infty\le\norm{\widehat W}_2\le\frac{\sqrt{|I_0||J_0|}}{D}\le1 .
\end{equation}
Thus $E_0+E_1\le\one$. If $J_0=\emptyset$, then $E_0+E_1=E_0=\Fm\mathsf M\Fm^*$ with $\mathsf M=(\Fm^*\Fm+T_x)^{-1}$, and steps $(a)$ and $(b)$ of \eqref{eq:positivity} with $\Xi=\Fm$ reduce the claim to $\Fm^*\Fm\le\Fm^*\Fm+T_x$, which is true; similarly $E_0+E_1=E_1\le \one$ if $I_0=\emptyset$. In all cases $0\le E_0+E_1\le\one$, so $(E_0,E_1,E_2)$ is a POVM and $E_0,E_1,E_2\le\one$.

\emph{Step 8: error budgets.} Let $i\in I_c$. Since $g_j\perp\cX\ni\tilde x_i$ for all $j\in J_0$, we have $\Gm^*\tilde x_i=0$ and hence $E_1\tilde x_i=0$ by \eqref{eq:povm}. Therefore
\begin{equation}\label{eq:err-x}
 \bra{x_i}E_1\ket{x_i}=\bra{x_i-\tilde x_i}E_1\ket{x_i-\tilde x_i}\le\norm{E_1}_\infty\norm{x_i-\tilde x_i}^2\overset{(a)}{\le}D\gamma\,\varepsilon_i\overset{(b)}{\le}\varepsilon_i ,
\end{equation}
where $(a)$ follows from $\norm{E_1}_\infty\le1$ (Step~7) and \eqref{eq:close}, and $(b)$ follows from \eqref{eq:consts}. For $i\notin I_c$ we have $\varepsilon_i=1\ge\bra{x_i}E_1\ket{x_i}$. Similarly we can show that $\bra{y_j}E_0\ket{y_j}\le\delta_j$ for all $j$.

\emph{Step 9: rejection budgets.} Let $i\in I_0$ and write $\Gamma_{\!f}\eqdef\Fm^*\Fm$ (the Gram matrix of the $f_l$, $l\in I_0$). For $l\in I_0$, $\ip{f_l}{\tilde x_i}=\ip{f_l}{f_i}+\ip{f_l}{p_i}=\ip{f_l}{f_i}$ because $p_i\in\cY$ and $f_l\perp\cY$. Hence $\Fm^*\tilde x_i=\Gamma_{\!f}\mathbf e_i$, where $\mathbf e_i$ is the $i$-th coordinate vector of $\bbC^{I_0}$, and
\begin{align}
 \bra{\tilde x_i}E_0\ket{\tilde x_i}
 &=\bigl(\Fm^*\tilde x_i\bigr)^*(\Gamma_{\!f}+T_x)^{-1}\bigl(\Fm^*\tilde x_i\bigr)
 =\bigl[\Gamma_{\!f}(\Gamma_{\!f}+T_x)^{-1}\Gamma_{\!f}\bigr]_{ii}\notag\\
 &\overset{(a)}{=}\bigl[\Gamma_{\!f}-T_x+T_x(\Gamma_{\!f}+T_x)^{-1}T_x\bigr]_{ii}
 \overset{(b)}{\ge}(\Gamma_{\!f})_{ii}-(T_x)_{ii}
 \overset{(c)}{=}\norm{\tilde x_i}^2-\norm{p_i}^2-2D\sqrt\gamma\,\mu_i,\label{eq:rej-1}
\end{align}
where $(a)$ follows from the identity $\Gamma_{\!f}(\Gamma_{\!f}+T_x)^{-1}\Gamma_{\!f}=(\Gamma_{\!f}+T_x-T_x)(\Gamma_{\!f}+T_x)^{-1}(\Gamma_{\!f}+T_x-T_x)=\Gamma_{\!f}-T_x+T_x(\Gamma_{\!f}+T_x)^{-1}T_x$; $(b)$ follows by dropping the positive semidefinite term; and $(c)$ follows from $(\Gamma_{\!f})_{ii}=\norm{f_i}^2=\norm{\tilde x_i}^2-\norm{p_i}^2$ (Pythagoras, since $f_i\perp p_i$) and \eqref{eq:TxTy}. Since $E_1\tilde x_i=0$ (Step 8),
\begin{equation}\label{eq:rej-2}
 \bra{\tilde x_i}E_2\ket{\tilde x_i}=\norm{\tilde x_i}^2-\bra{\tilde x_i}E_0\ket{\tilde x_i}\overset{(a)}{\le}\norm{p_i}^2+2D\sqrt\gamma\,\mu_i\overset{(b)}{\le}\bigl(4D\gamma+2D\sqrt\gamma\bigr)\mu_i ,
\end{equation}
where $(a)$ follows from \eqref{eq:rej-1} and $(b)$ follows from \eqref{eq:p-bound}. Finally, with $\norm{E_2^{1/2}(u+v)}^2\le2\norm{E_2^{1/2}u}^2+2\norm{E_2^{1/2}v}^2$ and $E_2\le\one$,
\begin{align}
 \bra{x_i}E_2\ket{x_i}&\le2\bra{\tilde x_i}E_2\ket{\tilde x_i}+2\norm{x_i-\tilde x_i}^2
 \overset{(a)}{\le}\bigl(8D\gamma+4D\sqrt\gamma\bigr)\mu_i+2D\gamma\,\varepsilon_i\notag\\
 &\overset{(b)}{\le}\bigl(10D\gamma+4D\sqrt\gamma\bigr)\mu_i\overset{(c)}{\le}\mu_i,\label{eq:rej-3}
\end{align}
where $(a)$ follows from \eqref{eq:close} and \eqref{eq:rej-2}, $(b)$ follows from $\varepsilon_i\le\mu_i$, and $(c)$ follows from \eqref{eq:consts}, which gives $10D\gamma+4D\sqrt\gamma\le\frac1{40}+\frac15<1$. For $i\notin I_0$ the budget is $\mu_i=1$. Similarly we can show that $\bra{y_j}E_2\ket{y_j}\le\nu_j$ for all $j$.
\end{proof}

\begin{remark}[Structure of the measurement]\label{rem:structure}
The operator $E_0$ is a regularized projection onto the span of the vectors $f_i$, $i\in I_0$. These are the eigenvectors $x_i$ of the null hypothesis that carry a rejection constraint, first freed from those components along $y_j$ whose overlap can be charged to the type-I error budget of $x_i$, and then made \emph{exactly} orthogonal to every (modified) alternative eigenvector on which a type-II error is constrained. The operator $E_1$ is built in the same way from the $y_j$, and everything else is rejected. The overlaps $w_{ij}$ that remain between the two accepted families are at most $2\sqrt{\gamma\mu_i\nu_j}$ in modulus by \eqref{eq:w-bound}, which is what the regularizations $T_x$, $T_y$, and thereby the rejection budgets, can absorb. The proof shows slightly more than \eqref{eq:budgets}, namely $\bra{x_i}E_1\ket{x_i}\le D\gamma\,\varepsilon_i$ for $i\in I_c$, $\bra{x_i}E_2\ket{x_i}\le\frac{9}{40}\mu_i$ for $i\in I_0$, $\bra{y_j}E_0\ket{y_j}\le D\gamma\,\delta_j$ for $j\in J_c$ and $\bra{y_j}E_2\ket{y_j}\le\frac{9}{40}\nu_j$ for $j\in J_0$; for the remaining indices the corresponding budget equals one and only the trivial bound holds. The constants have not been optimized.
\end{remark}

\begin{remark}[Tightness and the ordering assumption]\label{rem:ordering}
If a POVM with \eqref{eq:budgets} exists, then Lemma~\ref{lem:nec} (with $m=1$, $\bar e_{x_i}(E_0)\le1$ and $\bar e_{y_j}(E_1)\le1$) gives $\abs{U_{ij}}\le\sqrt{\delta_j}+\sqrt{\varepsilon_i}+\sqrt{\mu_i\nu_j}$ and hence $\abs{U_{ij}}^2\le9\max\{\varepsilon_i,\delta_j,\mu_i\nu_j\}$ for \emph{all} pairs; for pairs outside $(I_0\times J_c)\cup(I_c\times J_0)$ the maximum equals one and this is void. Thus, for budgets ordered as in \eqref{eq:ordering}, the solvability of \eqref{eq:budgets} is characterized, up to the factor $9/\gamma_D=3600D^4$, by the condition $\abs{U_{ij}}^2\le\kappa\max\{\varepsilon_i,\delta_j,\mu_i\nu_j\}$ on the pairs of $(I_0\times J_c)\cup(I_c\times J_0)$: it is sufficient for $\kappa=\gamma_D$ (Lemma~\ref{lem:suff}) and necessary for $\kappa=9$ (Lemma~\ref{lem:nec}).

The ordering assumption \eqref{eq:ordering} cannot be dropped. Let $D=2$, let $\{x_1,x_2\}$ be an orthonormal basis of $\bbC^2$, let $0<\gamma\le\gamma_2$, choose $\vartheta\in(0,\frac\pi2)$ with $\omega\eqdef\sin^2\vartheta\le\gamma/2$, put $y_1\eqdef\sin\vartheta\,x_1+\cos\vartheta\,x_2$, $y_2\eqdef\cos\vartheta\,x_1-\sin\vartheta\,x_2$ and $\tau\eqdef\omega/140$, and take the budgets
\begin{equation*}
 (\varepsilon_1,\mu_1)=\bigl(\tfrac12,\tfrac\omega8\bigr),\qquad(\varepsilon_2,\mu_2)=(\delta_1,\nu_1)=(\tau,1),\qquad(\delta_2,\nu_2)=(1,1),
\end{equation*}
which violate \eqref{eq:ordering} only in $\varepsilon_1>\mu_1$. Then $I_c=\{1,2\}$, $I_0=\{1\}$, $J_c=\{1\}$, $J_0=\emptyset$, the only pair in \eqref{eq:hyp} is $(1,1)$, and \eqref{eq:hyp} holds because $\abs{U_{11}}^2=\omega\le\gamma\varepsilon_1$. Suppose that a POVM satisfied \eqref{eq:budgets}. Using $\sum_kE_k=\one$, $y_1=\sqrt\omega\,x_1+\sqrt{1-\omega}\,x_2$, the Cauchy--Schwarz inequality $\abs{\bra uE\ket v}\le\sqrt{\bra uE\ket u\bra vE\ket v}$ for $E\ge0$, $\sqrt{1-\omega}\le1$, and the budgets,
\begin{align*}
 \sqrt\omega=\abs{\ip{x_1}{y_1}}
 &\le\abs{\bra{x_1}E_0\ket{y_1}}+\sqrt\omega\,\bra{x_1}E_1\ket{x_1}+\sqrt{1-\omega}\,\abs{\bra{x_1}E_1\ket{x_2}}+\abs{\bra{x_1}E_2\ket{y_1}}\\
 &\le\sqrt{1\cdot\tau}+\tfrac12\sqrt\omega+\sqrt{\tfrac12\cdot\tau}+\sqrt{\bra{x_1}E_2\ket{x_1}\cdot1}.
\end{align*}
Hence, with $\tau=\omega/140$, $\sqrt{\bra{x_1}E_2\ket{x_1}}\ge\sqrt\omega\,\bigl(\frac12-(1+\frac1{\sqrt2})/\sqrt{140}\bigr)>0.355\sqrt\omega$, so that $\bra{x_1}E_2\ket{x_1}>0.126\,\omega>\frac\omega8=\mu_1$, a contradiction.
\end{remark}

\section{Proofs of the Main Results}\label{sec:proofs}

Theorem~\ref{thm:pair} and the geometric equivalence of Proposition~\ref{prop:single} were proved in Section~\ref{sec:sectors}. We now connect this geometry to tests. The finite-blocklength sector inequality and the one-shot construction give $\cZ\subseteq\cA$. Pairwise necessity gives the all-sector condition for every point of $\cA$, and Proposition~\ref{prop:single} then yields $\cA\subseteq\cZ$.

\subsection{Finite-blocklength preparation}\label{sec:finite-preparation}

The geometric equivalence in Proposition~\ref{prop:single} concerns limiting sectors. The construction below uses the exact finite-blocklength support inequality \eqref{eq:support-each}. We first record the shift estimate that supplies a polynomial margin for the one-shot lemma.

\begin{lemma}[Shift estimate for $\Phi$]\label{lem:pairshift}
Let $a,b,K,L,\drho,\dsig\in\bbR$ and $\eta\ge0$. If $\drho<a-\eta$, $\dsig<b-\eta$, and at least one of $\drho<K-\eta$ or $\dsig<L-\eta$ holds, then
\begin{equation}\label{eq:shift}
 \Phi_{a,b,K,L}(\drho,\dsig)
 \ge\Phi_{a-\eta,b-\eta,K-\eta,L-\eta}(\drho,\dsig)+\eta.
\end{equation}
\end{lemma}

\begin{proof}
The first two assumptions give
\[
 (a-\drho)_+=(a-\eta-\drho)_++\eta,\qquad
 (b-\dsig)_+=(b-\eta-\dsig)_++\eta.
\]
If $\drho<K-\eta$, then $(K-\drho)_+=(K-\eta-\drho)_++\eta$, while $(L-\dsig)_+\ge(L-\eta-\dsig)_+$. The same conclusion holds with the two rejection terms exchanged if $\dsig<L-\eta$. Each of the three terms defining the first minimum in \eqref{eq:shift} is therefore at least the corresponding term defining the second minimum plus $\eta$.
\end{proof}

\begin{lemma}[Finite-blocklength sector inequality]\label{lem:sector-ineq}
Let $(a,b,K,L)\in\cZ$. Then
\begin{equation}\label{eq:sector-n}
 \Phi_{a,b,K,L}\bigl(\drho(\theta),\dsig(\theta)\bigr)\ \le\ 2\chi(\theta)\qquad\text{for all } n\ge1\ \text{and all }\theta\in\Theta_n,
\end{equation}
and consequently the all-sector condition \eqref{eq:sector} holds.
\end{lemma}

\begin{proof}
Suppose that \eqref{eq:sector-n} fails for some $n$ and some $\theta\in\Theta_n$, and write $(\drho,\dsig,\chi)\eqdef\bv(\theta)$; recall $\chi\ge0$ from \eqref{eq:sector-bounds}. By Lemma~\ref{lem:pairfn}, $\bv(\theta)\in \cV_1\cup\cV_2\cup\cV_3$.

\emph{Case $\bv(\theta)\in\cV_1$.} Let $s\in[0,1]$ be a witness of \eqref{eq:P1}. Then
\begin{align*}
 s\min\{a,K\}+(1-s)\,b
 &\overset{(a)}{>}s\,(\drho+2\chi)+(1-s)\,(\dsig+2\chi)\\
 &=s\,\drho+(1-s)\,\dsig+2\chi
 \overset{(b)}{\ge}\psi(s,1),
\end{align*}
where $(a)$ follows from the two strict inequalities defining $\cV_1$ and the fact that the weights $s,1-s$ are non-negative and not both zero, and $(b)$ follows from \eqref{eq:support-each} with $(\alpha,z)=(s,1)$. This contradicts the choice of $s$.

\emph{Case $\bv(\theta)\in\cV_2$.} In the same way, with a witness $s$ of \eqref{eq:P2}: $s\,a+(1-s)\min\{b,L\}>s\,\drho+(1-s)\,\dsig+2\chi\ge\psi(s,1)$, a contradiction.

\emph{Case $\bv(\theta)\in\cV_3$.} Let $(\alpha,z)\in\cD$ be a witness of \eqref{eq:F}, and put $w_1\eqdef\alpha+z-1$, $w_2\eqdef z-\alpha$, $w_3\eqdef1-z$. By \eqref{eq:triangle}, $w_1,w_2,w_3\ge0$, and
\begin{equation}\label{eq:weights}
 w_1+w_3=\alpha,\qquad w_2+w_3=1-\alpha,\qquad w_1+w_2+w_3=z\ge\tfrac12 .
\end{equation}
Therefore
\begin{align*}
 &(\alpha+z-1)\,a+(z-\alpha)\,b+(1-z)\,(K+L)\\
 &\qquad\overset{(a)}{>}w_1(\drho+2\chi)+w_2(\dsig+2\chi)+w_3(\drho+\dsig+2\chi)\\
 &\qquad\overset{(b)}{=}\alpha\,\drho+(1-\alpha)\,\dsig+2z\,\chi
 \overset{(c)}{\ge}\psi(\alpha,z),
\end{align*}
where $(a)$ follows from the three strict inequalities defining $\cV_3$ and the fact that the non-negative weights $w_1,w_2,w_3$ are not all zero, $(b)$ follows from \eqref{eq:weights}, and $(c)$ follows from \eqref{eq:support-each}. This contradicts the choice of $(\alpha,z)$.

Hence \eqref{eq:sector-n} holds. If $(\drho,\dsig,\chi)\in\cS$, there are $n_1<n_2<\cdots$ and $\theta_k\in\Theta_{n_k}$ with $\bv(\theta_k)\to(\drho,\dsig,\chi)$ (Definition~\ref{def:S}), and by \eqref{eq:sector-n} and the continuity of $\Phi$,
\begin{equation*}
 \Phi_{a,b,K,L}(\drho,\dsig)=\lim_{k\to\infty}\Phi_{a,b,K,L}\bigl(\drho(\theta_k),\dsig(\theta_k)\bigr)\le\lim_{k\to\infty}2\chi(\theta_k)=2\chi .
\end{equation*}
\end{proof}

Lemma~\ref{lem:sector-ineq} uses only the right inequality of the sector representation \eqref{eq:support-n}, which holds without any correction term. This is the reason why the achievability bound below holds for every $n$ and contains no error terms other than the polynomial factor $c_n$.

\subsection{Achievability: proof of Theorem~\ref{thm:main}(ii)}\label{sec:proof-ach}

Let $(a,b,K,L)\in\cZ$ and fix $n\ge1$.

\emph{Step 1: reduction to $K\le a$ and $L\le b$.} Put
\begin{equation}\label{eq:KLbar}
 (\bar K,\bar L)\eqdef\begin{cases}(K,L),&\text{if } K<a \text{ and } L<b,\\(a,b),&\text{otherwise.}\end{cases}
\end{equation}
Then $\bar K\le a$, $\bar L\le b$, and $(a,b,\bar K,\bar L)\in\cZ$. In the first case this is the assumption. In the second case, $K\ge a$ or $L\ge b$, so $(a,b)\in\cH$ by \eqref{eq:no-help}, and $(a,b)\in\cH\subseteq\cZ(a,b)$ by \eqref{eq:inclusions}, i.e., $(a,b,a,b)\in\cZ$.

\emph{Step 2: constants.} Let
\begin{equation}\label{eq:gamma-n}
\begin{aligned}
 &\gamma_n\eqdef\frac1{400\,N_n^2},\qquad\eta_n\eqdef\frac1n\log\frac1{\gamma_n},\qquad\text{so that}\\
 &e^{-n\eta_n}=\gamma_n\qquad\text{and}\qquad\gamma_n\le\frac1{400\,D_\lambda^4}=\gamma_{D_\lambda}\quad(\lambda\in\Lambda_n),
\end{aligned}
\end{equation}
where the last inequality is $D_\lambda^2\le N_n$ from \eqref{eq:block-bounds}. Put
\begin{equation}\label{eq:primes}
\begin{aligned}
 &a'\eqdef a-\eta_n,\qquad b'\eqdef b-\eta_n,\qquad K'\eqdef\bar K-\eta_n,\qquad L'\eqdef\bar L-\eta_n,\\
 &\text{so that}\qquad K'\le a',\qquad L'\le b'.
\end{aligned}
\end{equation}
(These four numbers may be negative.)

\emph{Step 3: budgets.} Fix $\lambda\in\Lambda_n$. Let $\{x_i\}_{i=1}^{D_\lambda}$ be an orthonormal eigenbasis of $\rho_\lambda$ with eigenvalues $\varrho_i$, and $\{y_j\}_{j=1}^{D_\lambda}$ an orthonormal eigenbasis of $\sigma_\lambda$ with eigenvalues $\varsigma_j$, and let
\begin{equation}\label{eq:block-budgets}
 \varepsilon_i\eqdef\min\Bigl\{1,\frac{e^{-na'}}{m_\lambda\varrho_i}\Bigr\},\quad
 \mu_i\eqdef\min\Bigl\{1,\frac{e^{-nK'}}{m_\lambda\varrho_i}\Bigr\},\quad
 \delta_j\eqdef\min\Bigl\{1,\frac{e^{-nb'}}{m_\lambda\varsigma_j}\Bigr\},\quad
 \nu_j\eqdef\min\Bigl\{1,\frac{e^{-nL'}}{m_\lambda\varsigma_j}\Bigr\}.
\end{equation}
By \eqref{eq:primes}, $e^{-na'}\le e^{-nK'}$ and $e^{-nb'}\le e^{-nL'}$; hence $0<\varepsilon_i\le\mu_i\le1$ and $0<\delta_j\le\nu_j\le1$, which is the ordering \eqref{eq:ordering}. For a pair $(i,j)$ let $P_i$ be the projection onto the eigenspace of $\rho_\lambda$ that contains $x_i$, let $Q_j$ be the projection onto the eigenspace of $\sigma_\lambda$ that contains $y_j$, and let $\theta_{ij}\eqdef(\lambda,P_i,Q_j)$. By \eqref{eq:sector-exp}, $m_\lambda\varrho_i=e^{-n\drho(\theta_{ij})}$ and $m_\lambda\varsigma_j=e^{-n\dsig(\theta_{ij})}$. Since $\min\{1,e^{-nt}\}=e^{-nt_+}$ for real $t$, the budgets \eqref{eq:block-budgets} are
\begin{equation}\label{eq:budget-exp}
\begin{aligned}
 \varepsilon_i&=e^{-n\pos{a'-\drho(\theta_{ij})}}, &\qquad \mu_i&=e^{-n\pos{K'-\drho(\theta_{ij})}},\\
 \delta_j&=e^{-n\pos{b'-\dsig(\theta_{ij})}}, &\qquad \nu_j&=e^{-n\pos{L'-\dsig(\theta_{ij})}},
\end{aligned}
\end{equation}
so that, by \eqref{eq:pairfn},
\begin{equation}\label{eq:budget-max}
 \max\bigl\{\varepsilon_i,\ \delta_j,\ \mu_i\nu_j\bigr\}=\exp\Bigl(-n\,\Phi_{a',b',K',L'}\bigl(\drho(\theta_{ij}),\dsig(\theta_{ij})\bigr)\Bigr),
\end{equation}
and the index sets \eqref{eq:index-sets} are described by (note that $\drho(\theta_{ij})$ depends only on $i$ and $\dsig(\theta_{ij})$ only on $j$)
\begin{equation}\label{eq:index-exp}
\begin{aligned}
 i\in I_c&\iff\drho(\theta_{ij})<a', &\qquad i\in I_0&\iff\drho(\theta_{ij})<K',\\
 j\in J_c&\iff\dsig(\theta_{ij})<b', &\qquad j\in J_0&\iff\dsig(\theta_{ij})<L'.
\end{aligned}
\end{equation}

\emph{Step 4: the hypothesis of Lemma~\ref{lem:suff} holds with $\gamma=\gamma_n$.} Let $(i,j)\in(I_0\times J_c)\cup(I_c\times J_0)$ and write $\theta=\theta_{ij}$. If $U_{ij}=\ip{x_i}{y_j}=0$, \eqref{eq:hyp} is trivial. Otherwise $0<\abs{U_{ij}}=\abs{\bra{x_i}P_iQ_j\ket{y_j}}$, so that $P_iQ_j\ne0$, i.e., $\theta\in\Theta_n$. By \eqref{eq:primes} and \eqref{eq:index-exp},
\begin{equation}\label{eq:case-check}
\begin{aligned}
 (i,j)\in I_0\times J_c&\ \Longrightarrow\ \drho(\theta)<K'=\bar K-\eta_n\le a-\eta_n\quad\text{and}\quad\dsig(\theta)<b'=b-\eta_n,\\
 (i,j)\in I_c\times J_0&\ \Longrightarrow\ \drho(\theta)<a'=a-\eta_n\quad\text{and}\quad\dsig(\theta)<L'=\bar L-\eta_n\le b-\eta_n .
\end{aligned}
\end{equation}
In both cases the assumptions of Lemma~\ref{lem:pairshift} are satisfied for the parameters $(a,b,\bar K,\bar L)$ and $\eta=\eta_n$. Therefore
\begin{align}
 \abs{U_{ij}}^2
 &\overset{(a)}{\le}e^{-2n\chi(\theta)}
 \overset{(b)}{\le}\exp\Bigl(-n\,\Phi_{a,b,\bar K,\bar L}\bigl(\drho(\theta),\dsig(\theta)\bigr)\Bigr)\notag\\
 &\overset{(c)}{\le}\exp\Bigl(-n\bigl[\Phi_{a',b',K',L'}\bigl(\drho(\theta),\dsig(\theta)\bigr)+\eta_n\bigr]\Bigr)
 \overset{(d)}{=}\gamma_n\,\max\bigl\{\varepsilon_i,\ \delta_j,\ \mu_i\nu_j\bigr\},\label{eq:hyp-chain}
\end{align}
where $(a)$ follows from \eqref{eq:sector-overlap}; $(b)$ follows from Lemma~\ref{lem:sector-ineq}, applied to $(a,b,\bar K,\bar L)\in\cZ$ (Step~1); $(c)$ follows from Lemma~\ref{lem:pairshift} together with \eqref{eq:primes}; and $(d)$ follows from \eqref{eq:budget-max} and $e^{-n\eta_n}=\gamma_n$. This is \eqref{eq:hyp}.

\emph{Step 5: the test.} By Steps~3 and~4 and $\gamma_n\le\gamma_{D_\lambda}$, Lemma~\ref{lem:suff} provides, for every $\lambda\in\Lambda_n$, a POVM $(E_0^\lambda,E_1^\lambda,E_2^\lambda)$ on $\bbC^{D_\lambda}$ that satisfies \eqref{eq:budgets} with the budgets \eqref{eq:block-budgets}. Let
\begin{equation}\label{eq:assembled}
 \Pi_{k}\eqdef\bigoplus_{\lambda\in\Lambda_n}E^\lambda_{k}\otimes\one_{m_\lambda},\qquad k=0,1,2 .
\end{equation}
Then $\Pi_k\ge0$ and $\Pi_0+\Pi_1+\Pi_2=\bigoplus_\lambda\one_{D_\lambda}\otimes\one_{m_\lambda}=\one$, so $T_n\eqdef(\Pi_0,\Pi_1,\Pi_2)$ is a test. We have
\begin{align}
 \alpha_n(T_n)=\tr\bigl[\rhon\Pi_1\bigr]
 &\overset{(a)}{=}\sum_{\lambda\in\Lambda_n}m_\lambda\tr\bigl[\rho_\lambda E_1^\lambda\bigr]
 \overset{(b)}{=}\sum_{\lambda\in\Lambda_n}\sum_{i=1}^{D_\lambda}m_\lambda\varrho_i\bra{x_i}E_1^\lambda\ket{x_i}
 \overset{(c)}{\le}\sum_{\lambda\in\Lambda_n}\sum_{i=1}^{D_\lambda}m_\lambda\varrho_i\,\varepsilon_i\notag\\
 &\overset{(d)}{\le}\sum_{\lambda\in\Lambda_n}\sum_{i=1}^{D_\lambda}e^{-na'}
 =\Bigl(\sum_{\lambda\in\Lambda_n}D_\lambda\Bigr)e^{-na'}\notag\\
 &\overset{(e)}{\le}N_n\,e^{n\eta_n}\,e^{-na}
 \overset{(f)}{=}400\,N_n^3\,e^{-na}=\tfrac12\,c_n\,e^{-na},\label{eq:ach-alpha}
\end{align}
where $(a)$ follows from \eqref{eq:block-states} with $t=1$ and \eqref{eq:assembled}; $(b)$ follows from the spectral decomposition $\rho_\lambda=\sum_i\varrho_i\ketbra{x_i}$; $(c)$ follows from \eqref{eq:budgets}; $(d)$ follows from $m_\lambda\varrho_i\varepsilon_i=\min\{m_\lambda\varrho_i,e^{-na'}\}\le e^{-na'}$, which holds by \eqref{eq:block-budgets}; $(e)$ follows from \eqref{eq:block-bounds} and \eqref{eq:primes}; and $(f)$ follows from $e^{n\eta_n}=1/\gamma_n=400N_n^2$ together with $N_n^3=(n+1)^{3(d^2-1)}$ and the definition of $c_n$ in Theorem~\ref{thm:main}. Replacing $(E_1^\lambda,\varepsilon_i,a')$ in \eqref{eq:ach-alpha} by $(E_2^\lambda,\mu_i,K')$, and using the eigenbasis $\{y_j\}$ of $\sigma_\lambda$ with $(E_0^\lambda,\delta_j,b')$ and $(E_2^\lambda,\nu_j,L')$, we obtain in the same way
\begin{equation}\label{eq:ach-rest}
 u_n(T_n)\le\tfrac12c_n\,e^{-n\bar K},\qquad\beta_n(T_n)\le\tfrac12c_n\,e^{-nb},\qquad v_n(T_n)\le\tfrac12c_n\,e^{-n\bar L} .
\end{equation}

\emph{Step 6: conclusion.} If $K<a$ and $L<b$, then $(\bar K,\bar L)=(K,L)$, and \eqref{eq:ach-alpha}, \eqref{eq:ach-rest} give \eqref{eq:finite-n}. Otherwise $(\bar K,\bar L)=(a,b)$, and we pass to the binary test $T_n'\eqdef(\Pi_0,\Pi_1+\Pi_2,0)$, for which, by \eqref{eq:ach-alpha} and \eqref{eq:ach-rest},
\begin{equation*}
\begin{aligned}
 \alpha_n(T_n')&=\alpha_n(T_n)+u_n(T_n)\le\tfrac12c_n\,e^{-na}+\tfrac12c_n\,e^{-n\bar K}=c_n\,e^{-na},\\
 \beta_n(T_n')&=\beta_n(T_n)\le\tfrac12c_n\,e^{-nb},\qquad u_n(T_n')=v_n(T_n')=0 .
\end{aligned}
\end{equation*}
Thus $T_n'$ is binary and satisfies \eqref{eq:finite-n}. This proves Theorem~\ref{thm:main}(ii). \qed

\subsection{The converse in sector form}\label{sec:proof-conv}

\begin{proposition}[Converse, sector form]\label{prop:converse}
If $(a,b,K,L)\in\cA$, then \eqref{eq:sector} holds.
\end{proposition}

\begin{proof}
Let $(T_n)_{n\ge1}$, $T_n=(\Pi_0,\Pi_1,\Pi_2)$, be tests satisfying \eqref{eq:achievable}, write $\alpha_n,\beta_n,u_n,v_n$ for their probabilities \eqref{eq:four-prob}, and let $(\drho,\dsig,\chi)\in\cS$. By Definition~\ref{def:S} there are $n_1<n_2<\cdots$ and sectors $\theta_k=(\lambda_k,P_k,Q_k)\in\Theta_{n_k}$ with $\bv(\theta_k)\to(\drho,\dsig,\chi)$.

\emph{Step 1: reduction to a pair of vectors.} Fix $k$ and write $n=n_k$ and $\theta=\theta_k=(\lambda,P,Q)$. By \eqref{eq:sector-overlap} there are unit vectors $x\in\ran P$ and $y\in\ran Q$ with
\begin{equation}\label{eq:conv-0}
 \abs{\ip xy}=\norm{PQ}_\infty=e^{-n\chi(\theta)} .
\end{equation}
Let $\iota_\lambda:\bbC^{D_\lambda}\otimes\bbC^{m_\lambda}\to\bigoplus_{\lambda'\in\Lambda_n}\bbC^{D_{\lambda'}}\otimes\bbC^{m_{\lambda'}}$ be the inclusion of the block $\lambda$, an isometry, and let $E'_{j}\eqdef\iota_\lambda^*\,\Pi_{j}\,\iota_\lambda$ for $j=0,1,2$. Then $E_j'\ge0$ and $E_0'+E_1'+E_2'=\iota_\lambda^*\iota_\lambda=\one$, so $(E_0',E_1',E_2')$ is a POVM on $\bbC^{D_\lambda}\otimes\bbC^{m_\lambda}$. Since $x$ is a unit vector in $\ran P$, we have $\rho_\lambda\ge\varrho_PP\ge\varrho_P\ketbra x$, and with \eqref{eq:block-states} for $t=1$,
\begin{equation*}
 \rhon=\bigoplus_{\lambda'\in\Lambda_n}\rho_{\lambda'}\otimes\one_{m_{\lambda'}}\ \ge\ \iota_\lambda\bigl(\rho_\lambda\otimes\one_{m_\lambda}\bigr)\iota_\lambda^*\ \ge\ \varrho_P\,\iota_\lambda\bigl(\ketbra x\otimes\one_{m_\lambda}\bigr)\iota_\lambda^* .
\end{equation*}
Therefore, for $j=0,1,2$, with the notation \eqref{eq:ebar} and $m=m_\lambda$,
\begin{equation}\label{eq:conv-1}
\begin{aligned}
 \tr\bigl[\rhon\,\Pi_{j}\bigr]&\overset{(a)}{\ge}\varrho_P\tr\bigl[\iota_\lambda(\ketbra x\otimes\one_{m_\lambda})\iota_\lambda^*\,\Pi_j\bigr]
 \overset{(b)}{=}\varrho_P\tr\bigl[(\ketbra x\otimes\one_{m_\lambda})E'_{j}\bigr]\\
 &\overset{(c)}{=}m_\lambda\varrho_P\,\bar e_x(E'_{j})\overset{(d)}{=}e^{-n\drho(\theta)}\,\bar e_x(E'_{j}),
\end{aligned}
\end{equation}
where $(a)$ follows from the operator inequality above together with $\Pi_j\ge0$, $(b)$ follows from the cyclicity of the trace and the definition of $E_j'$, $(c)$ follows from \eqref{eq:ebar}, and $(d)$ follows from \eqref{eq:sector-exp}. In the same way, $\tr[\sigman\Pi_{j}]\ge e^{-n\dsig(\theta)}\,\bar e_y(E'_{j})$. With \eqref{eq:four-prob} and $\bar e_x,\bar e_y\le1$ this gives
\begin{equation}\label{eq:conv-2}
\begin{aligned}
 \bar e_x(E'_1)&\le\min\bigl\{1,\alpha_n\,e^{n\drho(\theta)}\bigr\}, &\qquad \bar e_x(E'_2)&\le\min\bigl\{1,u_n\,e^{n\drho(\theta)}\bigr\},\\
 \bar e_y(E'_0)&\le\min\bigl\{1,\beta_n\,e^{n\dsig(\theta)}\bigr\}, &\qquad \bar e_y(E'_2)&\le\min\bigl\{1,v_n\,e^{n\dsig(\theta)}\bigr\}.
\end{aligned}
\end{equation}

\emph{Step 2: pairwise necessity.} By \eqref{eq:conv-0}, Lemma~\ref{lem:nec}, the bounds $\bar e_x(E_0')\le1$, $\bar e_y(E_1')\le1$, and \eqref{eq:conv-2},
\begin{align}
 e^{-n\chi(\theta)}=\abs{\ip xy}
 &\le\sqrt{\bar e_x(E'_0)\,\bar e_y(E'_0)}+\sqrt{\bar e_x(E'_1)\,\bar e_y(E'_1)}+\sqrt{\bar e_x(E'_2)\,\bar e_y(E'_2)}\notag\\
 &\le\sqrt{\bar e_y(E'_0)}+\sqrt{\bar e_x(E'_1)}+\sqrt{\bar e_x(E'_2)\,\bar e_y(E'_2)}\notag\\
 &\le\sqrt{\min\bigl\{1,\beta_ne^{n\dsig(\theta)}\bigr\}}+\sqrt{\min\bigl\{1,\alpha_ne^{n\drho(\theta)}\bigr\}}\notag\\
 &\qquad+\sqrt{\min\bigl\{1,u_ne^{n\drho(\theta)}\bigr\}\min\bigl\{1,v_ne^{n\dsig(\theta)}\bigr\}} .\label{eq:conv-3}
\end{align}

\emph{Step 3: exponents.} Fix $\xi>0$. By \eqref{eq:achievable} there is $n_\xi$ such that for all $n\ge n_\xi$,
\begin{equation*}
 \alpha_n\le e^{-n(a-\xi)},\qquad\beta_n\le e^{-n(b-\xi)},\qquad u_n\le e^{-n(K-\xi)},\qquad v_n\le e^{-n(L-\xi)} .
\end{equation*}
Since $\min\{1,e^{-nt}\}=e^{-nt_+}$, \eqref{eq:conv-3} yields, for $n=n_k\ge n_\xi$,
\begin{align}
 e^{-n\chi(\theta)}
 &\le e^{-\frac n2\pos{b-\xi-\dsig(\theta)}}+e^{-\frac n2\pos{a-\xi-\drho(\theta)}}+e^{-\frac n2\left[\pos{K-\xi-\drho(\theta)}+\pos{L-\xi-\dsig(\theta)}\right]}\notag\\
 &\le3\exp\Bigl(-\frac n2\,\Phi_{a-\xi,\,b-\xi,\,K-\xi,\,L-\xi}\bigl(\drho(\theta),\dsig(\theta)\bigr)\Bigr),\label{eq:conv-4}
\end{align}
because each of the three exponents is at least the minimum \eqref{eq:pairfn}. Taking logarithms in \eqref{eq:conv-4},
\begin{equation*}
 2\chi(\theta_k)\ \ge\ \Phi_{a-\xi,\,b-\xi,\,K-\xi,\,L-\xi}\bigl(\drho(\theta_k),\dsig(\theta_k)\bigr)-\frac{2\log3}{n_k}\qquad\text{for all } k \text{ with } n_k\ge n_\xi .
\end{equation*}
Letting $k\to\infty$ and then $\xi\downarrow0$, and using the continuity of $\Phi$ in both limits,
\begin{equation*}
 2\chi\ \ge\ \Phi_{a-\xi,\,b-\xi,\,K-\xi,\,L-\xi}(\drho,\dsig)\ \xrightarrow[\xi\downarrow0]{}\ \Phi_{a,b,K,L}(\drho,\dsig),
\end{equation*}
which is \eqref{eq:sector} at the point $(\drho,\dsig,\chi)$.
\end{proof}

\subsection{Completion of the proof of Theorem~\ref{thm:main}}\label{sec:proof-main}

If $(a,b,K,L)\in\cA$, then \eqref{eq:sector} holds by Proposition~\ref{prop:converse}. Then together with Proposition~\ref{prop:single}, we have that $(a,b,K,L)\in\cZ$ as desired. Hence Theorem~\ref{thm:main}(i) follows. Combining with the proof of Theorem~\ref{thm:main}(ii) in Section~\ref{sec:proof-ach}, we complete the proof of Theorem~\ref{thm:main}.
\qed

\subsection{Proof of Theorem~\ref{thm:fixed}}\label{sec:proof-fixed}

\emph{Step 1: $\cZ$ is closed in $(0,\infty)^4$.} Let $(a_k,b_k,K_k,L_k)\in\cZ$ converge to $(a,b,K,L)\in(0,\infty)^4$, with witnesses $s_k,s_k'\in[0,1]$ and $(\alpha_k,z_k)\in\cD$ of \eqref{eq:P1}, \eqref{eq:P2} and \eqref{eq:F}. Since $[0,1]$ and $\cD$ are compact, we may pass to a subsequence along which $s_k\to s$, $s_k'\to s'$ and $(\alpha_k,z_k)\to(\alpha,z)\in\cD$. Since $\psi$ is continuous (Proposition~\ref{prop:psi}(a)) and so is $(x,y)\mapsto\min\{x,y\}$,
\begin{align*}
 s\min\{a,K\}+(1-s)\,b&=\lim_{k\to\infty}\bigl[s_k\min\{a_k,K_k\}+(1-s_k)\,b_k\bigr]\le\lim_{k\to\infty}\psi(s_k,1)=\psi(s,1),\\
 s'a+(1-s')\min\{b,L\}&=\lim_{k\to\infty}\bigl[s_k'a_k+(1-s_k')\min\{b_k,L_k\}\bigr]\le\lim_{k\to\infty}\psi(s_k',1)=\psi(s',1),
\end{align*}
and, in the same way, $(\alpha+z-1)\,a+(z-\alpha)\,b+(1-z)(K+L)=\lim_k\bigl[(\alpha_k+z_k-1)\,a_k+(z_k-\alpha_k)\,b_k+(1-z_k)(K_k+L_k)\bigr]\le\lim_k\psi(\alpha_k,z_k)=\psi(\alpha,z)$. Hence $(a,b,K,L)\in\cZ$, and consequently every slice $\cZ(K,L)$ is closed in $(0,\infty)^2$.

\emph{Step 2: right inclusion in \eqref{eq:sandwich}.} By \eqref{eq:R-in-A} and Theorem~\ref{thm:main}(i),
\begin{equation*}
 (a,b)\in\cR(K,L)\ \Longrightarrow\ (a,b,K,L)\in\cA\ \Longrightarrow\ (a,b,K,L)\in\cZ\ \Longrightarrow\ (a,b)\in\cZ(K,L).
\end{equation*}

\emph{Step 3: left inclusion in \eqref{eq:sandwich}.} Let $\eta>0$ and $(a,b)\in\cZ(K+\eta,L+\eta)$. By Theorem~\ref{thm:main}(ii) there are tests $T_n$, $n\ge1$, with
\begin{equation*}
\begin{aligned}
 &\alpha_n(T_n)\le c_n\,e^{-na},\qquad\beta_n(T_n)\le c_n\,e^{-nb},\\
 &u_n(T_n)\le c_n\,e^{-n\eta}\,e^{-nK},\qquad v_n(T_n)\le c_n\,e^{-n\eta}\,e^{-nL}.
\end{aligned}
\end{equation*}
Since $\frac1n\log c_n\to0$, there is $n_1$ with $c_n\le e^{n\eta}$ for all $n\ge n_1$, and then $u_n(T_n)\le e^{-nK}$ and $v_n(T_n)\le e^{-nL}$ for $n\ge n_1$. For $n<n_1$ we replace $T_n$ by the test $(\one,0,0)$, for which $u_n=v_n=0$. The resulting sequence satisfies \eqref{eq:hard} for all $n\ge1$, and $\liminf_n-\frac1n\log\alpha_n\ge a$, $\liminf_n-\frac1n\log\beta_n\ge b$. Hence $(a,b)\in\cR(K,L)$.

\emph{Step 4: $\cR(K,L)$ is closed in $(0,\infty)^2$.} Let $(a_j,b_j)\in\cR(K,L)$, $j\ge1$, converge to $(a,b)\in(0,\infty)^2$. For each $j$ let $(T_n^{(j)})_{n\ge1}$ be a sequence of tests that satisfies \eqref{eq:hard} for all $n\ge1$ and has $\liminf_n-\frac1n\log\alpha_n(T_n^{(j)})\ge a_j$ and $\liminf_n-\frac1n\log\beta_n(T_n^{(j)})\ge b_j$. Choose integers $N_1<N_2<\cdots$ such that
\begin{equation*}
 \alpha_n\bigl(T_n^{(j)}\bigr)\le e^{-n(a_j-1/j)}\quad\text{and}\quad\beta_n\bigl(T_n^{(j)}\bigr)\le e^{-n(b_j-1/j)}\qquad\text{for all } n\ge N_j ,
\end{equation*}
and put $T_n\eqdef T_n^{(j)}$ for $N_j\le n<N_{j+1}$, and $T_n\eqdef(\one,0,0)$ for $n<N_1$. Every $T_n$ satisfies \eqref{eq:hard}. If $N_j\le n<N_{j+1}$, then $-\frac1n\log\alpha_n(T_n)\ge a_j-\frac1j$ and $-\frac1n\log\beta_n(T_n)\ge b_j-\frac1j$; since $j\to\infty$ as $n\to\infty$, we obtain $\liminf_n-\frac1n\log\alpha_n(T_n)\ge\lim_j(a_j-\frac1j)=a$ and, likewise, $\liminf_n-\frac1n\log\beta_n(T_n)\ge b$. Hence $(a,b)\in\cR(K,L)$.

\emph{Step 5: equality for $K+L\ne-2\log F(\rho,\sigma)$.} Let $\cZ^{+}\eqdef\bigcup_{\eta>0}\cZ(K+\eta,L+\eta)$ denote the left-hand side of \eqref{eq:sandwich}. We show that for every $(a,b)\in\cZ(K,L)$ and every $\zeta\in(0,\min\{a,b\})$ there is $\eta>0$ with
\begin{equation}\label{eq:perturb}
 (a-\zeta,\,b-\zeta)\in\cZ(K+\eta,L+\eta).
\end{equation}
Then $(a-\zeta,b-\zeta)\in\cZ^{+}$ for all small $\zeta>0$, so $(a,b)\in\cl\,\cZ^{+}$; together with Steps~1--4,
\begin{equation*}
 \cZ(K,L)\subseteq\cl\,\cZ^{+}\subseteq\cl\,\cR(K,L){}=\cR(K,L)\subseteq\cZ(K,L),
\end{equation*}
which proves the last assertion of the theorem and \eqref{eq:fixed-equality}. To prove \eqref{eq:perturb}, let $s$, $s'$ and $(\alpha,z)$ be witnesses of \eqref{eq:P1}, \eqref{eq:P2} and \eqref{eq:F} for $(a,b,K,L)$. For $s\in\{0,1\}$ the left-hand side of \eqref{eq:P1} is $b>0$ or $\min\{a,K\}>0$, whereas $\psi(0,1)=\psi(1,1)=0$ (Proposition~\ref{prop:psi}(d)); hence $s\in(0,1)$, and likewise $s'\in(0,1)$.

\emph{Condition (P1).} If $0<\eta\le\zeta\,(1-s)/s$, then
\begin{align*}
 s\min\{a-\zeta,K+\eta\}+(1-s)(b-\zeta)
 &\overset{(a)}{\le}s\bigl(\min\{a,K\}+\eta\bigr)+(1-s)(b-\zeta)\\
 &=\bigl[s\min\{a,K\}+(1-s)\,b\bigr]+s\,\eta-(1-s)\,\zeta\\
 &\overset{(b)}{\le}\psi(s,1),
\end{align*}
where $(a)$ follows from $\min\{a-\zeta,K+\eta\}\le\min\{a,K+\eta\}\le\min\{a,K\}+\eta$, and $(b)$ follows from the choice of $s$ and $s\,\eta\le(1-s)\,\zeta$.

\emph{Condition (P2).} If $0<\eta\le\zeta\,s'/(1-s')$, then in the same way
\begin{equation*}
\begin{aligned}
 s'(a-\zeta)+(1-s')\min\{b-\zeta,L+\eta\}
 &\le\bigl[s'a+(1-s')\min\{b,L\}\bigr]-s'\zeta+(1-s')\,\eta\\
 &\le\psi(s',1).
\end{aligned}
\end{equation*}

\emph{Condition (F).} If $K+L<-2\log F(\rho,\sigma)$ and $0<\eta\le\frac12\bigl(-2\log F(\rho,\sigma)-K-L\bigr)$, then \eqref{eq:F} holds for $(a-\zeta,b-\zeta,K+\eta,L+\eta)$ at the point $(\frac12,\frac12)$, where by \eqref{eq:special} it reads $\frac12(K+L+2\eta)\le-\log F(\rho,\sigma)$. If $K+L>-2\log F(\rho,\sigma)$, then \eqref{eq:F} for $(a,b,K,L)$ fails at $(\frac12,\frac12)$, so that $(\alpha,z)\ne(\frac12,\frac12)$; since $(\frac12,\frac12)$ is the only point of $\cD$ with $z=\frac12$, we have $z>\frac12$. If $z=1$ let $\eta>0$ be arbitrary, and if $z<1$ let $0<\eta\le\zeta\,(2z-1)/(2(1-z))$. Then
\begin{align*}
 &(\alpha+z-1)(a-\zeta)+(z-\alpha)(b-\zeta)+(1-z)(K+L+2\eta)\\
 &\qquad=\bigl[(\alpha+z-1)\,a+(z-\alpha)\,b+(1-z)(K+L)\bigr]-(2z-1)\,\zeta+2(1-z)\,\eta\ \le\ \psi(\alpha,z).
\end{align*}
Choosing $\eta>0$ so small that all three requirements are met gives \eqref{eq:perturb}. \qed

\subsection{Proof of the symmetric-pair specialization}\label{sec:proof-symmetric}

\begin{proof}[Proof of Corollary~\ref{cor:symmetric}]
\emph{Symmetry.} For every unitary $V$, $Q_{\alpha,z}(V\rho V^*\|V\sigma V^*)=Q_{\alpha,z}(\rho\|\sigma)$, because $(VAV^*)^t=VA^tV^*$ for positive definite $A$ and the trace is unitarily invariant. Hence, by the assumption on $V$ and \eqref{eq:Qnorm},
\begin{equation*}
 Q_{\alpha,z}(\rho\|\sigma)=Q_{\alpha,z}(V\sigma V^*\|V\rho V^*)=Q_{\alpha,z}(\sigma\|\rho)=Q_{1-\alpha,z}(\rho\|\sigma),
\end{equation*}
i.e., $\psi(\alpha,z)=\psi(1-\alpha,z)$. Substituting $s\mapsto1-s$ in \eqref{eq:Hoeffding-fn} gives $H_r(\sigma\|\rho)=H_r(\rho\|\sigma)$. By concavity (Proposition~\ref{prop:psi}(b)),
\begin{equation}\label{eq:sym-max}
 \psi(\alpha,z)=\tfrac12\psi(\alpha,z)+\tfrac12\psi(1-\alpha,z)\le\psi(\tfrac12,z)=\psi_{1/2}(z)\qquad\text{for all }\alpha\in\bbR,\ z>0 .
\end{equation}
In particular $C=\psi_{1/2}(1)=\max_{s\in[0,1]}\psi(s,1)$.

\emph{The constants.} The function $\psi_{1/2}$ is real-analytic, concave and non-decreasing on $(0,\infty)$ (Proposition~\ref{prop:psi}). Let
\begin{equation}\label{eq:varpi}
 \varpi(z)\eqdef\psi_{1/2}(z)-(z-\tfrac12)\,\psi_{1/2}'(z),\qquad\text{so that}\qquad\varpi'(z)=-(z-\tfrac12)\,\psi_{1/2}''(z)\ge0\quad\text{for } z\ge\tfrac12 .
\end{equation}
Then $\varpi$ is non-decreasing on $[\frac12,1]$, $\varpi(\frac12)=r_0$ and $\varpi(1)=r_{\rm c}$ by \eqref{eq:sym-constants}; hence $r_0\le r_{\rm c}$. Moreover $r_{\rm c}=C-\frac12\psi_{1/2}'(1)\le C$.

We also record the equality case:
\begin{equation}\label{eq:sym-threshold-equality}
 r_0=r_{\rm c}\quad\Longleftrightarrow\quad\rho\sigma=\sigma\rho,
 \qquad\text{in which case }r_0=r_{\rm c}=C.
\end{equation}
If $r_0=r_{\rm c}$, the tangent line to the concave function $\psi_{1/2}$ at $1$ agrees with it also at $\frac12$. Hence $\psi_{1/2}$ is affine on $[\frac12,1]$, and by real analyticity it is affine on $(0,\infty)$. Its slope is nonnegative. On the other hand, the supporting-plane inequality for $\psi$ at $(0,1)$ gives
\begin{equation*}
 \psi(\tfrac12,z)
 \le\psi(0,1)+\tfrac12\partial_\alpha\psi(0,1)
 +(z-1)\partial_z\psi(0,1)
 =\tfrac12\partial_\alpha\psi(0,1)\qquad(z>0),
\end{equation*}
where the equality uses $\psi(0,z)=0$. Thus $\psi_{1/2}$ is bounded above on $(0,\infty)$, so its slope is zero. In particular $r_0=r_{\rm c}=C$ and
$\tr\rho^{1/2}\sigma^{1/2}=\norm{\rho^{1/2}\sigma^{1/2}}_1$.
For $X=\rho^{1/2}\sigma^{1/2}=U|X|$, write $|X|=\sum_k t_k\ketbra{e_k}$ with $t_k>0$. The last equality gives
$\sum_k t_k\bra{e_k}U\ket{e_k}=\sum_k t_k$.
Since $|\bra{e_k}U\ket{e_k}|\le1$, taking real parts shows that every diagonal entry equals $1$; equality in the Cauchy--Schwarz inequality then gives $Ue_k=e_k$. Thus $U=\one$, $X$ is Hermitian, and $\rho^{1/2}\sigma^{1/2}=\sigma^{1/2}\rho^{1/2}$, which implies that $\rho$ and $\sigma$ commute. Conversely, for commuting states $Q_{\alpha,z}$ is independent of $z$, so $r_0=r_{\rm c}=C$.

\emph{Conditions (P1) and (P2).} For $(a,b,K,L)=(e,e,r,r)$, \eqref{eq:P2} is obtained from \eqref{eq:P1} by the substitution $s\mapsto1-s$, so the two conditions are equivalent, and by Lemma~\ref{lem:Hform} they are equivalent to $e\le\max\{H_e,H_r\}$. By \eqref{eq:H-region}, $e\le H_e$ if and only if $(e,e)\in\cH$, i.e., $e\le\psi(s,1)$ for some $s\in[0,1]$, i.e., $e\le C$. Hence
\begin{equation}\label{eq:sym-P}
 \text{\eqref{eq:P1} and \eqref{eq:P2} hold for }(e,e,r,r)\iff e\le\max\{C,H_r\}.
\end{equation}

\emph{Condition (F).} For $(e,e,r,r)$ the left-hand side of \eqref{eq:F} is $(2z-1)\,e+2(1-z)\,r$. By \eqref{eq:sym-max}, and since $(\frac12,z)\in\cD$ for $z\in[\frac12,1]$, \eqref{eq:F} is equivalent to
\begin{equation}\label{eq:sym-F}
 (2z-1)\,e+2(1-z)\,r\ \le\ \psi_{1/2}(z)\qquad\text{for some } z\in[\tfrac12,1].
\end{equation}
At $z=\frac12$ this reads $r\le r_0$. Hence, if $r\le r_0$, \eqref{eq:sym-F} holds for all $e$, and by \eqref{eq:sym-P} and $\cA=\cZ$ we get $e^\ast(r)=\max\{C,H_r\}$, which is (a). If $r>r_0$, \eqref{eq:sym-F} fails at $z=\frac12$, and it holds if and only if
\begin{equation}\label{eq:sym-F2}
 e\ \le\ \frac{\psi_{1/2}(z)-2(1-z)\,r}{2z-1}\qquad\text{for some } z\in(\tfrac12,1].
\end{equation}
We show below that the supremum $e_F(r)$ of the right-hand side over $z\in(\frac12,1]$ is attained. Then \eqref{eq:sym-F2} is equivalent to $e\le e_F(r)$, and (b) follows from \eqref{eq:sym-P}.

\emph{The function $e_F$.} Let $r>r_0$. For $y\in(0,1]$ and $z=\frac{1+y}2$ put $\varphi(y)\eqdef\psi_{1/2}(z)-(1-y)\,r$, so that the right-hand side of \eqref{eq:sym-F2} is $\varphi(y)/y$ and $\varphi'(y)=\frac12\psi_{1/2}'(z)+r$. Then
\begin{equation}\label{eq:phi-deriv}
 y^2\,\frac{d}{dy}\frac{\varphi(y)}{y}=y\,\varphi'(y)-\varphi(y)=(z-\tfrac12)\,\psi_{1/2}'(z)+y\,r-\psi_{1/2}(z)+(1-y)\,r=r-\varpi(z).
\end{equation}
If $r\ge r_{\rm c}$, then $r-\varpi(z)\ge r-\varpi(1)=r-r_{\rm c}\ge0$ for all $z\in(\frac12,1]$, so $\varphi(y)/y$ is non-decreasing on $(0,1]$, and
\begin{equation*}
 e_F(r)=\varphi(1)=\psi_{1/2}(1)=C .
\end{equation*}
If $r_0<r<r_{\rm c}$, then $\varpi(\frac12)<r<\varpi(1)$, and since $\varpi$ is continuous and non-decreasing there is $z_r\in(\frac12,1)$ with $\varpi(z_r)=r$, $\varpi\le r$ on $[\frac12,z_r]$ and $\varpi\ge r$ on $[z_r,1]$. By \eqref{eq:phi-deriv}, $\varphi(y)/y$ is non-decreasing on $(0,y_r]$ and non-increasing on $[y_r,1]$, where $y_r\eqdef2z_r-1$. Hence the supremum is attained at $y_r$, and, using $y_r\varphi'(y_r)=\varphi(y_r)$ and $r=\varpi(z_r)$,
\begin{equation*}
 e_F(r)=\frac{\varphi(y_r)}{y_r}=\varphi'(y_r)=\tfrac12\psi_{1/2}'(z_r)+\varpi(z_r)=\psi_{1/2}(z_r)+(1-z_r)\,\psi_{1/2}'(z_r).
\end{equation*}
Moreover, the derivative of $\varphi(y)/y$ at $y=1$ is $r-r_{\rm c}<0$ by \eqref{eq:phi-deriv}, so $\varphi(y)/y>\varphi(1)=C$ for $y<1$ close to $1$, and therefore $e_F(r)>C$. If $z\in(\frac12,1)$ is another solution of $\varpi(z)=r$, then \eqref{eq:eF-formula} gives the same value at $z$ as at $z_r$: indeed, $\varpi=r$ between $z$ and $z_r$, so that, by \eqref{eq:phi-deriv}, the function $\varphi(y)/y$ is constant between the corresponding values of $y$. This proves (c) and shows that the supremum in \eqref{eq:eF} is attained.

\emph{The last assertion.} If $r>r_0$ and $r\ge r_{\rm c}$, then $e^\ast(r)=\min\{\max\{C,H_r\},C\}=C$ by (b) and (c). The only remaining case with $r\ge r_{\rm c}$ is $r=r_{\rm c}=r_0$. By \eqref{eq:sym-threshold-equality}, $r_0=r_{\rm c}=C$. Since $\psi(s,1)\le C$ for every $s\in[0,1]$, \eqref{eq:Hoeffding-fn} gives $H_C\le C$, with equality at $s=\frac12$. Hence (a) gives $e^\ast(r_{\rm c})=C$. Let $0<r<r_{\rm c}$. Then $r<C$, and the choice $s=\frac12$ in \eqref{eq:Hoeffding-fn} gives $H_r\ge2\psi(\frac12,1)-r=2C-r>C$. If $r\le r_0$, (a) gives $e^\ast(r)=H_r>C$; if $r_0<r<r_{\rm c}$, (b) and (c) give $e^\ast(r)=\min\{H_r,e_F(r)\}>C$.

\emph{The limit $r\downarrow r_0$ and the criterion \eqref{eq:jump}.} Assume $r_0<r_{\rm c}$. We claim that $\varpi(z)>r_0$ for all $z\in(\frac12,1]$. Otherwise $\varpi(z_1)=r_0=\varpi(\frac12)$ for some $z_1\in(\frac12,1]$; then $\varpi$ is constant on $[\frac12,z_1]$ by monotonicity, hence constant on $[\frac12,1]$ because it is real-analytic, which contradicts $\varpi(1)=r_{\rm c}>r_0$. Now let $z_1\in(\frac12,1)$ and $r_0<r<\varpi(z_1)$. Every solution $z_r$ of $\varpi(z_r)=r$ satisfies $z_r<z_1$, because $\varpi\ge\varpi(z_1)>r$ on $[z_1,1]$. Hence $z_r\to\frac12$ as $r\downarrow r_0$, and by (c),
\begin{equation}\label{eq:eF-limit}
 \lim_{r\downarrow r_0}e_F(r)=\lim_{r\downarrow r_0}\bigl[r+\tfrac12\psi_{1/2}'(z_r)\bigr]=r_0+\tfrac12\psi_{1/2}'(\tfrac12).
\end{equation}
The function $r\mapsto H_r$ is finite and, as a supremum of affine functions of $r$, convex on $(0,\infty)$; hence it is continuous. By (a), (b) and \eqref{eq:eF-limit},
\begin{equation*}
\begin{aligned}
 &e^\ast(r_0)=\max\{C,H_{r_0}\}=\lim_{r\uparrow r_0}e^\ast(r),\\
 &\lim_{r\downarrow r_0}e^\ast(r)=\min\bigl\{\max\{C,H_{r_0}\},\ r_0+\tfrac12\psi_{1/2}'(\tfrac12)\bigr\}.
\end{aligned}
\end{equation*}
Thus $e^\ast$ is discontinuous at $r_0$ if and only if \eqref{eq:jump} holds. If the reverse strict inequality holds, then by \eqref{eq:eF-limit} and the continuity of $H_r$ we have $e_F(r)>\max\{C,H_r\}$ for all $r>r_0$ sufficiently close to $r_0$, and then $e^\ast(r)=\max\{C,H_r\}$ by (b).
\end{proof}

\section{Discussion and Future Work}\label{sec:discussion}

\subsection{Three ways to pay for an overlap}

Theorem~\ref{thm:pair} and its proof suggest the following picture. On the exponential scale, the pair $(\rhon,\sigman)$ can be analyzed through pairs $(x,y)$ of unit eigenvectors of the block matrices $\rho_\lambda$ and $\sigma_\lambda$, with multiplicity-weighted eigenvalues $e^{-n\drho}=m_\lambda\varrho_P$ and $e^{-n\dsig}=m_\lambda\varsigma_Q$ and with an overlap of at most $e^{-n\chi}$. A test has to pay for the overlap of every such pair in one of three ways (Lemma~\ref{lem:nec}): by a type-I error on $x$, by a type-II error on $y$, or by rejecting on both $x$ and $y$, in which case the cost is the geometric mean of the two rejection probabilities. Conversely, if the budgets are ordered as in \eqref{eq:ordering} and every pair on which the relevant constraints are active can be paid for with a polynomial margin, a test exists (Lemma~\ref{lem:suff}).

For binary tests only the first two options are available. They involve the sector exponents only through the combinations $\drho+2\chi$ and $\dsig+2\chi$, which are the exponents of $m_\lambda\varrho_P\norm{PQ}_\infty^2$ and $m_\lambda\varsigma_Q\norm{PQ}_\infty^2$. These have the form of the Nussbaum--Szko\l a weights \cite{NussbaumSzkola09}, with the block multiplicity included and the squared overlap maximized over the two eigenspaces. In terms of the support function \eqref{eq:support} only the directions $(s,1-s,2)$, that is, the Petz quantities $\psi(s,1)$, occur, and one recovers the Hoeffding bound (Remark~\ref{rem:hoeffding}). The third option involves $\drho+\dsig+2\chi$, i.e., the direction $(1,1,2)=2\,(\frac12,\frac12,1)$, which corresponds to $(\alpha,z)=(\frac12,\frac12)$ and hence to the fidelity. Mixtures of the three options produce exactly the directions $(\alpha,1-\alpha,2z)$ with $(\alpha,z)\in\cD$; see \eqref{eq:s-to-az}. This is how the two-parameter family of $\alpha$--$z$ quantities enters the support-function characterization. Restricting its witnesses to the Petz or sandwiched edges can lose part of the achievable region, as the example in Section~\ref{sec:example} demonstrates. The two rejection budgets enter through their product, so that in the direction $(1,1,2)$ the probability exponents $\drho+\dsig$ carry twice the weight, relative to the overlap exponent, that they carry in the Petz directions $(s,1-s,2)$.

The same picture explains the threshold $K+L=-2\log F(\rho,\sigma)$. By \eqref{eq:support-each} at $(\alpha,z)=(\frac12,\frac12)$, every sector satisfies
\begin{equation}\label{eq:disc-fid}
 \tfrac12\drho(\theta)+\tfrac12\dsig(\theta)+\chi(\theta)\ \ge\ \psi(\tfrac12,\tfrac12)=-\log F(\rho,\sigma).
\end{equation}
Hence, if $K+L\le-2\log F(\rho,\sigma)$, then every sector satisfies
\begin{equation}\label{eq:disc-rej}
 \bigl(K-\drho(\theta)\bigr)+\bigl(L-\dsig(\theta)\bigr)\ \le\ -2\log F(\rho,\sigma)-\drho(\theta)-\dsig(\theta)\ \overset{(a)}{\le}\ 2\chi(\theta),
\end{equation}
where $(a)$ follows from \eqref{eq:disc-fid}. Thus a pair on which both rejection constraints are active ($\drho<K$ and $\dsig<L$) can always be paid for by rejection, and the set $\cV_3$ of \eqref{eq:B-sets} contains no sector. Only $\cV_1$ and $\cV_2$ remain. On a sector with $\dsig\ge L$ the rejection budget of $y$ equals one, the third term of \eqref{eq:pairfn} reduces to $\pos{K-\drho}$, and the pair is paid for if and only if
\begin{equation*}
 \min\Bigl\{\pos{a-\drho},\ \pos{b-\dsig},\ \pos{K-\drho}\Bigr\}=\min\Bigl\{\pos{\min\{a,K\}-\drho},\ \pos{b-\dsig}\Bigr\}\ \le\ 2\chi ,
\end{equation*}
that is, by the type-II error budget of $y$ or by the larger of the two budgets of $x$ (for the type-I error and for rejection). This is the binary condition with $a$ replaced by $\min\{a,K\}$, and $\cV_1$ consists of the points violating it; indeed, every sector in $\cV_1$ has $\dsig>L$, because it satisfies $\drho+2\chi<K$ and hence, by \eqref{eq:disc-rej}, $L-\dsig\le2\chi-(K-\drho)<0$. This is the content of (P1), and (P2) arises in the same way from $\cV_2$. If $K+L>-2\log F(\rho,\sigma)$, sectors for which \eqref{eq:disc-fid} is (nearly) an equality can no longer be paid for by rejection alone, and condition (F) describes the resulting constraint.

\subsection{Relation to previous bounds}

The converse of \cite{JiRegula25} is $\cA\subseteq\{\text{(P1) and (P2)}\}$, and Theorem~\ref{thm:main} adds condition (F). The achievability result \cite[Proposition~6]{JiRegula25} is obtained by asymptotic pinching \cite{Hayashi02,MosonyiOgawa15,HayashiTomamichel16} and is expressed through $\psi(s,s)$, $\psi(1-s,s)$ and their reverse versions; the corresponding region is contained in $\cZ$, as it must be, and in the example of Section~\ref{sec:example} it is strictly smaller (Table~\ref{tab:example}, Fig.~\ref{fig:regions}).

Let $D(\rho\|\sigma)=\tr\rho\,(\log\rho-\log\sigma)$ denote the quantum relative entropy. Since $\psi(\cdot,1)$ is concave with $\psi(0,1)=\psi(1,1)=0$, the quotient $\psi(s,1)/s$ is non-increasing and $\psi(s,1)/(1-s)$ is non-decreasing in $s\in(0,1)$, so that, as $r\downarrow0$,
\begin{equation*}
\begin{aligned}
 H_r(\rho\|\sigma)&\ \uparrow\ \sup_{0<s<1}\frac{\psi(s,1)}{s}=\lim_{s\downarrow0}\frac{\psi(s,1)}{s}=\partial_\alpha\psi(0,1)=D(\sigma\|\rho),\\
 H_r(\sigma\|\rho)&\ \uparrow\ \sup_{0<s<1}\frac{\psi(s,1)}{1-s}=\lim_{s\uparrow1}\frac{\psi(s,1)}{1-s}=-\partial_\alpha\psi(1,1)=D(\rho\|\sigma).
\end{aligned}
\end{equation*}
Hence, by Lemma~\ref{lem:Hform} and Remark~\ref{rem:petz}, every pair $(a,b)$ with $a<D(\sigma\|\rho)$ and $b<D(\rho\|\sigma)$ belongs to $\cZ(K,L)$ for all sufficiently small $K,L>0$, and no pair with $a>D(\sigma\|\rho)$ or $b>D(\rho\|\sigma)$ belongs to any $\cZ(K,L)$. Thus the union of the regions $\cZ(K,L)$ over all $K,L>0$ is, up to its boundary, the rectangle in which both error exponents are bounded by the respective Stein exponents. This is consistent with the results of \cite{JiRegula25} for the regime in which the probability of the inconclusive outcome vanishes more slowly than exponentially.

\subsection{The optimal tests}

The tests constructed in Section~\ref{sec:proof-ach} are block diagonal with respect to the decomposition of Lemma~\ref{lem:blocks}, and inside each block they are given by the explicit formulas \eqref{eq:labels}--\eqref{eq:povm} in terms of eigenvectors of $\rho_\lambda$ and $\sigma_\lambda$ and of the budgets \eqref{eq:block-budgets}; see Remark~\ref{rem:structure}. They are collective measurements. The same construction covers all quadruples of $\cZ$, and for $K\ge a$ or $L\ge b$ it produces binary tests that attain the quantum Hoeffding bound. The polynomial factor $c_n$ in \eqref{eq:finite-n} comes from the factor $400D_\lambda^4$ of Lemma~\ref{lem:suff}, which is at most $400N_n^2$ by $D_\lambda^2\le N_n$, and the sum over at most $\sum_\lambda D_\lambda\le N_n$ eigenvectors in \eqref{eq:ach-alpha}. These give $400N_n^3$; Step~6 of Section~\ref{sec:proof-ach} contributes the remaining factor $2$. The constants have not been optimized. Whether a dimension-free pairwise bound holds for tests with rejection, analogous to the binary bound of \cite{Audenaert07} and the recent multiple-hypothesis bounds of \cite{ChengLiu26,LeeLami26}, is an interesting question. The earlier construction of \cite{Li16} is a structural precedent, but its prefactor depends on the number of distinct eigenvalues of the states, and hence in general on the dimension \cite[Theorem~2]{Li16}. A sharper rejection bound would reduce $c_n$; obtaining second-order asymptotics would additionally require a refined analysis of the sector fluctuations.

\subsection{Future Work}

\begin{enumerate}[leftmargin=*,label=\arabic*)]
\item \emph{Rejection constraints without polynomial slack.} Theorem~\ref{thm:main} determines the set $\cA$ completely, including its boundary. For the regions $\cR(K,L)$, in which the rejection probabilities must not exceed $e^{-nK}$ and $e^{-nL}$ for every $n$, Theorem~\ref{thm:fixed} shows that $\cR(K,L)=\cZ(K,L)$ when $K+L\ne-2\log F(\rho,\sigma)$. The general noncommuting fixed-budget region on the hypersurface $K+L=-2\log F(\rho,\sigma)$ is not identified by the stated bounds. Exact constraints in this case require an argument beyond the polynomial-slack construction.
\item \emph{Other regimes and tasks.} For the regime in which the rejection probability does not vanish exponentially \cite{JiRegula25}, characterizing its finite length asymptotics of two type of error exponents requires delicate analysis and new techniques. Other settings such as multiple quantum states discrimination, composite or correlated quantum hypotheses testing, and classical-quantum channel coding with an erasure option in the sense of Forney \cite{Forney68} are natural problems to which the sector method may apply.
\end{enumerate}

\section*{Acknowledgments}
This project was initiated while the author was a postdoctoral researcher at the National University of Singapore, working with Marco Tomamichel. The author is grateful to Marco Tomamichel for his encouragement, early discussions, and support during the initial stage of this work.

\section*{Use of artificial intelligence}
The author used ChatGPT 5.6 Sol (OpenAI) to assist with mathematical derivations and proof checking, numerical calculations, language editing, and LaTeX preparation. The author takes responsibility for the final manuscript, including the correctness of its mathematical results and the accuracy of its references.

\appendix
\section{Proof of Lemma~\ref{lem:Hform}}\label{app:Hform}

Let $a_K\eqdef\min\{a,K\}>0$. By Proposition~\ref{prop:psi}(d), $\psi(0,1)=\psi(1,1)=0$, so the inequality in \eqref{eq:P1} fails at $s=0$, where it reads $b\le0$, and at $s=1$, where it reads $a_K\le0$. For $s\in(0,1)$ we divide by $1-s>0$:
\begin{equation}\label{eq:appA}
 \text{\eqref{eq:P1}}\iff s\,a_K+(1-s)\,b\le\psi(s,1)\ \text{for some } s\in(0,1)\iff b\le f(s)\ \text{for some } s\in(0,1),
\end{equation}
where
\begin{equation*}
 f(s)\eqdef\frac{\psi(s,1)-s\,a_K}{1-s},\qquad\sup_{0<s<1}f(s)=H_{a_K}(\sigma\|\rho)
\end{equation*}
by \eqref{eq:Hoeffding-fn}. If the right-hand side of \eqref{eq:appA} holds, then $b\le\sup_{0<s<1}f(s)=H_{a_K}(\sigma\|\rho)$. Conversely, let $b\le H_{a_K}(\sigma\|\rho)$. The function $f$ is continuous on $(0,1)$, $f(s)\to\psi(0,1)=0$ as $s\downarrow0$, and $f(s)\to-\infty$ as $s\uparrow1$, because the numerator tends to $\psi(1,1)-a_K=-a_K<0$. Since $\sup f\ge b>0$, there are $0<s_1<s_2<1$ such that $f<b/2$ outside $[s_1,s_2]$. Hence the supremum is attained at some $s\in[s_1,s_2]$, and $f(s)=H_{a_K}(\sigma\|\rho)\ge b$, which is the right-hand side of \eqref{eq:appA}. Thus \eqref{eq:P1} is equivalent to $b\le H_{a_K}(\sigma\|\rho)$. Finally, $r\mapsto H_r(\sigma\|\rho)$ is non-increasing, so that
\begin{equation*}
 H_{\min\{a,K\}}(\sigma\|\rho)=\max\bigl\{H_a(\sigma\|\rho),H_K(\sigma\|\rho)\bigr\}.
\end{equation*}
The statement about \eqref{eq:P2} follows in the same way: with $b_L\eqdef\min\{b,L\}$, \eqref{eq:P2} holds if and only if $a\le\bigl(\psi(s,1)-(1-s)\,b_L\bigr)/s$ for some $s\in(0,1)$, i.e., if and only if $a\le H_{b_L}(\rho\|\sigma)=\max\{H_b(\rho\|\sigma),H_L(\rho\|\sigma)\}$. \qed

\end{document}